%% file: SGSOWhitePaper.tex
\documentclass[11pt,twoside,openright,a4paper]{report}
\usepackage{amsmath}
\usepackage{amssymb}

\usepackage{geometry} 
\usepackage{booktabs} 
\usepackage{array} 
\usepackage{paralist} 
\usepackage{verbatim} 
\usepackage{graphicx}
\usepackage[colorlinks,
            linkcolor=blue,
            citecolor=blue,
            filecolor=blue,
            urlcolor=blue]{hyperref} 
\usepackage{tabularx} 
\usepackage[table]{xcolor} 
\usepackage{wasysym} 
\usepackage{subfigure} 
\usepackage{enumitem}
\usepackage{xspace}

\usepackage[compress,numbers]{natbib}

\usepackage{titlesec, blindtext, color}

\titleformat{\chapter}[frame]
{\normalfont}
{\filright
\footnotesize
\enspace PART \thechapter\enspace} 
{8pt}
{\LARGE\bfseries\filcenter}

\usepackage{fancyhdr} 
\begin{document}
\newcommand{\HESS}{H.E.S.S.\xspace}
\newcommand{\degree}{$^{\circ}$\xspace}



\pagestyle{fancy}
\thispagestyle{empty}


\begin{center}
 \fontsize{50}{30}\selectfont
 Science Case for a Wide Field-of-View Very-High-Energy Gamma-Ray Observatory in the Southern Hemisphere\\
\end{center}

\vspace*{1cm}


\input{authors.tex}

\vfill
\subsection*{Abstract}
We outline the science motivation for SGSO, the Southern Gamma-Ray Survey Observatory. SGSO will be a next-generation wide field-of-view gamma-ray survey instrument, sensitive to gamma-rays in the energy range from 100 GeV to hundreds of TeV. Its science topics include unveiling galactic and extragalactic particle accelerators, monitoring the transient sky at very high energies, probing particle physics beyond the Standard Model, and the characterization of the cosmic ray flux. SGSO will consist of an air shower detector array, located in South America. Due to its location and large field of view, SGSO will be complementary to other current and planned gamma-ray observatories such as HAWC, LHAASO, and CTA.


\cleardoublepage
\tableofcontents


\chapter{Introduction and Aims of this Document} 
The exploration of the gamma-ray sky at Very High Energies (VHE; $\sim$100 GeV -- $\sim$100 TeV) has advanced dramatically over the last decade. Photons in this energy range are detected indirectly from the ground by observing the particle cascades they produce in the atmosphere, the so-called Extensive Air Showers (EASs). Other cosmic ray particles, such as electrons and atomic nuclei, also generate EASs and therefore induce a background rate against which the gamma-ray induced showers need to be discriminated. Currently, there are two ground-based techniques used in gamma-ray astronomy to observe VHE photons: 

\begin{itemize}
\item[-] One or more Imaging Atmospheric Cherenkov Telescopes (IACT) observe the Cherenkov light produced by the charged particles in the EASs while they propagate through the atmosphere before reaching the ground.  
\item[-] An Array of Shower Particle Detectors measures the particles of the EAS at ground level.
\end{itemize}
The merits and current status of these different techniques, together with direct detection measurements by satellites, will be briefly reviewed in Section \ref{sec:Instrumentation} of this document.\\

This white paper outlines the wide science program that would be enabled by the construction of a next-generation gamma-ray observatory using ground-level particle detectors in the Southern Hemisphere: the Southern Gamma-ray Survey Observatory\footnote{\url{https://www.sgso-alliance.org}} (SGSO). This science case rests on four main pillars: {\bf unveiling galactic and extragalactic particle accelerators, monitoring the transient sky at very high energies, probing particle physics beyond the Standard Model, and the characterization of the cosmic ray flux}.

 Several currently ongoing efforts to advance the technical performance of such an observatory will briefly be discussed in Section \ref{sec:design}. However, the focus of this document will be on the broad spectrum of astrophysical questions SGSO will address, rather than the details of the technical implementation of a detector. To assess the potential to specific astrophysical applications, a ``straw man" detector design is used. This design provides an increase in effective area, number of channels, and a lower energy threshold 
 compared to existing observatories such as the HAWC gamma-ray observatory. Intrinsic figures of merit that quantify the potential performance of SGSO (like angular resolution and background rates) are based on the demonstrated performance of HAWC~\cite{HAWC_CRAB}. Details on this design and its performance are given in Section \ref{sec:straw_perf}. This approach is chosen as a balance to obtain both an ambitious and feasible performance, suitable to a wide range of applications, while keeping the assumptions about the technical implementation to a minimum. It is foreseen that some applications could benefit from enhanced performance over the straw man design in certain energy ranges; the feasibility and need of reaching enhanced performance will be discussed case by case.

The outcome of the studies presented in this document will give an overview of the scientific potential of a next-generation gamma-ray observatory in the southern hemisphere and provide guidance to the finalization of the design of such an observatory. 

\cleardoublepage
\chapter{Instrumental Context}\label{sec:Instrumentation} 

\begin{figure}[h!]
  \centerline{
  \includegraphics[width=0.53\linewidth]{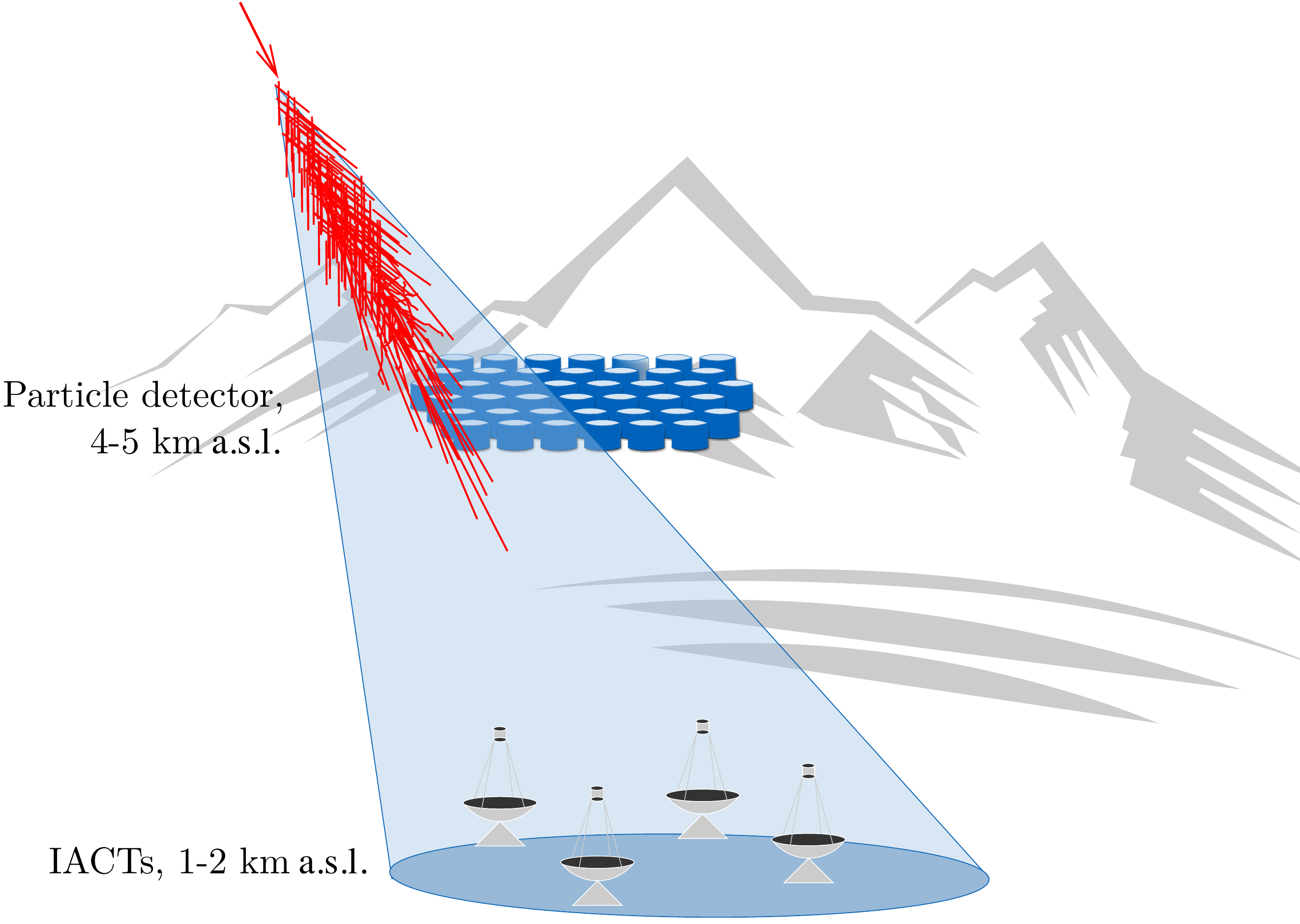}
  \hfill
    \includegraphics[width=0.47\linewidth]{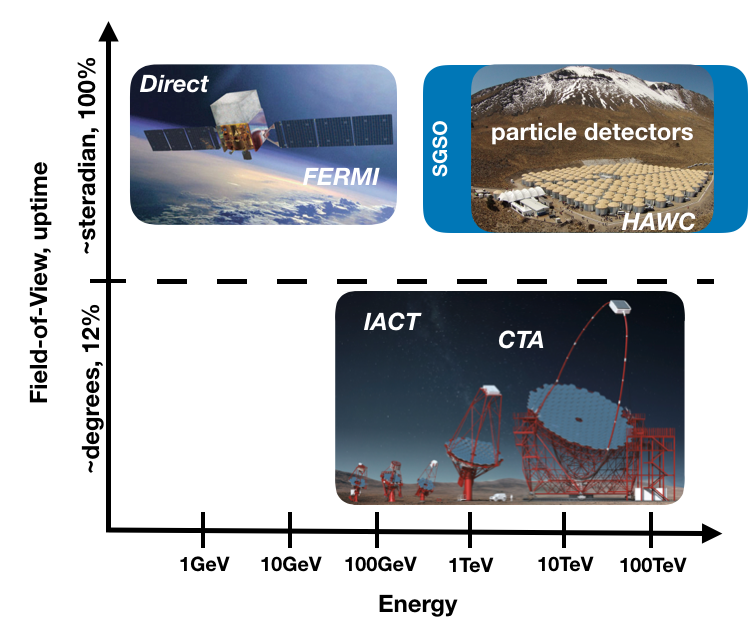}
}
  \caption{Illustration of the various complementary detection techniques of high-energy gamma rays.}
  \label{fig:Techniques}
\end{figure}

Up to $\sim$100~GeV, satellite-based instrumentation is extremely effective for the purpose of gamma-ray astronomy. At higher (VHE) energies, ground-based instrumentation is necessary due to the declining flux emitted by cosmic sources. Ground-based detectors provide collection areas up to km$^{2}$-scale by utilizing the phenomenon of extensive air showers (EASs), with the Earth's atmosphere acting as part of the detector. Two primary approaches exist: i) the detection of Cherenkov light produced by air-shower particles (using Imaging Atmospheric Cherenkov Telescopes: IACTs), and ii) the direct detection of air-shower particles at ground level. Figure~\ref{fig:Techniques} summarizes the different approaches to gamma-ray detection in terms of technology, energy range and field of view. The instruments operating in the GeV domain have very wide fields of view and high duty cycles, allowing the monitoring of the entire sky at these energies on a daily basis. At higher energies the ground-based instruments are yet to provide this coverage, with existing and under-construction wide-field instrumentation only present in the Northern hemisphere and hence with almost no sensitivity to the center of our own Galaxy and the rest of the southern sky. Limited by the telescope optics and the size of the camera, IACTs are currently limited to fields of view $<10$\degree. Concepts exist to cover larger patches of the sky with Cherenkov detectors~\cite{MACHETE,TAIGA,CTAdivergent} but such solutions imply considerable investment and remain limited by the need for dark and cloud-free conditions, resulting in a typical IACT duty cycle of 15\%. Ground-level particle detection is possible with 100\% duty cycle and is inherently wide-field in nature, with shower directions established through nanosecond-level accuracy measurement of particle arrival times. On the other hand, the precision and instantaneous sensitivity achievable with IACTs is typically much greater than ground-level particle detectors. The common challenge of both ground-based approaches is the rejection of the huge background of EASs initiated by charged, close to isotropic, cosmic rays. The complementarity of both techniques is summarized in Table~\ref{tab:comparison}.

\noindent

Until recently, IACTs dominated the field of ground-based gamma-ray astronomy with three major facilities providing coverage of both hemispheres: \HESS, MAGIC and VERITAS (see e.g.~\cite{NewAstRevHinHof}). IACTs provide a view of the whole shower in the atmosphere, enabling a calorimetric shower energy estimate and extremely effective background rejection through image parameters~\cite{HillasOrig} and image template fitting~\cite{lebohec}. Historically, the ground-particle approach has been hampered by limited access to the domain below $\sim$100~TeV, where gamma-ray source emission is established, and by the difficulty of background rejection. Classical ground-particle arrays had fill-factors (ratio of sensitive instrumented area to total array area) of $\ge$1\%. The breakthrough in the use of this approach for gamma-ray astronomy is the combination of high fill-factors ($>$50\%) and high altitude ($>$4~km), to allow detection at $\sim$1~TeV energies and provide background rejection power through the measurement of shower substructure at the ground and/or identification of muons. HAWC is the first truly competitive instrument based on this technique for TeV gamma-ray astronomy as demonstrated by the recent survey of the northern sky with the first year of HAWC data~\cite{HAWCsurvey}. Building on the success of HAWC (which followed the pioneering efforts of MILAGRO \cite{MILAGRO}), a major facility for cosmic ray studies and gamma-ray astronomy is now under construction in the Sichuan province of China (LHAASO~\cite{LHAASO}). Simultaneously, the global IACT community is engaged with the construction of the Cherenkov Telescope Array (CTA), which brings an order of magnitude sensitivity improvement with respect to current IACTs, greater precision, and access to both hemispheres~\cite{CTA_ScienceTDR}.

\begin{table}[!thp]
\caption{Comparison of typical performance of current and planned IACT arrays and ground particle arrays for gamma-ray astronomy. For IACTs, 50 hours/year are assumed to be available to target a specific point-like source. For the IACT duty cycle, the range given corresponds to astronomical darkness at a modest quality site to observations up to full moon at an excellent site.}
\begin{center}
\begin{tabular}{|l|c|c|}
\hline
 & IACT Arrays & Ground-particle Arrays \\
\hline
Field of view & 3\degree--10\degree  & 90\degree \\
Duty cycle & 10\%--30\% & $>$95\%\\
Energy range &30~GeV -- $>$100~TeV& $\sim$500~GeV -- $>$100~TeV \\
Angular resolution & 0.05\degree--0.02\degree & 0.4\degree--0.1\degree \\
Energy resolution & $\sim$7\% & 60\%--20\%\\
Background rejection & $>$95\% & 90\%--99.8\%\\
\hline
\end{tabular}
\end{center}
\label{tab:comparison}
\end{table}%

\cleardoublepage

\chapter{Straw Man Design for Science Case Studies}
\label{sec:straw_perf}
In order to perform case studies for different scientific objectives, the performance of a simulated observatory is used. The code used in the following calculations is made publicly available\footnote{\url{https://github.com/harmscho/SGSOSensitivity}}. The aim of this next generation observatory is to achieve roughly an order of magnitude higher sensitivity over the current generation instruments like HAWC.

The improvement by this {\it{straw man}} design is obtained mainly by increasing the size, density, and altitude of the observatory, limited by assumptions about the technical implementation. However, it is foreseen that specific hardware developments might have a positive effect on gamma-ray detection efficiency, background rejection power and angular resolution. These different concepts will be briefly discussed in Section~\ref{sec:design}.
We take a pragmatic approach in assessing the sensitivity of the {\it{straw man}} design of the instrument. For this, the published HAWC performance figures \cite{HAWC_CRAB}, like angular resolution, gamma and hadron cut passing rates, are taken. To assess the performance of an observatory of different design, we are using air showers simulated using CORSIKA \cite{corsika}, and a toy-detector design. We generated a database of gamma ray and proton induced air showers distributed in the energy range from 20\,GeV up to 500\,TeV. The particles from these air showers are evaluated at ground-level by an array of idealized detector units. These detector units integrate the total energy carried by photons, electrons, and positrons that pass through a squared area. We will refer to this energy as the electromagnetic energy. In addition, we count units that are hit by one or more muons. Using these toy detector units, the following properties are derived for each simulated air shower:
\begin{description}
\item[Detected energy, $E_{\text{det}}$] The sum of electromagnetic energy observed by all units that passed a unit detection threshold.  
\item[Number of detection units, $N_{\text{det}}$] The sum of the number of the units that have recorded electromagnetic energy above a certain threshold or have at least one muon passing through. 
\end{description}
The performance figures of HAWC were provided as a function of analysis (i.e. energy proxy) bins, which correspond to the fraction of total number of sensors that recorded a signal. Each of these analysis bins contains a distribution of possible gamma-ray energies. To link these analysis bins to a typical gamma ray energy, we selected the most likely value from the distributions published in \cite{HAWC_CRAB}. For these typical gamma-ray energies, we find the average $E_{\text{det}}$ for a toy-detector with properties (fill-factor, area, and elevation) the same as the HAWC observatory. To assess the performance of an observatory with different configurations, the performance figures of HAWC expressed as a function of $E_{\text{det}}$ are applied. Since the straw man design has significantly more instrumented area and is higher in elevation than HAWC, it will detect more energy for a given gamma ray. This is a conservative approach, in which the details of the detector unit design do not impact the overall performance of the observatory. To explore more optimistic (and probably more realistic) scenarios, additional scaling factors on background rejection and angular resolution can be applied.

\begin{figure}[ht]
  \begin{center}	
    \includegraphics[width=0.38\linewidth]{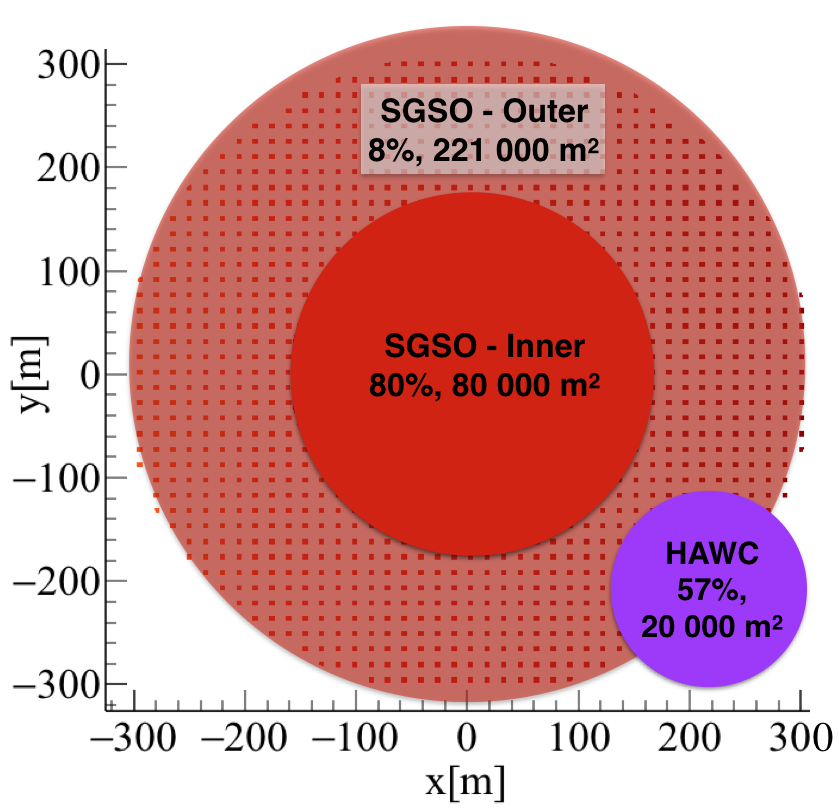}
    \includegraphics[width=0.49\linewidth]{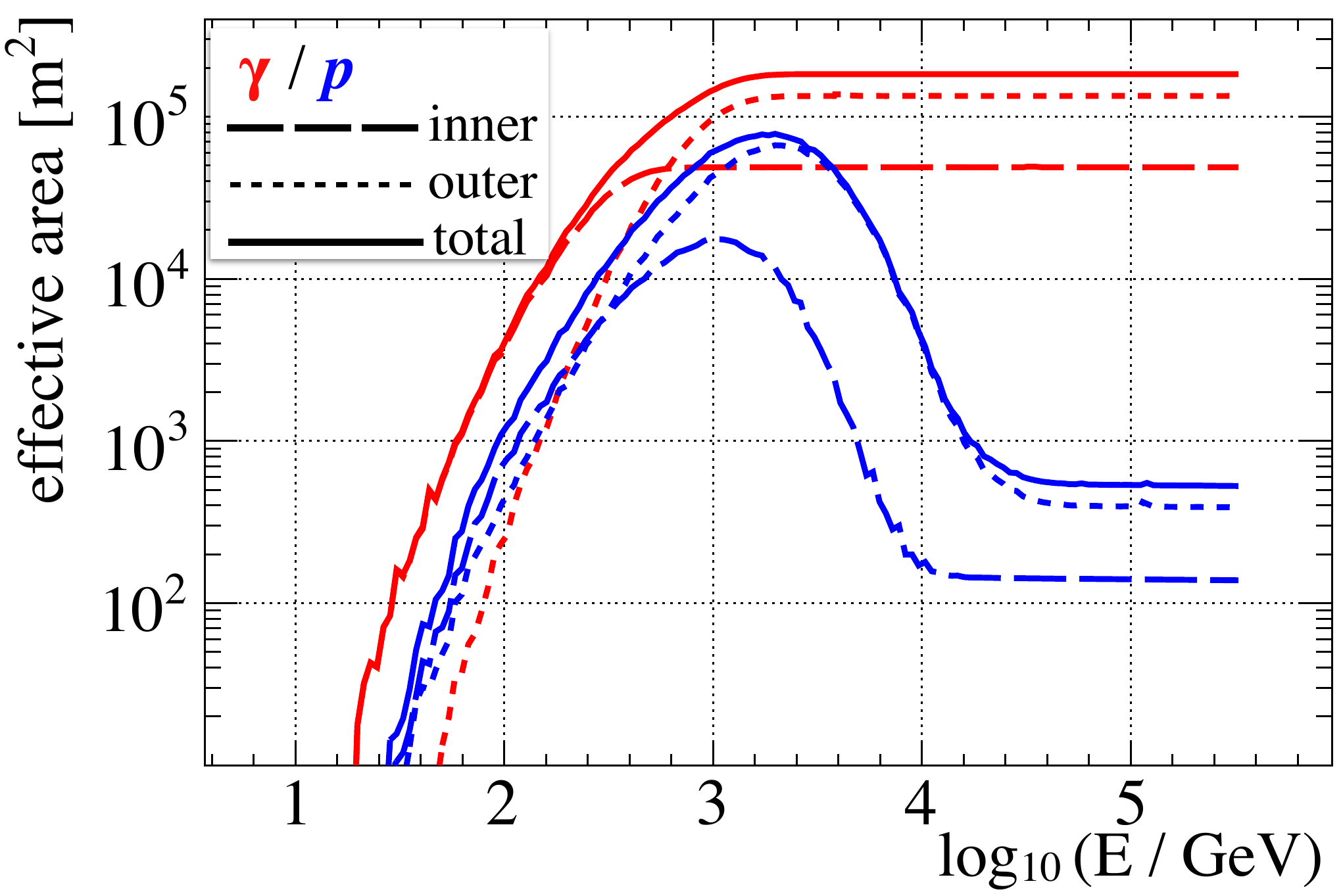}
   \end{center}
  \caption{{\em Left:} Straw man design of SGSO indicating instrumented areas and their fill factors. {\em Right:} Effective area as a function of energy of the primary particle (gamma ray or proton) after applying gamma-hadron separation and trigger multiplicity cuts. }
  \label{fig:SGSOScale}
\end{figure}
\noindent
To explore the science case for SGSO, a toy observatory has been designed with the following parameters:
\begin{itemize}
\item 5000\,m elevation above sea level. 
\item latitude 25$^{\circ}$ South
\item 4\,m$\times$4\,m units, with an electromagnetic energy threshold of 50\,MeV.
\item A dense array with 4000 units covering an area of 80,000\,m$^2$ with a fill-factor of 80\%.
\item A sparse array with 1000 units covering an area of 221,000\,m$^2$ with a fill-factor of 8\% .
\end{itemize}
A sketch of the configuration, and size comparison with the HAWC-like toy detector, is shown in  the left panel of Figure~\ref{fig:SGSOScale}. The total number of electronic channels is roughly a factor of four higher than HAWC (we assumed one channel in a smaller detector unit, while HAWC has four in a bigger unit). In the right panel of Figure~\ref{fig:SGSOScale}, the effective area after selection cuts for gamma-ray like events is shown for gamma ray and proton induced air showers. 

\begin{figure}[!ht]
  \begin{center}	
    	\includegraphics[width=0.47\linewidth]{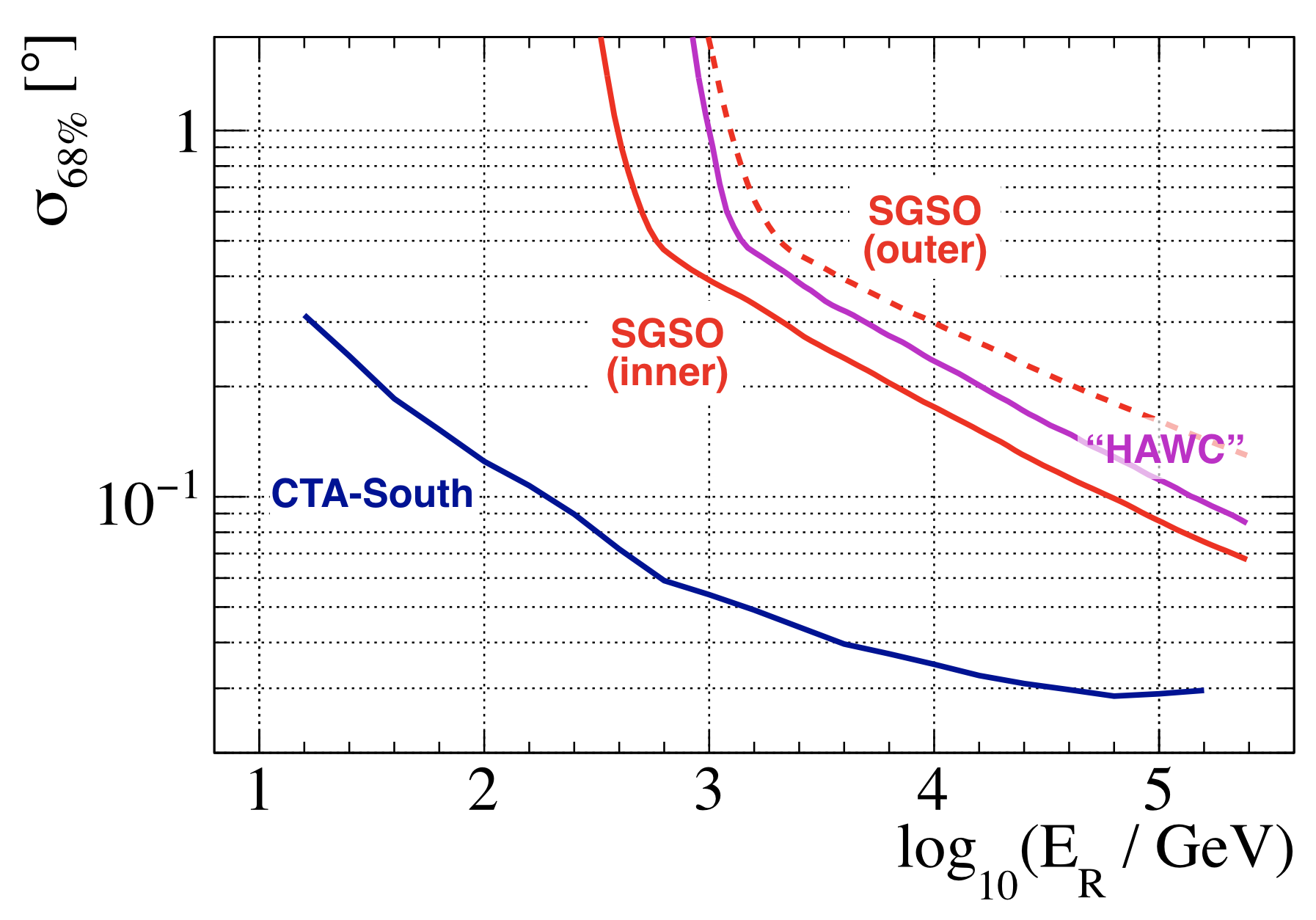}
	    \includegraphics[width=0.49\linewidth]{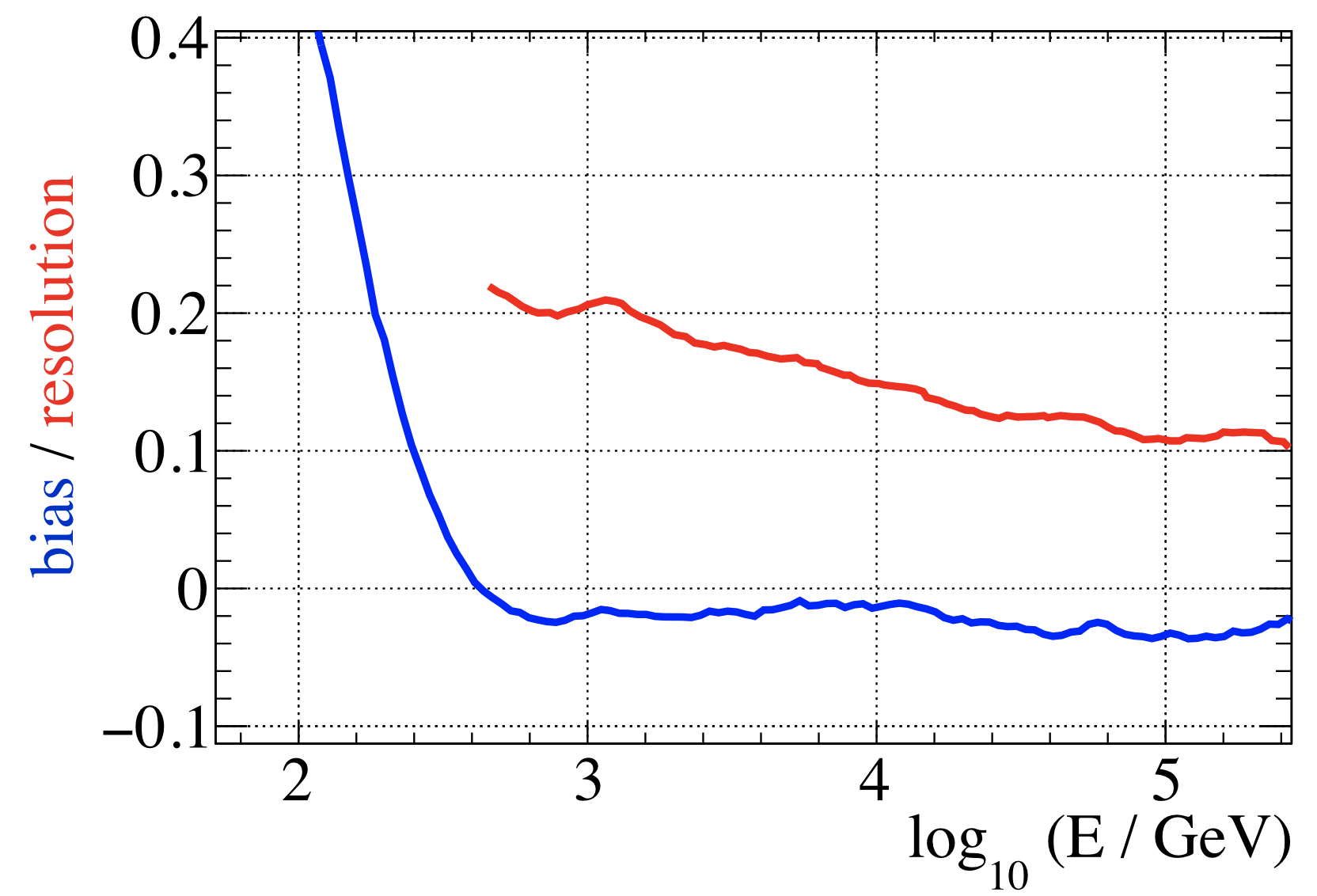}
   \end{center}
  \caption{{\em Left:} SGSO angular resolution, defined as the 68\% containment area of gamma rays from a point source, compared to CTA and HAWC. {\em Right:} Energy bias and resolution, where bias is defined as mean value of $\Delta = (\log_{10}{E_{\text{R}}} - \log_{10}{E_{\text{T}}}$) (with reconstructed energy  $E_\text{R}$ and true energy $E_\text{T}$), while the resolution in taken as the root mean square of $\Delta$.}
  \label{fig:SGSOperformance}
\end{figure}
\noindent
The assumed angular resolution and energy resolution of the straw man design are given in Figure~\ref{fig:SGSOperformance}. The improvement on angular resolution versus the HAWC-like observatory comes from the increase of detected energy for a given gamma ray. For the energy resolution, shown in the right panel of Figure~\ref{fig:SGSOperformance}, a very simple energy estimator is used that relates the amount of detected energy directly to the gamma-ray energy. This simple estimate might not be the most optimal energy resolution, but is comparable in performance to algorithms currently under development within the HAWC collaboration~\cite{HAWC_Energy}.  

\begin{figure}[!t]
  \begin{center}	
    	\includegraphics[width=0.58\linewidth]{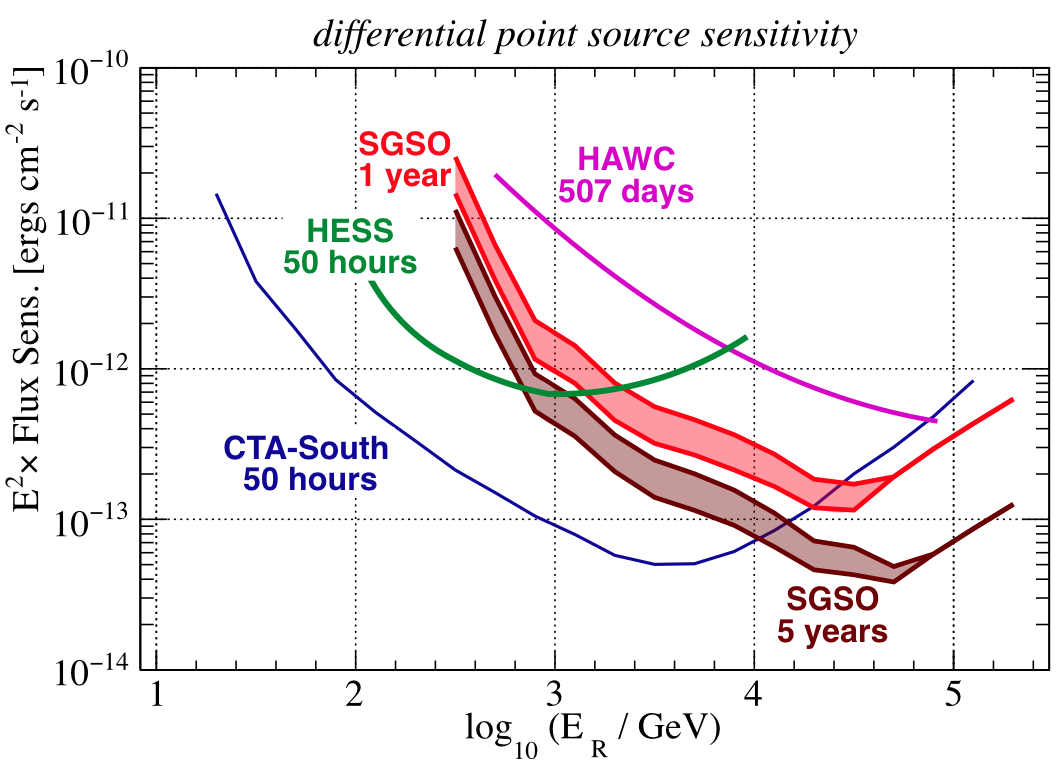}
	    \includegraphics[width=0.41\linewidth]{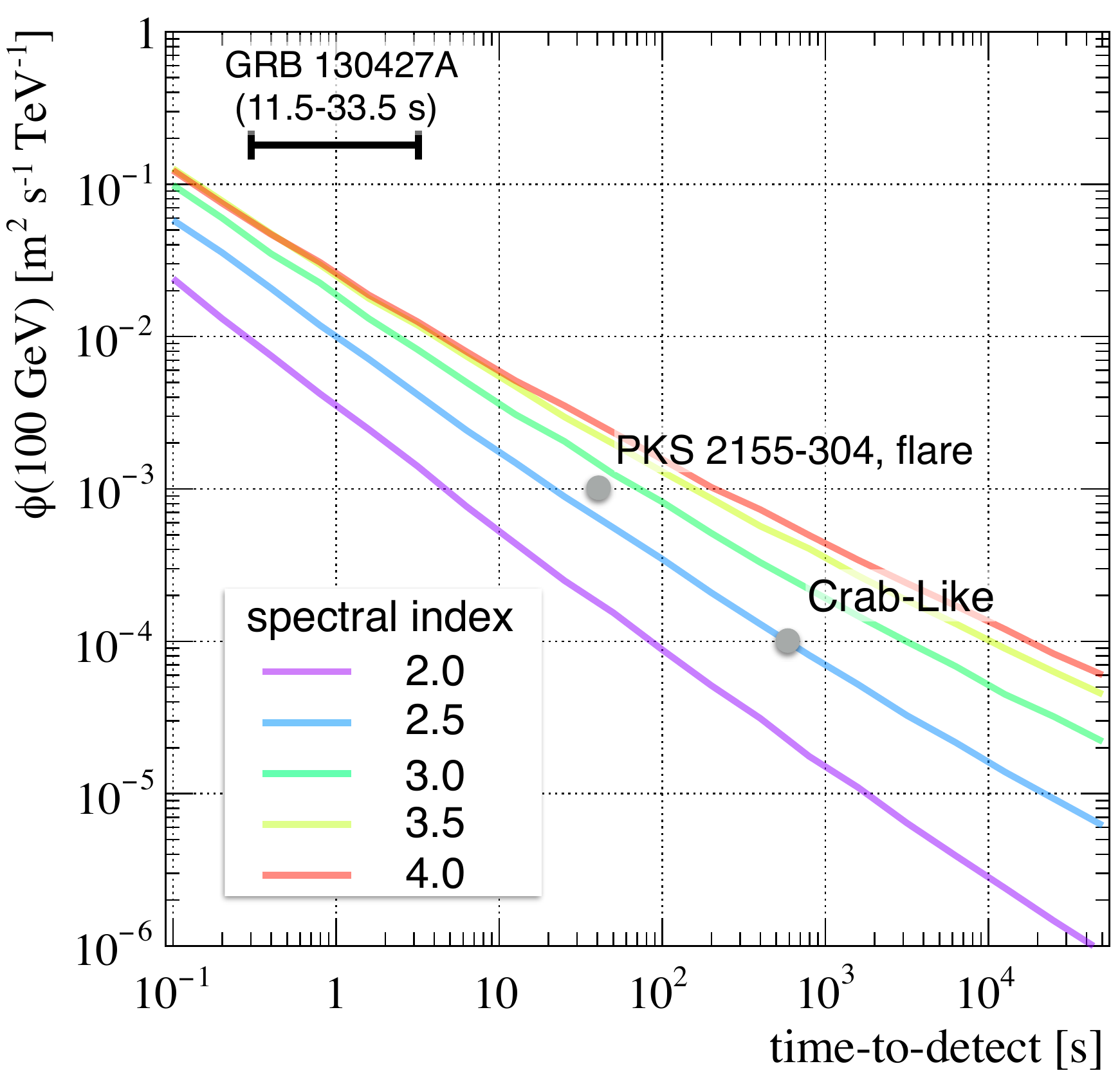}
   \end{center}
  \caption{{\em Left:} Comparison of differential point-source sensitivity as a function of reconstructed gamma-ray energy for several ground-based gamma-ray observatories in the southern hemisphere (see text for details). {\em Right:} Time needed for a $5\sigma$ detection of a point-source with a given flux (evaluated at 100~GeV). The lines indicate the detection times for sources  whose spectra follow a simple power-law behavior. In addition, the time-to-detect for a source with a Crab-like like spectrum, a bright flaring active galactic nuclei (PKS~2155-304   \cite{2007ApJ...664L..71A}), and the brightest \emph{Fermi}-LAT detected gamma-ray burst (GRB~130427A~\cite{highenergyphoton}) are indicated (see text for more details).}
  \label{fig:SGSOSens}
\end{figure}

 In the left panel of Figure~\ref{fig:SGSOSens}, the differential point source sensitivity of the straw man design is compared to that of CTA, HAWC and H.E.S.S. The sensitivity is estimated for a steady point source at a zenith angle of 20\degree and the assumption that we can observe it for 6 hours per day. This roughly corresponds to the scenario of a source transiting straight overhead. For the SGSO sensitivity we show a band bracketing two scenarios: the top of the band corresponds to the established background rejection and angular resolution of the HAWC observatory, while the bottom of the curve corresponds to a  gamma-ray angular resolution scaled by a factor 0.8, and the hadron-shower passing rate by 0.5 compared to the default HAWC-like performance. The energy-range of the point-source sensitivity is limited on the lower side due to loss of energy resolution at few hundred GeV (Figure~\ref{fig:SGSOperformance}). However, from the effective area curves in Figure~\ref{fig:SGSOScale} it is clear that we still have significant gamma-ray effective area below the energy at which the differential sensitivity figure stops. This low energy performance is important to observe transient and flaring objects, therefore we show additionally in the right panel of Figure~\ref{fig:SGSOSens} the time needed for a $5\sigma$ detection under different spectral assumptions. We show the flux normalization at 100\,GeV that is needed to yield a detection within a given duration for sources that follow a simple power law spectrum (as indicated by the lines). In addition we calculate the time to detection for three examples of bright gamma-ray sources with different spectra:
 \begin{enumerate}
     \item A bright steady source with a spectrum like the Crab Nebula.
     \item A flare with the spectrum corresponding to the bright flare of PKS~2155-304 \cite{2007ApJ...664L..71A} (here modeled with a hard cutoff in the spectrum above 1~TeV).
     \item An example of a gamma-ray burst extrapolated from the \emph{Fermi}-LAT spectrum of GRB~130427A~\cite{highenergyphoton} measured between 11.5--33.5~s after the start of the burst. The attenuation of the gamma-ray flux due to its interaction with the extragalactic background light has been modeled assuming a redshift of z = 1~\cite{2011MNRAS.410.2556D}. The indicated time range needed to detect the burst reflects a range of assumptions, from nominal performances to a PSF worse by a factor of 2 and no background rejection (see Section~\ref{sec:GRB}). 
 \end{enumerate}
 
\cleardoublepage

\chapter{Unveiling Galactic Particle Accelerators} 

The study of the Galactic Plane with VHE gamma rays is one of the most powerful methods available to search for astrophysical particle accelerators. Objects such as supernova remnants and pulsar wind nebulae are likely to be the sources of the very high-energy electrons, positrons, and hadrons which fill the Galaxy. An unknown number of these sources (or a collective behaviour of an ensemble of them) are pevatrons, able to accelerate hadrons up to the ``knee'' of the cosmic-ray spectrum. Complementary evidence for particle acceleration may also be found by searching for VHE counterparts to the \emph{Fermi} bubbles, or by looking for VHE gamma rays produced when Galactic cosmic rays interact with giant molecular clouds.

\begin{figure}[t!]
  \centerline{
  \includegraphics[width=0.95\linewidth]{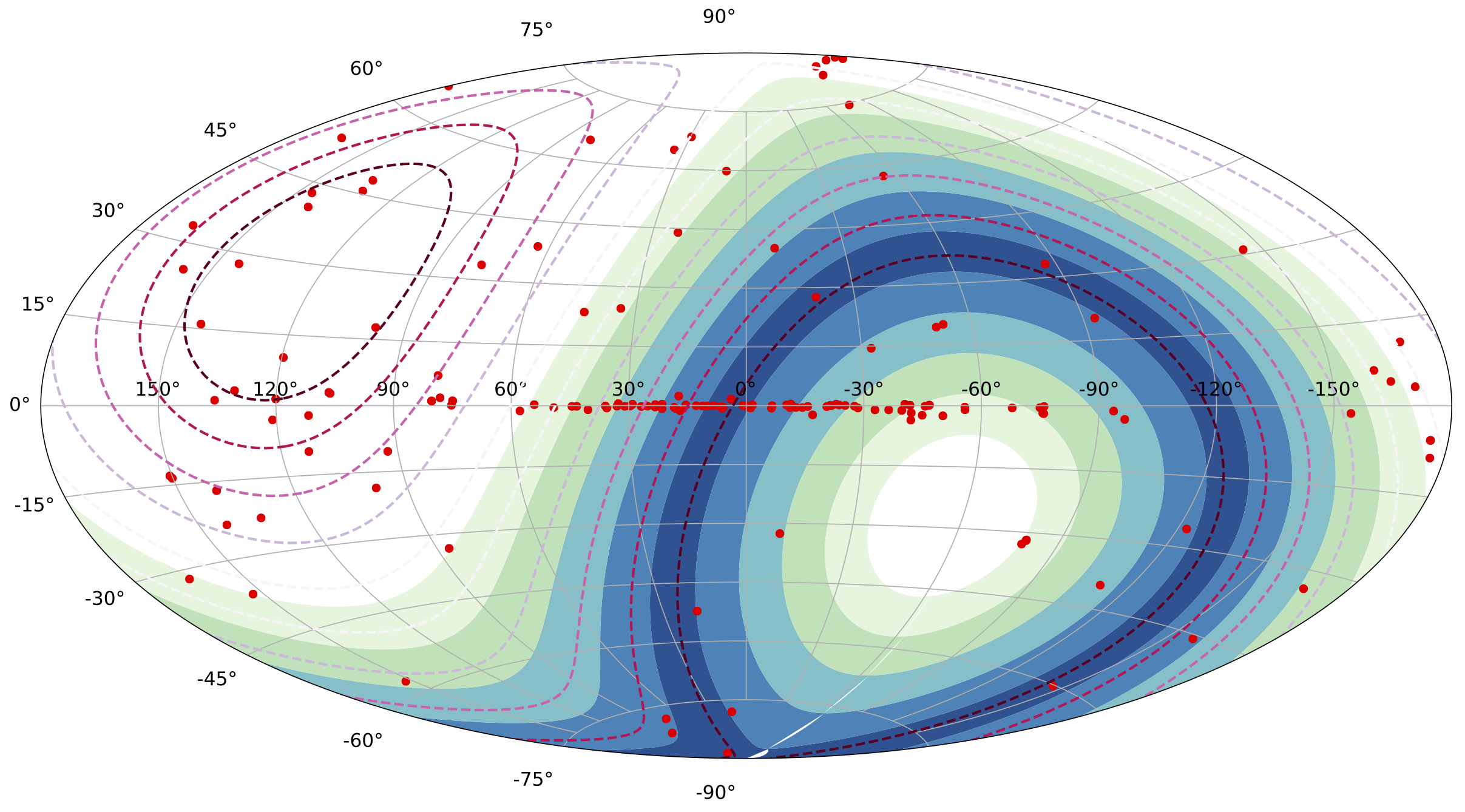}
}
  \caption{Sky-map in galactic coordinates showing the complementarity of the visibility ranges between HAWC and SGSO. The color bands correspond to 10$^{\circ}$  in zenith angle (up to 45$^{\circ}$) for SGSO at latitude 25$^{\circ}$ South. The dashed contours show the same for HAWC, with the darkest line marking the edge of the field-of-view at 45$^{\circ}$ from zenith. The planned LHAASO experiment\cite{LHAASO} (not indicated) will be 10\degree further north compared to HAWC). The red markers correspond to TeV gamma-ray sources discovered by \HESS, MAGIC and VERITAS (data from \url{http://gamma-sky.net}).
  }
  \label{fig:SkyVisibility}
\end{figure}

Searches for VHE gamma rays in the Galaxy are complicated by considerable observational difficulties, including source confusion, diffuse gamma-ray backgrounds, and spatially extended regions of emission. The design of SGSO is well suited to handle these complications. With its high uptime and wide field of view, the instrument is capable of making precision measurements of the background of isotropic cosmic rays as well as very extended gamma-ray emission, complementing the sensitivity to point sources from IACTs. And with its excellent sensitivity above 10~TeV, SGSO will isolate Galactic pevatrons from the background of softer gamma-ray sources. The detector will identify hard-spectrum sources for long-term observations with CTA, and will complement the coverage of the Northern Hemisphere with HAWC (Figure~\ref{fig:SkyVisibility}).  

\noindent

\section{Pulsar emissions to measure diffusion coefficients and constrain the positron flux at the Earth} 

Positrons are a part of the cosmic-ray sea of particles that strike the Earth. They are thought to be secondaries produced in cosmic ray interactions with the interstellar medium. Their relative abundance with respect to electrons is supposed to decrease with energy, contrary to what has been measured by different experiments like PAMELA~\cite{PhysRevLett.111.081102} and AMS~\cite{PhysRevLett.110.141102} above a few GeV. This anomaly, or positron excess, can be explained if there are nearby sources injecting primary positrons. Pulsars have been proposed to be the origin of this excess. Since e$^\pm$ are charged particles, they are deflected by magnetic fields and their arrival directions do not point back to their source. On the other hand, neutral particles like gamma rays are expected to be co-produced at the acceleration site of the positrons. Studying gamma rays from potential positron sources therefore provides crucial input to the positron excess puzzle.

Amongst the known pulsars, Geminga and PSR~B0656+14 are two of the best candidates due to their age and distance to Earth, which favour the production of a particle halo that can reach the observatory. However, the all-sky observation mode of SGSO will also allow testing large halos from other plausible energetic pulsars, including millisecond pulsars~\cite{2015ApJ...807..130V}. HAWC has recently demonstrated that under the assumption of an isotropic and uniform diffusion coefficient from these pulsars to the Earth, they are unlikely to be the main contributors of this excess~\cite{Geminga_HAWC}. 

\begin{figure}[t!]
\begin{center}
\includegraphics[width=0.75\linewidth]{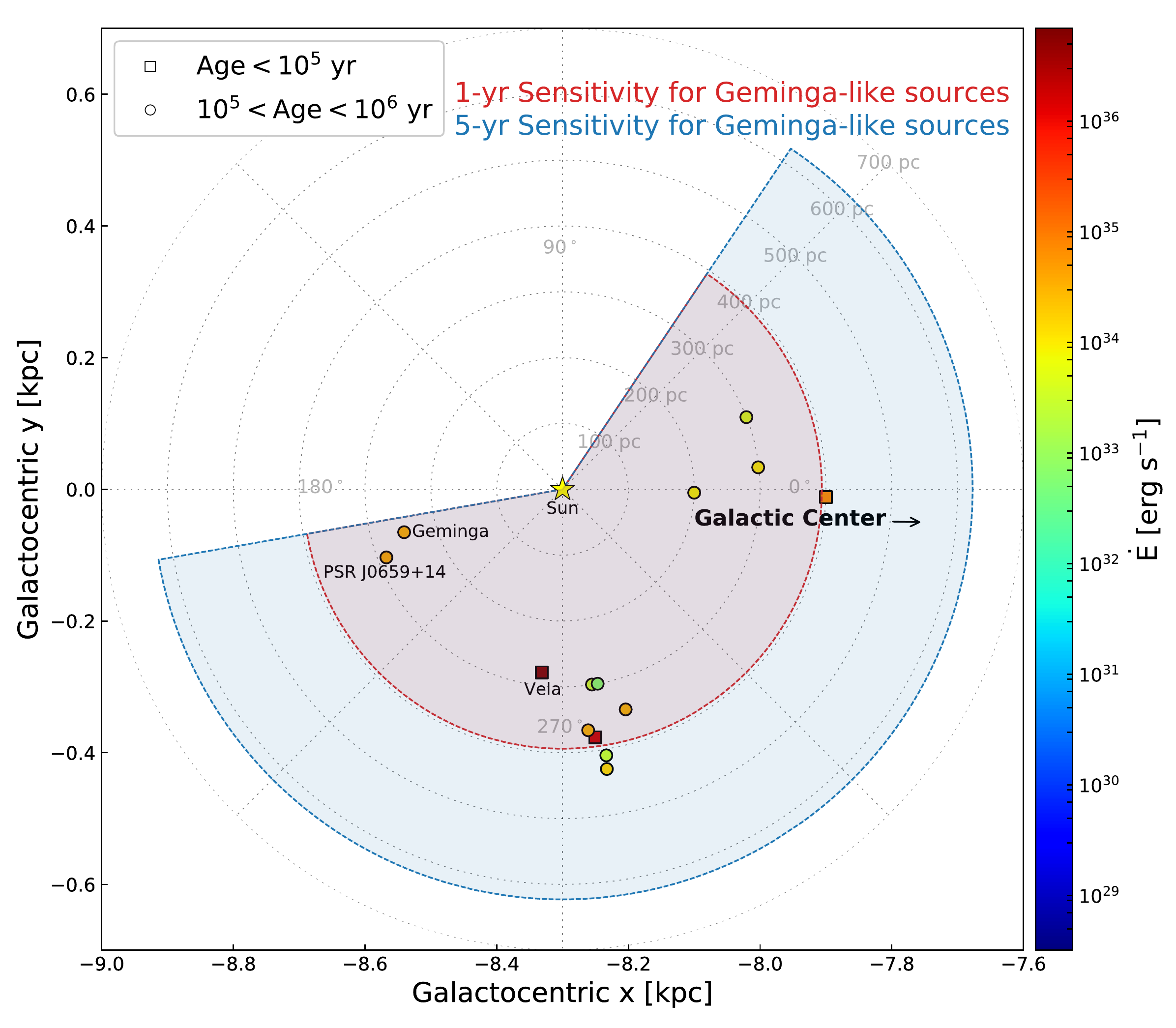}    
  \caption{Pulsars with age $<10^6$ yr within 500 pc. The color scale indicates the spin-down power of the pulsar and the shape of the marker its age. The red shaded area indicates the SGSO 1 year sensitivity for sources with the same luminosity as Geminga and the blue shaded one the 5-year sensitivity.}  \label{fig:pulsars_southern}
  \end{center}
\end{figure}

The characterization of the mechanisms involved in particle propagation in these regions is also very important to unveil the properties of the accelerators and the medium around them. This can be done by measuring the surface brightness profiles of these extended regions. On the other hand, establishing that pulsars are not the origin of the positron excess could point to more exotic processes such as dark matter particle annihilation or decay. To unveil the origin of the positron excess is thus one of the most important questions to be solved in astroparticle physics nowadays.

If we assume propagation with the typical diffusion coefficient in the vicinity of the Earth, the highest-energy positrons detected by AMS should come from sources located at a distance $<$500~pc. If we assume that these sources are pulsars, it is reasonable to make a cut on an age $<$10$^6$~yr for them to still produce enough electrons and positrons.
The only known pulsars visible from the Northern Hemisphere with age $<$10$^6$~yr and located within a distance $<$500~pc from the Earth are the two that HAWC observed. For an observatory on the Southern Hemisphere, a larger number of pulsars with these characteristics is accessible due to the exposure to a much larger part of the Galaxy. In Figure \ref{fig:pulsars_southern} we show all known pulsars with age $<$10$^6$~yr in galactocentric coordinates. The figures also illustrates the sensitivity of a Southern array to Geminga-like sources. This sensitivity was extrapolated from that reached by HAWC in the Geminga observations and is the one needed to reach a $5\sigma$ significance for a source of the same extension as Geminga. This calculation does not take into account the declination of the sources, which would change the sensitivity due to the different exposure of the source at different declination.

\begin{figure}[t!]
  \centerline{
  \includegraphics[width=0.75\linewidth]{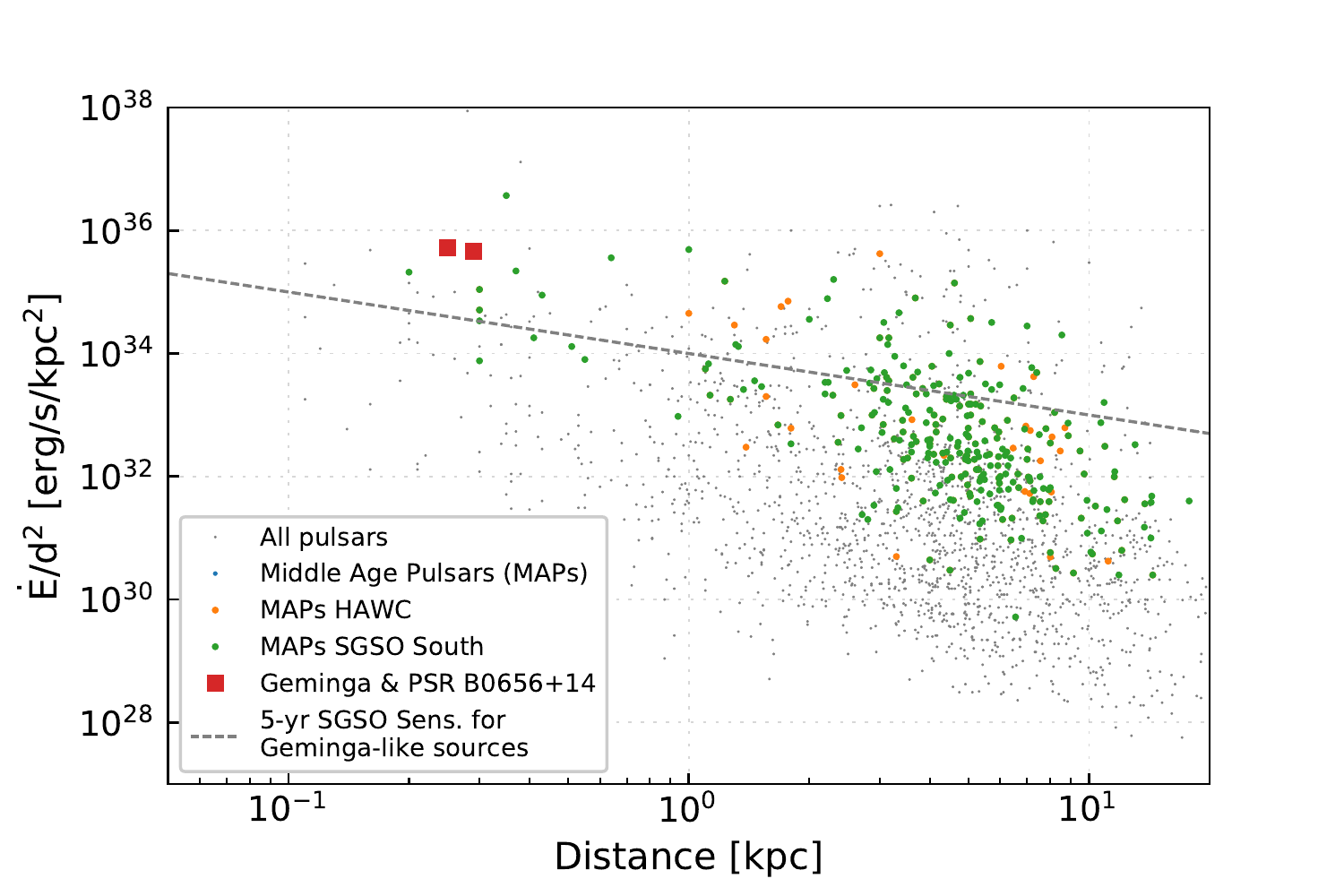}
}
  \caption{Pulsar spin-down power divided by d$^2$ as a function of the distance. Green points are the MAPs accessible by SGSO and orange points those accessible by HAWC. Grey points indicate the rest of the pulsars from ATNF. The gray line indicates the sensitivity to detect these sources if they produce a TeV halo similar to that produced by Geminga. This sensitivity does not take into account the limiting angular resolution of the instrument.}
  \label{fig:sensitivity_distance_pulsars}
\end{figure}

In Figure \ref{fig:sensitivity_distance_pulsars} we show SGSO's sensitivity to Geminga-like sources as a function of their distance. We include all the pulsars from the ATNF catalog \cite{Manchester:2004bp} and mark in green the middle-aged pulsars (MAPs, age $<10^6$ yr) in the reach of a Southern gamma-ray observatory.  We assume that the luminosity of these \emph{Geminga-like} sources scales linearly with their spin-down power, and that their intrinsic size is that of Geminga. Since the gamma-ray and background fluxes are reduced by the distance ($d$) squared, the SGSO  sensitivity scales like $\dot{E}/d^2 \times d =\dot{E}/d$. This scaling works out to distances where the halo size is large compared to the angular resolution of SGSO.

SGSO will be able to complete the survey of known nearby pulsars with age $<10^6$ yr. If all these pulsars have an emission similar to Geminga and PSR B0656+14, we will be able to increase the number of similar-detected sources from two to about ten. Also important may be the sources whose pulsars are not pointing towards the Earth as it may be the case for the source (or sources) explaining the positron excess at the Earth \cite{lopez18}. The emission from these sources will also be detected by SGSO~\cite{linden17}.

\section{\emph{Fermi} bubbles} 

\begin{figure}[t!]
  \centerline{
  \includegraphics[width=0.5\linewidth]{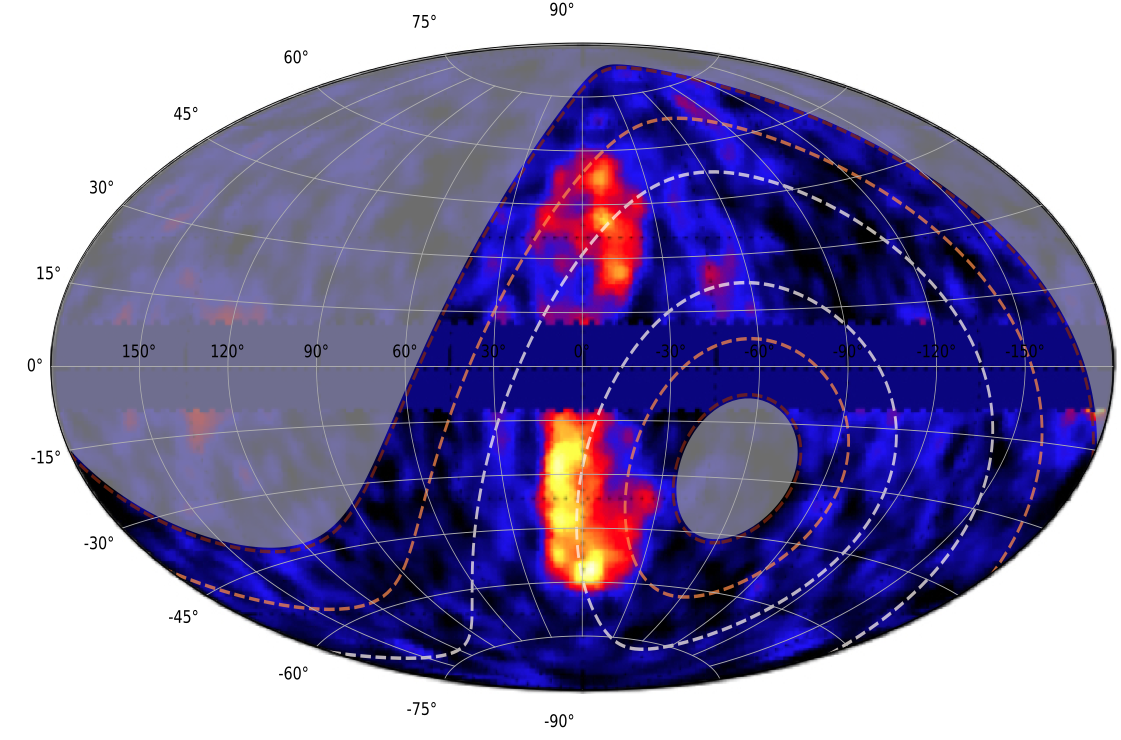}
  \includegraphics[width=0.45\linewidth]{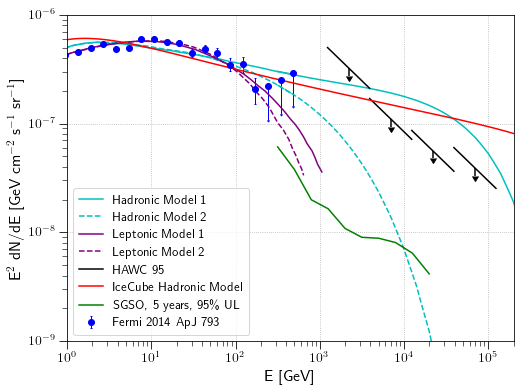}
}
  \caption{{\em Left:} Field-of-view of SGSO overlaid on the \emph{Fermi} bubbles (Credit:\emph{Fermi}-LAT team, A. Franckowiak, D. Malyshev), the dotted lines indicate steps of 15$^\circ$ from zenith. {\em Right:} SGSO differential sensitivity from the straw man design to detect the Northern \emph{Fermi} Bubble compared to different emission scenarios ~\cite{HAWC_FermiBubbles,Fermi-LAT:2014sfa,Lunardini:2014prd}.}
  \label{fig:FermiBubbles}
\end{figure}

Bubble-like structures, both above and below the Galactic plane (as illustrated in Figure~\ref{fig:FermiBubbles} (left)), have been detected by both radio/microwave \cite{Carretti} and gamma-ray \cite{Fermi-LAT:2014sfa} instruments, in the energy bands $10^{-5}-10^{-4}$~eV and $10^{8}-2\times 10^{11}$~eV, respectively.

These giant structures extend away from the Galactic plane in the latitude range $\sim 10^{\circ}-50^{\circ}$, with an approximately constant surface brightness at latitudes above $10^{\circ}$. The edges of the bubbles, and the rapidity with which they spatially drop off at the edge region, remain somewhat unclear, being constrained (at best) to cut-off within an angular range of $\sim 5^{\circ}$. Provided that there are sufficiently bright features in the bubble at very high energies, SGSO's anticipated angular resolution (less than half a degree above 1\,TeV), will be sufficient to characterize them considerably more accurately than that achievable by \emph{Fermi}-LAT presently, aiding future follow-up observations of the edge regions with CTA.
The latitude-averaged spectrum of the bubbles is shown in Figure~\ref{fig:FermiBubbles} (right), together with sensitivity of the SGSO straw man design, HAWC flux upper-limits and several model predictions. In the case that a spectral softening, or cut-off, occurs above $\sim$500\,GeV, the straw-man design has the sensitivity to characterize this spectral feature.  
With good sensitivity down to 300~GeV and  $\sim$70\% of the bubble regions transiting daily within 25$^\circ$ from zenith, SGSO will probe the continuation of the bubble spectrum up to higher energies. Furthermore, this fortuitous energy and spatial coverage of the bubble regions will also allow a deeper probe of their structures. 
At lower latitudes, the connection of these bubbles to the brighter Galactic nucleus (from where they are expected to originate) remains unclear \cite{Casandjian:2015ura}. However, tentative evidence of a brighter, harder spectral behaviour without indications of a cutoff (up to 1 TeV) at these latitudes have been reported \cite{Fermi-LAT:2014sfa}. Furthermore, the edges of the Southern bubble at VHE may be studied in order to investigate its morphology and the edge sharpness in this energy range. A key challenge for the study of the \textit{Fermi} Bubbles with SGSO will be the characterization of the background over large angular scales in order to accurately quantify the gamma-ray excess. 
From the experience of the HAWC observatory \cite{HAWC_FermiBubbles}, three main factors determine the background discrimination at these large scales. The first one is the ability to distinguish between hadronic cosmic rays and gamma rays. The second is to measure the isotropic cosmic ray and gamma ray fluxes. The third is removing effects from the large-scale cosmic ray anisotropy which can still be present due to imperfect gamma-hadron separation (See section \ref{subsec:cranis}). It is anticipated that SGSO will achieve significant improvement in background rejection and characterization over HAWC. 

\section{Pevatrons, cosmic rays and supernova remnants} 

The search and study of pevatrons, i.e. sources capable of accelerating particles up to $10^{15}$~eV, is a major key science project of the very-high-energy community. 
The study of pevatrons is tightly connected to the study of cosmic rays (CRs). CRs are charged particles, filling the Galaxy with an energy density comparable to the energy density of the interstellar magnetic field or the thermal ISM energy density, but whose origin is still unknown. 

Several decades of measurements showed that, up to the \emph{knee} of the cosmic-ray spectrum ($\approx$ 1--3 PeV~\cite{antoni2005}) the CRs arriving at the Earth are mainly protons, and their differential spectrum follows a power-law in energy very close to $E^{-2.7}$ before steepening to $E^{-3.1}$ at energies above the \textit{knee}. 
Natural questions of CR physics are: What sources can produce the observed features of the CR spectrum? and moreover: What sources can accelerate protons up to PeV energy, therefore being \textit{pevatrons}? 

The study of spatial and temporal confinement of CRs indicates that up to the $\textit{knee}$, the CRs must be of Galactic origin, and, that to sustain the level measured, the sources must inject a typical power of the order of $10^{41}$ erg/s~\cite{hillas2005}. Supernova remnants (SNRs) have for long been proposed as sources of Galactic CRs, because they can somewhat satisfy these requirements. Indeed, the power requirement can be satisfied assuming that some fraction (typically $\sim$10\%) of the kinetic energy of SNR shockfronts is converted into CRs. In addition, diffusive shock acceleration mechanisms are compatible with the measured differential spectrum of CRs~\cite{drury1983}, and finally, magnetic field amplification, expected in diffusive shock acceleration theoretically allows for the acceleration of protons up to the knee~\cite{bell2004}. 

It has been understood that the acceleration of protons to relativistic energies should be accompanied by the production of gamma rays~\cite{drury1994} due to the decays of neutral pions, produced in interactions of cosmic ray protons with protons of the ISM. The detection of several SNRs in TeV gamma rays thus came as a necessary condition to the SNR hypothesis~\cite{HESSSNRCAT}. However, these detections cannot be seen as a definitive proof of proton acceleration in SNRs, because the observed TeV gamma-ray emission can also be produced by relativistic electrons interacting with soft photons via inverse Compton scattering. 

Protons at PeV energies should produce gamma rays of up to $\sim$ 100 TeV close to the acceleration sites. At these energies, inverse Compton scattering becomes inefficient. The detection of 100 TeV gamma rays from SNRs would be a direct evidence of proton acceleration at SNR shocks. Instruments sensitive to gamma rays up to the 100 TeV range, such as SGSO, would therefore come as the most natural instruments to study pevatrons. 

If SNRs are indeed accelerating PeV particles, they are expected to do it at the early phases of their evolution, until at most a few hundred years~\cite{gabici2007,aharonian2013,schure2013}. The number of candidates is therefore reduced, and simple estimates indicate that at most up $\approx 10$ SNRs could be active pevatrons accessible to current instruments, but that in any case, it would be extremely challenging to identify them as pevatrons with instruments optimized in the TeV range~\cite{cristofari2018}. On the other hand, instruments optimized in the 100 TeV range, and with a wide field, such as SGSO, would be efficient in an all-sky search of the potential SNR pevatron candidates.  

SGSO will help survey a portion of the sky (Figure~\ref{fig:SkyVisibility}) with unprecedented sensitivity above a few tens of TeV (See Figure \ref{fig:SGSOperformance}). During its survey, SGSO will extend the high-energy range for known (bright) sources with extremely hard spectra, like for example the  Galactic center region (Section~\ref{sec:J1745}). In addition, it might be able to discover sources with hard spectra that are to faint to be detected with the current generation of IACTs. One interesting object is  G1.9+0.3 (Decl. -27$^\circ$), the youngest known Galactic SNR, and thus a natural excellent pevatron candidate~\cite{gabici2016}, given its young age of about 100 years~\cite{reynolds2008}. So far, only upper limits were set by current TeV instruments~\cite{G19HESS}, but the more than one order of magnitude in sensitivity improvement that SGSO will bring above 10 TeV could potentially bring evidence of the first SNR pevatron.

In addition, while a number of SNRs have been spatially resolved by IACTs and space telescopes~\cite{HESSSNRCAT,FERMISNRCAT}, the IACTs have limited FOVs and in this sense, SGSO, by design, deals more easily with the very large source extensions. As an example, we mention the source G150.3+4.5, identified as a good target for HAWC due to its large size of $2.5 \times 3~\mathrm{deg}$~\cite{G150}. Similar candidates are expected in the southern sky and would be prime targets for SGSO.

\section{Unbiased survey of the Galactic Plane} 

One can summarize the open questions in the search of the origin of the most energetic cosmic rays as follows:

\begin{enumerate}

\item
Is the TeV emission from young shell-type SNRs hadronic or leptonic in nature?

The observed young SNR shells accelerate particles up to hundreds of TeV. Electron acceleration proceeds in young shell-type SNRs up to hundreds of TeV as testified by the detection of non-thermal hard X-rays associated to the TeV emission of these young SNRs. Protons are likely to be accelerated through the same mechanism up to hundreds of TeV. Observational evidence in favor of hadronic acceleration is given by the strongly amplified magnetic fields, a likely effect of cosmic ray streaming, detected in the shells of young SNRs. It is however not clear what fraction of the energy input of the SN explosion is used to accelerate electrons and what fraction goes into ultra-relativistic hadrons. 

\item
Are SNRs the major contributors to Galactic CRs up to the knee? 

If SNRs are the major CR contributors, the vast majority of SNRs have to inject roughly $10^{50}$ erg in accelerated protons to sustain the cosmic ray population. The energy input in protons above few hundreds GeV of each SNR detected at TeV can in principle be estimated using the luminosity in gamma-rays and the gas density in the shell. Due to the uncertainty in the measurements of gas density in SN shells, it is however difficult to estimate the actual fraction of the energy of the SN explosion used to accelerate very high energy protons. Furthermore, different scenarios are possible, where only some classes of SNRs input the required 10$^{50}$ erg in accelerated protons or maybe none of them produces this power.

\item How are proton and nuclei accelerated to PeV? 

For SNRs, Diffusive Shock Acceleration has been proposed to account for particle acceleration (see e.g. \cite{drury1983,bell2004} and references therein), but several details of the physics involved are still not understood and are being investigated actively by the theoretical community; see e.g. \cite{bykov2017,blasi2017,caprioli2018}. For other potential accelerators, the situation is even less clear.

\item How do particles leave the accelerator and impact the surrounding the ISM? 

As illustrated by the recent results of HAWC~\cite{Geminga_HAWC}, large FOV instruments have potential to bring valuable estimates on the surroundings of accelerators, as for example values of the diffusion coefficient in the vicinity of sources. SGSO will also be able to study the gamma-ray emission from giant molecular clouds (GMCs) embedded in the \textit{sea} of cosmic rays, which gives an opportunity to study the flux of cosmic rays in terms of their propagation and distribution throughout the Galaxy. Even more generally, SGSO will characterize the large scale diffuse gamma-ray emission along the Galactic Plane. 

\item
A new population of pevatrons in the Galaxy?

A recent analysis of TeV data from the extended diffuse emission around the Galactic Center (GC), done by the \HESS, has suggested the presence of a powerful pevatron in the Center of the Milky Way. However, the GC pevatron itself would not be able to sustain the CR population close to the knee unless one assumes that this source was much more powerful in the past. Furthermore, a number of hard spectrum gamma-ray sources extending beyond 10-20 TeV without evidence of cutoffs is now emerging from the \HESS and HAWC surveys of the Galactic Plane. The question is now: Do these recently discovered sources represent a new population of Galactic pevatrons? What are the counterparts of these sources at other wavelengths? Are these sources molecular clouds illuminated by PeV particles escaping nearby SNRs or maybe some more mysterious astrophysical source? From a theoretical point of view, other classes of sources, such as collective stellar winds and SNR shocks in clusters and associations of massive stars, have long been suggested as potential alternative or additional Galactic pevatrons. 

\item What is the distribution of pevatrons in the Galaxy? 

With only one pevatron detection (i.e.\ the Galactic Center) so far, this question is hard to answer. Systematic surveys are needed to help understand this topic and confront it with theoretical models.

\end{enumerate}

Until now, the approach to the question of the CR origin has been to select a possible class of astrophysical sources and decide if this class can be a CR factory. However, different astrophysical sources might contribute at different energies and different levels to the CR population. Therefore ideally one would like to be able to count all possible CR contributors.
An unbiased survey of the Galactic Plane in the crucial energy range of several tens of TeV, corresponding to photons emitted by CRs close to PeV, can help solve the above mentioned question. SGSO, capable of observing large fields of view, is the ideal survey instrument both to map a substantial part of the Galactic disk efficiently and to study the extended sources. Its significant sensitivity at high energy ($>$10 TeV) is an obvious advantage. Combined with the expected energy resolution (cf. Figure~\ref{fig:SGSOperformance}), it will allow measurements of the spectral shape up to extreme gamma-ray energies. Conducting large surveys requires significant efforts for pointing instruments such as IACTs, which at the same time have limited sensitivity studying extended sources above 10 TeV. Detailed spectroscopy beyond 10 TeV gamma-ray energy, which is so crucial to probe the acceleration of PeV particles, requires in fact investing several hundreds of hours of observations and this is feasible only if one knows where to point. In this regard, SGSO can provide important input to other observatories and will be very useful and complementary for example to CTA.  

In addition to the unbiased survey mentioned in this  section, we note that several sources detected in the southern sky by current TeV instruments, and currently listed as unidentified sources, i.e. the type of astrophysical object to which they belong is unclear, display spectrum in the TeV range hard enough to make them pevatron candidates.  This is for example the case of  HESS J1641-463, HESS J1741-302, and HESS J1826-130~\cite{anguner2017}.

 \section{The Galactic center}\label{sec:J1745} 
\begin{figure}[!t]
   \centerline{
     \includegraphics[width=0.65\linewidth]{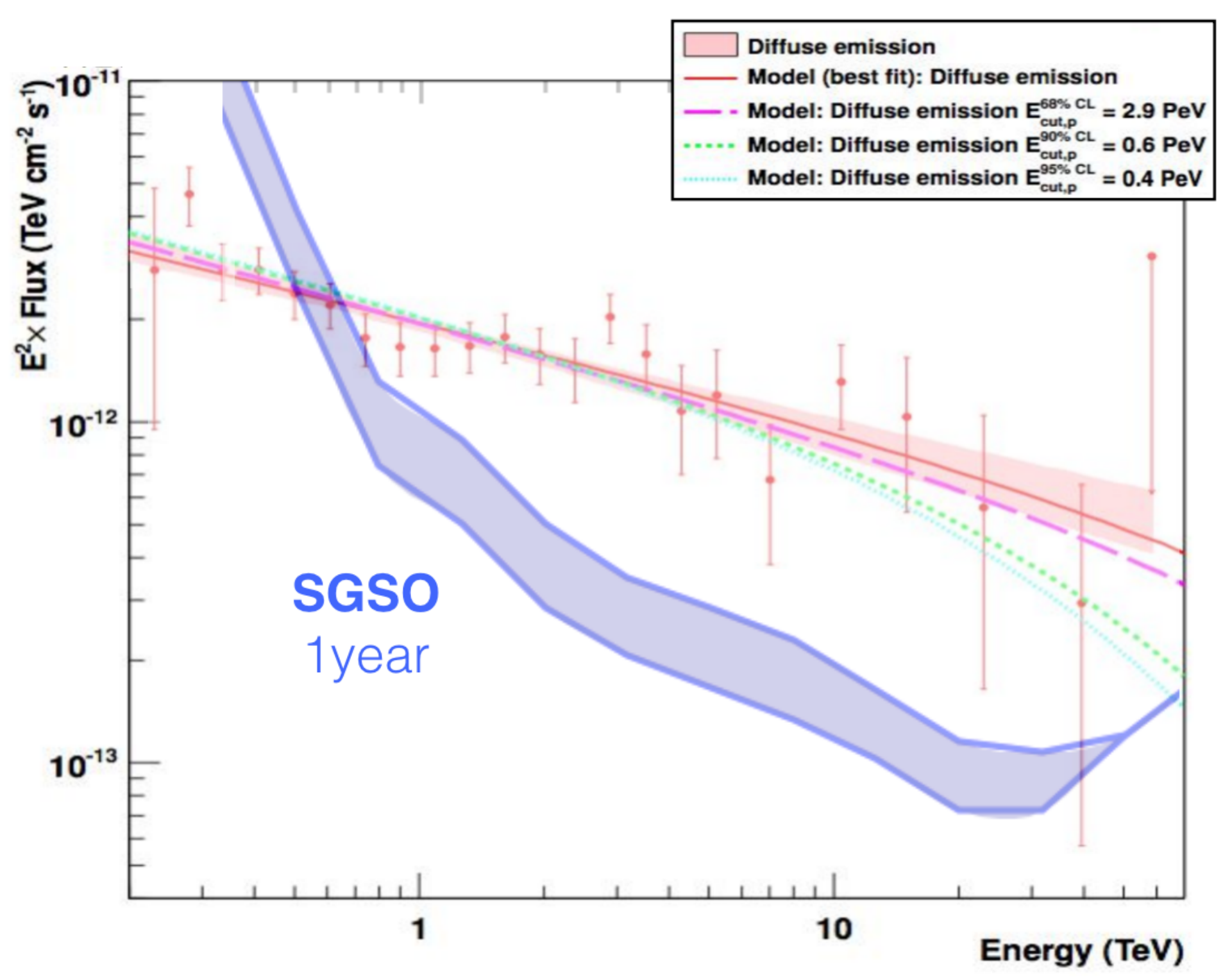}}
   \caption{Diffuse VHE gamma-ray flux from the central molecular zone in the Galactic center region measured by \HESS in comparison with the sensitivity after 1~year of SGSO operations. Modified from~\cite{HESSGC}.}\label{fig:J1745}
 \end{figure}

 Evidence for acceleration of PeV protons in the Galactic center was presented by the \HESS collaboration~\cite{HESSGC}. This claim is supported by the detection of gamma rays of several tens of TeV, with a hard spectrum and no sign of cut--off in the spectrum at the highest energy, indicating the production of gamma rays up to at least $\approx$100 TeV, and therefore, the acceleration of PeV protons (cf. Figure~\ref{fig:J1745}). Observations helped identifying the presence of PeV protons in the central $\sim$10 parsec of the Galaxy, but there is still no unequivocal identification of the source of these PeV protons. The most plausible explanation is that the supermassive black hole Sagittarius A$^{*}$ is somewhat involved in the acceleration process~\cite{HESSGC,fujita2017,guo2017}. Other hypotheses have been, and are still being investigated, such as for example the increased SN rate in the Galactic center~\cite{jouvin2017}, the role of massive stars~\cite{aharonian2018}, or  the contribution from millisecond pulsars~\cite{guepin2018}, among others. 

As illustrated in Figure~\ref{fig:J1745}, the Galactic center region is an excellent target for instruments optimized in the TeV range, such as CTA, and for a highly sensitive instrument in the tens of TeV range, such as SGSO. Especially the highest energy data will allow to understand the production of PeV particles at the first known Galactic PeV accelerator in detail.

 \section{Star-forming regions} 

So far, only a few star-forming regions have been clearly detected as gamma-ray sources by current TeV instruments, but they constitute a major key science target for the next generation of gamma-ray observatories. Star-forming regions can be powerful gamma-ray emitters. In the VHE domain, the measured gamma rays trace back to the populations of accelerated particles, and can therefore help understand the distribution, acceleration  mechanisms and propagation of particles in star-forming regions. 

The problem of stellar formation is still an open topic, with many questions still unanswered. Among these open questions, the role played by CRs on the formation of structures is a major one. It has been proposed that CRs can significantly enhance or reduce stellar formation, emphasizing the importance of understanding the CR physics in star-forming regions. 

Recently, observations of prominent Galactic clusters, such as Westerlund 1, Westerlund 2, or Cyg OB2 have shown a decrease in the CR density around these clusters, scaling with the inverse of the distance. Such a density profile is seen as a result of continuous injection of CRs and ensuing diffusion in the ISM~\cite{aharonian2018}. In addition, gamma-ray observations suggest evidence of the acceleration of protons up to the PeV range. The possibility of efficient energy conversion at stellar winds has made stellar clusters good candidates for the sources of Galactic CRs. 
SGSO will provide an unprecedented chance to study star-forming regions, both Galactic and extragalactic, with improved sensitivity in the 100 TeV range. 

Natural excellent candidates for observation are objects already detected in gamma rays, such as for example Westerlund 1. Westerlund 1 is one of the most massive Galactic clusters. It has already been detected in the VHE range by \HESS~\cite{westerlundHESS}, but an accurate understanding of the spectral properties and the morphology of the region is still missing. Future instruments such as CTA, with an improved angular resolution, are expected to be efficient in constraining the spatial origin of the emission, and SGSO, in constraining the spectral properties, especially in the 100 TeV range. Westerlund 1 is located close to the Galactic center, and is therefore a target of choice for SGSO, whereas observations with HAWC are limited. Furthermore, the distribution of CRs near Westerlund 1 derived from archival H.E.S.S data~\cite{aharonian2018} follows a $1/r$ distribution, thus pointing to CR diffusion. However, such a conclusion requires an accurate measurement of gamma-ray emission profile over large angular scales, which is difficult for the current and future IACTs due to their limited FOV.
In this regard, SGSO has a unique advantage due to its large FOV. The study of Westerlund 1 with SGSO will complement studies of other well known star-forming regions such as Eta Carina~\cite{etaHESS} or the Cygnus region~\cite{cygnusfermi}, and the Andromeda Galaxy already targeted by HAWC~\cite{M31HAWC}.

\section{Galactic diffuse emission}

\begin{figure}[!t]
   \centerline{
     \includegraphics[width=0.75\linewidth]{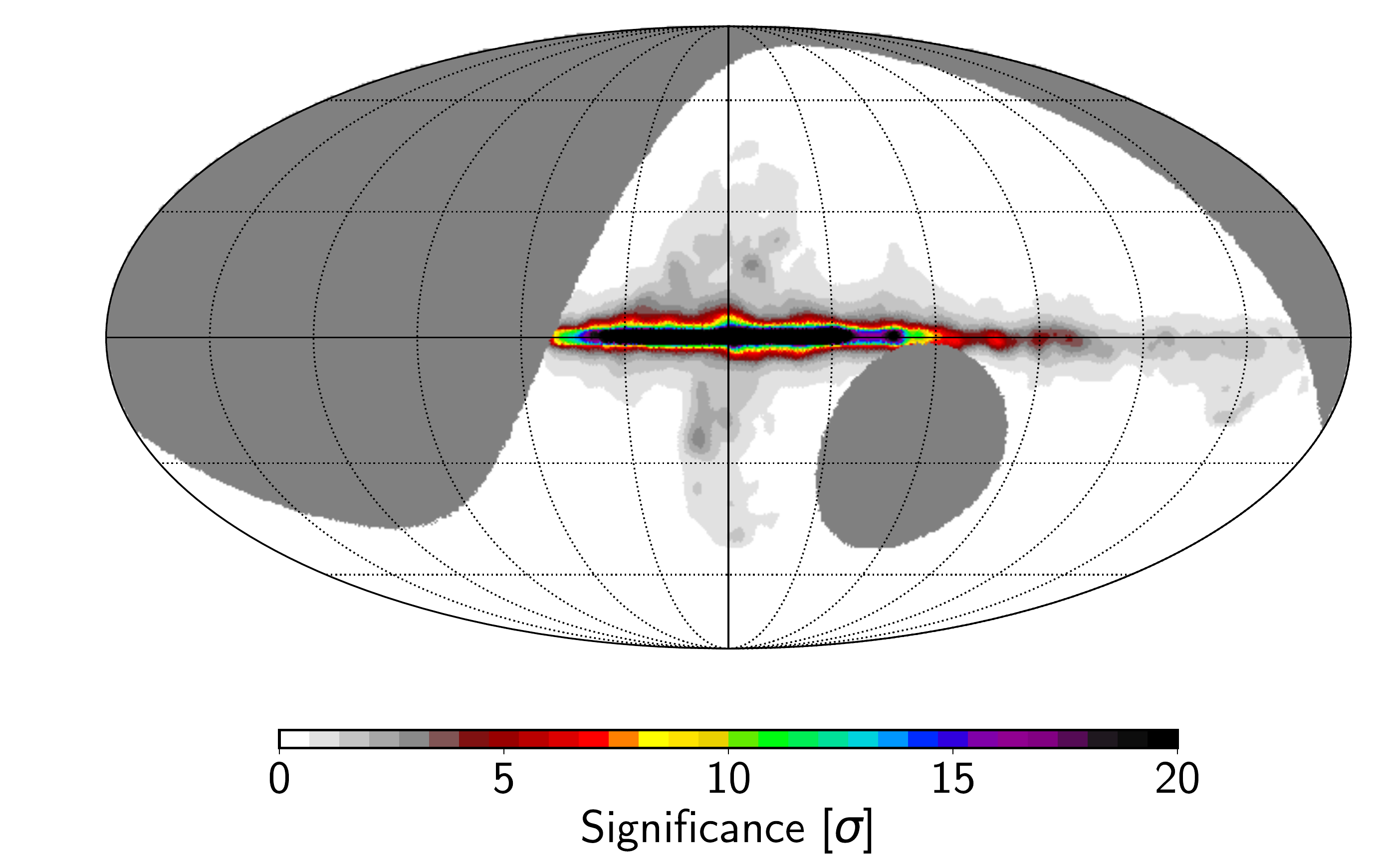}}
   \caption{Projected significance for the observation of the Galactic diffuse emission with one year of SGSO based on an extrapolation of the \emph{Fermi} diffuse model from~\cite{FermiDiffuse}. A detector latitude of $-24.2^{\circ}$ has been assumed, and the sky map is restricted to a zenith angle below $40^{\circ}$.The skymap has been smoothed with a $2^{\circ}$ top-hat circular kernel to increase the sensitivity to extended and diffuse emission in the plane.}\label{fig:galdiffuse}
 \end{figure}

The large field of view and good angular resolution of SGSO will allow it to perform a detailed study of the diffuse gamma-ray emission from the Galaxy. This is a measurement that is particularly challenging for current and even future IACTs given their small field of view and the background estimation techniques used, which can average out significant gamma-ray backgrounds if they appear as slow flux gradients over a large sky region.
A detailed study of the Galactic diffuse emission would constrain the cosmic ray propagation parameters in the galactic medium, which are used as input to propagation codes such as GALPROP~\cite{GALPROP}. 

One of the main challenges for this study is the issue of source confusion, as large sources extending over many degrees (as already observed by HAWC) may be indistinguishable from the ``true'' diffuse emission produced by energetic electrons and protons propagating in the interstellar medium. 

To evaluate the capability of SGSO to observe the diffuse galactic background, we make use the \emph{Fermi} diffuse model\footnote{\href{https://fermi.gsfc.nasa.gov/ssc/data/access/lat/BackgroundModels.html}{https://fermi.gsfc.nasa.gov/ssc/data/access/lat/BackgroundModels.html}} {\texttt gll\_iem\_v06} which is valid up to $\sim 500$ GeV~\cite{FermiDiffuse}. Based on this model, we extrapolate the flux using a single power-law per pixel based on the flux points between 50 and 500 GeV up to an energy of 2 TeV. We note that this is a fairly optimistic model as some regions, most notably the \emph{Fermi} ``bubbles'' show curved spectra that would produce lower fluxes at high energies than a simple power-law extrapolation would predict. Folding these gamma-ray spectra with the effective areas in Fig.~\ref{fig:SGSOScale}, and doing similarly for the expected cosmic-ray background after gamma/hadron cuts are applied, we can calculate a significance for the detection of the diffuse emission signature above 100 GeV up to 2 TeV using a simple $S/\sqrt{B}$ argument.  

The resulting skymap for a 1-year exposure with SGSO is shown in Fig.~\ref{fig:galdiffuse} which shows a high-significance detection using a suboptimal $2^{\circ}$ smoothing of the skymap. The high statistics that SGSO will be able to collect will allow it to disentangle the diffuse emission from other sources near the Galactic plane and allow spectral studies as a function of Galactic longitude and latitude. The geographic latitude of SGSO will enable a detailed study of both the extended inner 60$^{\circ}$ of the Galactic plane, and both \emph{Fermi} bubbles as already shown in Fig.~\ref{fig:FermiBubbles}.

\cleardoublepage
\chapter{Monitoring the Transient Sky} 
The wide field of view, high duty cycle and lower energy threshold compared to current air shower arrays render SGSO well suited for monitoring transient sources at very high energies. Primary targets include active galactic nuclei (AGNs), gamma-ray bursts (GRBs), Galactic transients such as binaries and microquasars, and transients whose underlying physics is poorly understood, such as fast radio bursts (FRBs). The variety of morphologies and energy spectra, even within the same type of phenomenon, has prevented full understanding so far. Thus, an unbiased increase on the number of observations will allow us to go from an individual description to a complete and general understanding of their nature. For instance, it would provide additional valuable information for constraining the mechanisms that accelerate particles at the highest energies and produce very high energy gamma rays, neutrinos or even gravitational waves.

In the case of extragalactic objects, pair production on the extragalactic background light (EBL) limits the range at which sources can be detected. An illustration of this effect is given in Figure~\ref{fig:GRBExample}. Sensitivity at low energies, from 100 GeV to 300 GeV, where the resulting absorption of VHE gamma rays is limited, is consequently of paramount importance for extragalactic studies. SGSO, with its order of magnitude increase in sensitivity over existing ground arrays at these energies, will be a prime instrument providing useful observations complementary to other upcoming experiments such as CTA. Its performance will allow SGSO to both capture very high energy views of transient phenomena and issue alerts to other experiments for multi-wavelength and multi-messenger follow-up, thus enabling the construction of an increasingly complete picture of the underlying phenomena. 

\section{Active Galactic Nuclei} 
\label{subsec:AGN}

Active galactic nuclei are extremely luminous galactic cores powered by accretion onto a supermassive black hole (see, e.g.~\cite{1978PhyS...17..265B,1984RvMP...56..255B}).
Many exhibit relativistic jets that transport plasma via bulk outflows with high Doppler factors~\cite{1978A&A....70L..71S,2012rjag.book.....B}.
The spectra of AGNs are extremely broadband, ranging from radio to TeV energies, and their short variability timescales across a wide range of wavelengths is indicative of small emitting regions.
The emission from AGNs in the GeV and TeV gamma-ray bands is non-thermal and commonly associated with their jets or lobes~\cite{1992Natur.358..477P,2010Sci...328..725A}.

Several different models exist to explain the spectra of AGNs from radio wavelengths to very high energy gamma rays. The simplest of these is the synchrotron self-Compton (SSC) model~\cite{1967MNRAS.137..429R,1969ARA&A...7..375G,1996ApJ...461..657B,1998ApJ...509..608T}, which explains the emission via a single population of relativistic electrons, and is often chosen due to its relative simplicity. Extensions such as the external Compton model~\cite{1992A&A...256L..27D,1994ApJ...421..153S} or including contributions from accelerated hadrons~\cite{1989A&A...221..211M,1993A&A...269...67M,2001APh....15..121M,2003ApJ...586...79A,2014MNRAS.441.1209F,2015APh....70...54F,2015APh....71....1F} are often introduced when the SSC model provides an unsatisfactory description of the data. In general, first-order Fermi acceleration in shocks, again due to its simplicity, is invoked to explain the acceleration of particles in AGNs. Recently, alternative models including second-order Fermi acceleration~\cite{2011ApJ...740...64L}, magnetic reconnection~\cite{2000ApJ...530L..77B,2010MNRAS.408L..46G,2015MNRAS.450..183S,2016MNRAS.462.3325P}, or the stochastic acceleration of cosmic rays~\cite{2013A&A...558A..19W} have been advanced to explain unexpectedly hard spectra, extremely rapid variations, and other surprising aspects of AGN emission. What is clear is that additional data, both from high and low flux states, is necessary for unraveling the mechanisms at work in these extreme environments.

Blazars (radio-loud AGNs with a jet oriented at a small angle with respect to our line of sight) dominate the extragalactic VHE sky, with 70 of the 78 extragalactic VHE sources being identified as blazars\footnote{See \url{http://tevcat.uchicago.edu}}.
The power spectra of blazars typically show two broad bumps, one that peaks at optical to X-ray energies and is due to the synchrotron radiation of relativistic electrons, and another that peaks in gamma rays with an origin that is still debated, either due to inverse Compton scatterings of electrons in the jet, the decay of neutral pions from hadronic interactions, or a combination of both processes.
Approximately 50 of the known VHE blazars belong to the sub-class of high-frequency peaked BL Lac objects (HBLs), for which the synchrotron radiation peaks at frequencies above $10^{15}$ Hz.

Although blazars represent by far the most common VHE extragalactic source class, a handful of AGNs with unaligned jets have been detected in the VHE band.
These radio galaxies exhibit extended lobes and large-scale off-axis jets with large viewing angles.
VHE detections in this source class include Centaurus A~\cite{2009ApJ...695L..40A}, M87~\cite{2006Sci...314.1424A}, NGC 1275~\cite{2014A&A...564A...5A}, IC 310~\cite{2014A&A...563A..91A}, and the recently announced 3C 264~\cite{2018ATel11436....1M}.

\subsection{SGSO detection capabilities: known VHE Blazars}
\begin{figure}[!t]
  \begin{center}
    \includegraphics[width=0.45\linewidth]{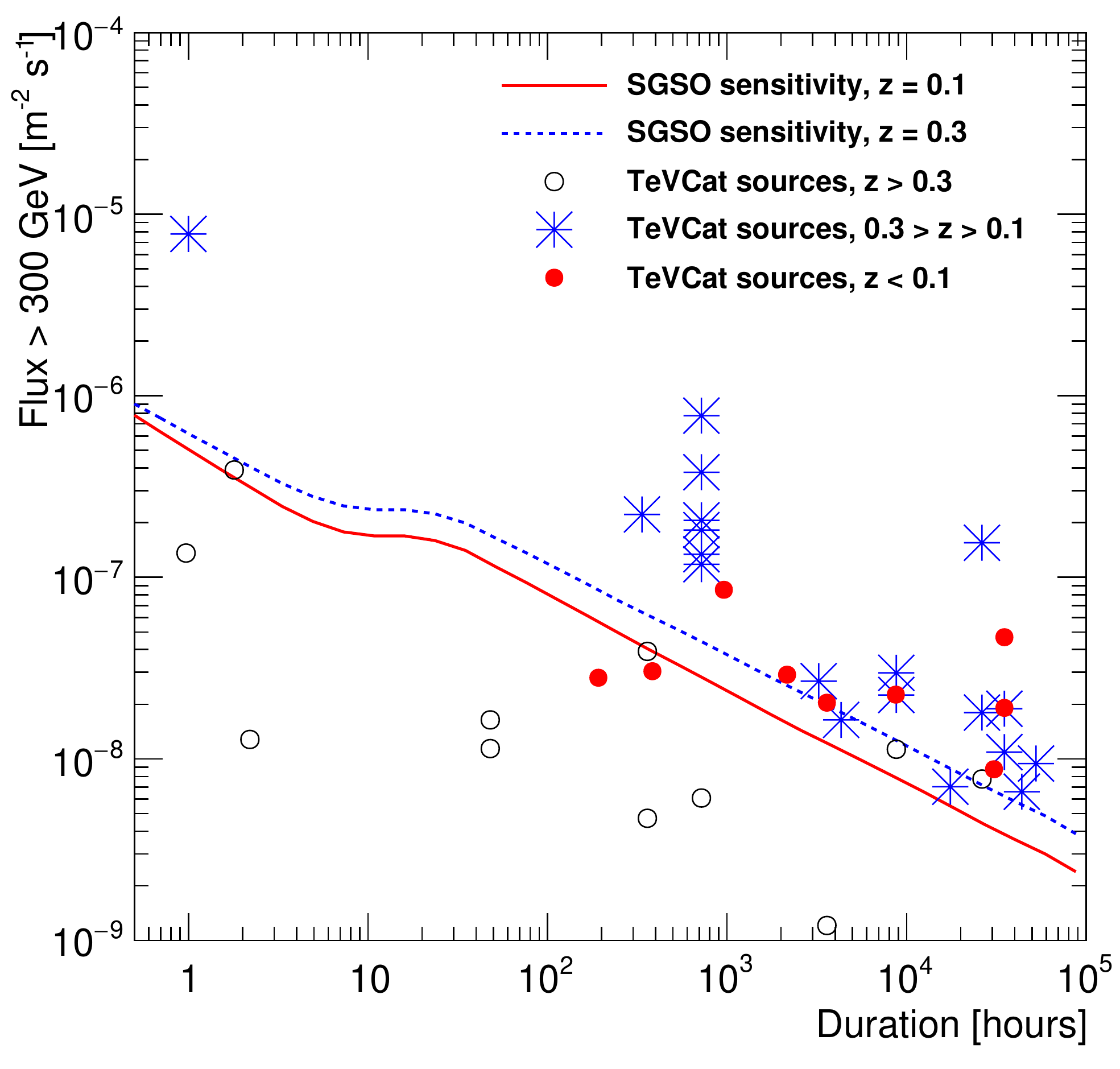}
    \includegraphics[width=0.45\linewidth]{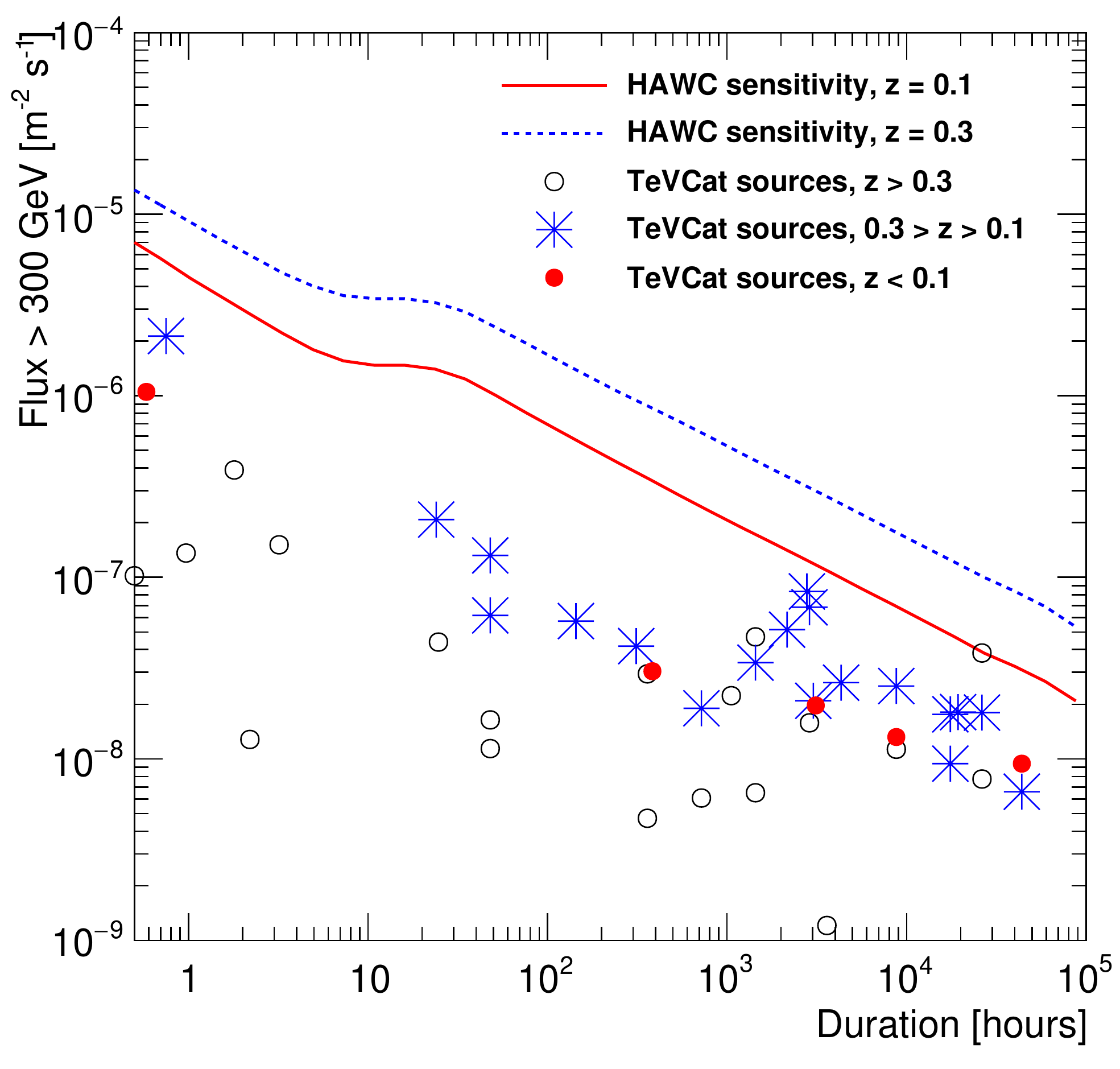}
   \end{center}
  \caption{Integral flux above 300 GeV versus duration (see text) for TeVCat sources with published spectra. The left plot shows sources in the declination range from -54$^\circ$ to +6$^\circ$ (lying within 30$^\circ$ of one of the potential SGSO sites, i.e. $\approx 24^\circ$ S latitude), along with the assumed SGSO sensitivities. The sources are divided into three redshift ranges as indicated. The right plot is the same, but for the HAWC sensitivity and a declination range from -11$^\circ$ to +49$^\circ$, corresponding to the HAWC latitude of 19$^\circ$ N.}\label{Fig:AGN-TeVCat-spectra}
\end{figure}

The detection of blazars in the VHE band is particularly challenging due to the attenuation of their spectra by pair production interactions with the extragalactic background light (EBL).
Figure~\ref{Fig:AGN-TeVCat-spectra} summarizes the published spectra from known VHE blazars in the context of their detection by HAWC and SGSO. The plot is made as a function of duration, which is taken to be the actual reported duration of the observations in the case of flares, or the total time span over which data were collected in the case of a long-term average over many observations. In the left panel, we plot the integral flux above 300 GeV for the published VHE spectra of blazars that would be located within the sky coverage of SGSO, assuming one of the potential sites, i.e.\ a latitude around 24$^\circ$ S (cf. Sec.~\ref{subsec:site}). We divide the blazars into three categories based on their redshift: $z<0.1$, $0.1<z<0.3$, and $z>0.3$, and we show the SGSO sensitivity for our baseline proposed detector for a source with an intrinsic spectrum with a power-law index of 2 at $z=0.1$ and $z=0.3$. For the sensitivity calculation, we attenuate the intrinsic spectrum using the EBL model from~\cite{2012MNRAS.422.3189G}. For the data, we conservatively assume a hard cutoff at the maximum reported data point in the spectrum. We note that the IACT data used for this figure is in part influenced by the biased observations obtained with the current IACTs, which often react to flaring episodes reported at other wavelengths. In general the figure shows that SGSO will be capable of detecting several previously reported VHE blazars at both short and long time scales. Combined with its ability to perform these searches in an unbiased way over the full southern sky, we expect SGSO to provide contributions to extragalactic VHE science, as detailed in the rest of this section.

The right panel of Figure~\ref{Fig:AGN-TeVCat-spectra} shows the same results for the presently operational HAWC detector. The nearby blazars Markarian 421 and Markarian 501\footnote{Hereafter, we refer to Markarian with the designation Mrk.}, which are already detected by HAWC at high significance, are excluded from this plot. It is clear that no other published VHE spectra cross the HAWC sensitivity curves. Therefore, it is not surprising that Mrk 421 and Mrk 501 are the only extragalactic sources detected by HAWC to date.

\begin{table}[!t]
\caption{Blazars used to construct Figure~\ref{Fig:AGN-TeVCat-spectra}. The redshifts and declinations are taken from \url{http://tevcat.uchicago.edu.}}
  \scriptsize
\begin{center}
\begin{tabular}{rcccc}
\hline
Blazar & Redshift & Declination & SGSO & HAWC \\
\hline
PKS 2005-489 & 0.071 & -48.8$^\circ$ & Y & N \\
PKS 0447-439 & 0.343 & -43.8$^\circ$ & Y & N \\
1ES 1312-423 & 0.102 & -42.6$^\circ$ & Y & N \\
PKS 1440-389 & 0.065 & -39.1$^\circ$ & Y & N \\
PKS 0625-35 & 0.0549 & -35.5$^\circ$ & Y & N \\
PKS 0548-322 & 0.069 & -32.3$^\circ$ & Y & N \\
1RXS J101015.9-311909 & 0.1426 & -31.3$^\circ$ & Y & N \\
H 2356-309 & 0.165 & -30.6$^\circ$ & Y & N \\
PKS 2155-304 & 0.116 & -30.2$^\circ$ & Y & N \\
AP Librae & 0.049 & -24.4$^\circ$ & Y & N \\
PKS 0301-243 & 0.2657 & -24.1$^\circ$ & Y & N \\
1ES 1101-232 & 0.186 & -23.5$^\circ$ & Y & N \\
SHBL J001355.9-185406 & 0.095 & -18.9$^\circ$ & Y & N \\
1ES 0347-121 & 0.188 & -12.0$^\circ$ & Y & N \\
PKS 1510-089 & 0.361 & -9.1$^\circ$ & Y & Y \\
3C 279 & 0.5362 & -5.8$^\circ$ & Y & Y \\
1ES 0414+009 & 0.287 & +1.1$^\circ$ & Y & Y \\
RGB J0152+017 & 0.08 & +1.8$^\circ$ & Y & Y \\
TXS 0506+056 & 0.3365 & +5.7$^\circ$ & Y & Y \\
PG 1553+113 & 0.5 & +11.2$^\circ$ & N & Y \\
1ES 1440+122 & 0.1631 & +12.0$^\circ$ & N & Y \\
RBS 0413 & 0.19 & +18.8$^\circ$ & N & Y \\
1ES 1741+196 & 0.084 & +19.5$^\circ$ & N & Y \\
OJ 287 & 0.3056 & +20.1$^\circ$ & N & Y \\
1ES 0229+200 & 0.1396 & +20.3$^\circ$ & N & Y \\
PKS 1222+21 & 0.432 & +21.4$^\circ$ & N & Y \\
W Comae & 0 & +28.2$^\circ$ & N & Y \\
1ES 1215+303 & 0.131 & +30.1$^\circ$ & N & Y \\
1ES 1218+304 & 0.182 & +30.2$^\circ$ & N & Y \\
B3 2247+381 & 0.1187 & +38.4$^\circ$ & N & Y \\
BL Lacertae & 0.069 & +42.3$^\circ$ & N & Y \\
3C 66A & 0.34 & +43.0$^\circ$ & N & Y \\
\hline
\end{tabular}
\end{center}
\label{table:TeVCat-AGN-table}
\end{table}

Table~\ref{table:TeVCat-AGN-table} shows the blazars used to construct Figure~\ref{Fig:AGN-TeVCat-spectra}. The redshifts and declinations from \url{http://tevcat.uchicago.edu} are shown in the table, along with whether the source is observable by SGSO and/or HAWC.

\subsection{Searching for New VHE Blazars}
\begin{figure}[!t]
\centering
\includegraphics[width=0.8\hsize]{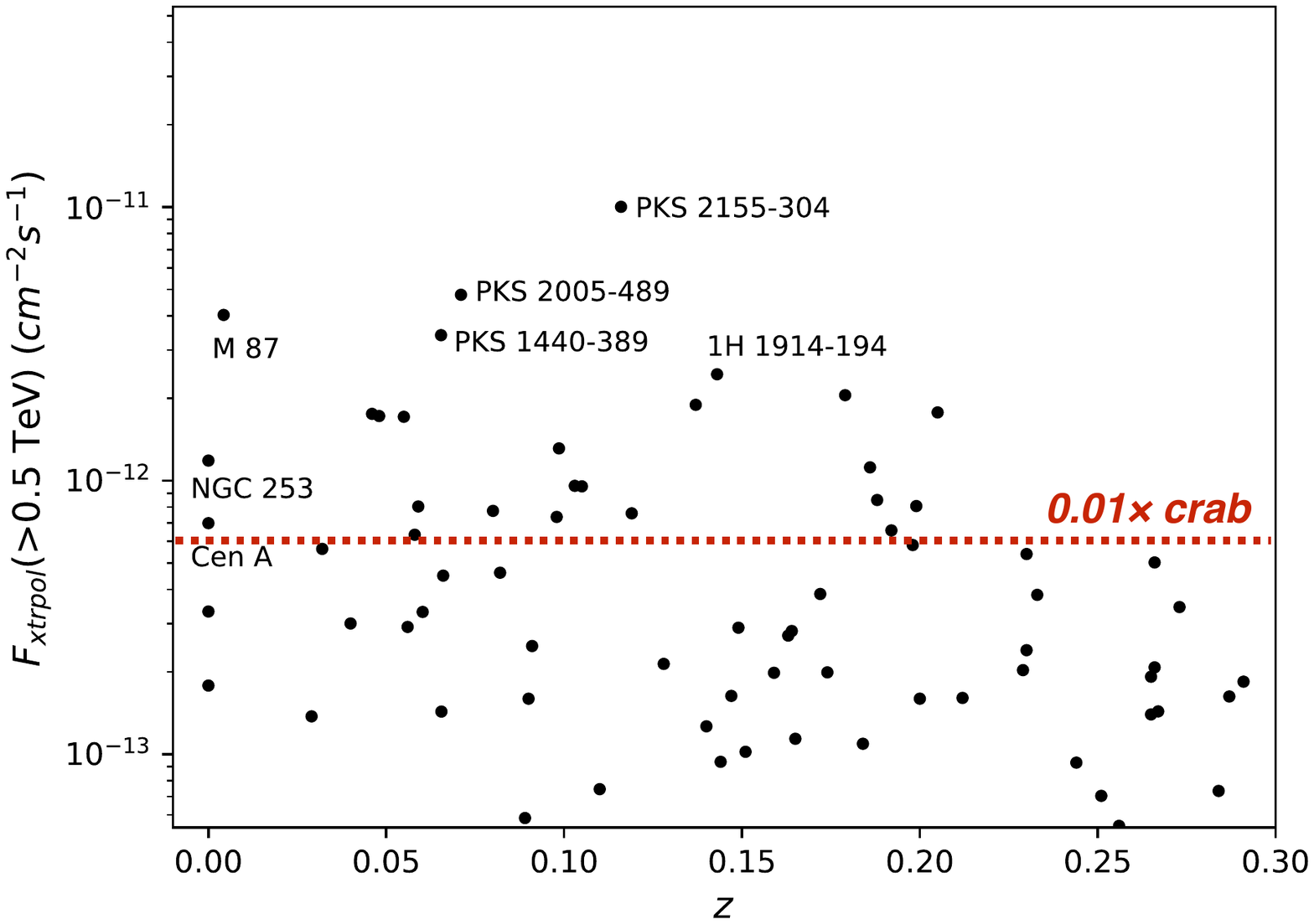}
\caption{Extrapolation of 3FHL spectra above 0.5~TeV, with EBL absorption taken into account~\cite{2011MNRAS.410.2556D}. For scale, the dotted red line marks 1\% of the Crab flux above 0.5\,TeV, this corresponds roughly to sensitivity of 1 year of SGSO observations for a source passing through zenith with spectrum similar to the Crab Nebula.}
\label{fig:3fhl-xtrpol}
\end{figure}

SGSO also offers the possibility of conducting an unbiased VHE survey of more than half the sky with unprecedented sensitivity in order to detect new VHE blazars. The sensitivity, presented in Figure~\ref{Fig:AGN-TeVCat-spectra}, shows that SGSO is expected to detect the long-term emission from several known VHE blazars, supporting the notion that several other, so far unknown VHE blazars will be accessible to SGSO.

A good starting point for searching for new VHE blazars is the third catalog of hard {\it Fermi}-LAT sources above 10 GeV (3FHL), which contains 1231 objects identified or associated with active galaxies, with the majority of these being BL Lac-type blazars~\cite{2017ApJS..232...18A}. A total of 108 3FHL objects with known redshifts $z\leq0.3$ are located at declinations within $40^\circ$ of the latitude of one of the potential SGSO site ($\mathrm{Lat} \approx 24^\circ$S). Of these, 36 sources show emission in the 50 to 150 GeV band with a significance greater than $5\sigma$. The extrapolation of the 3FHL spectra into the VHE range, coupled with attenuation on the EBL~\cite{2011MNRAS.410.2556D}, provides a sizable number of objects that are expected to have integral fluxes accessible with SGSO after a short period of observations. In Figure~\ref{fig:3fhl-xtrpol}, the expected flux above 0.5~TeV is shown in comparison to the SGSO sensitivity after one year of data taking for a Crab-like source passing through zenith. We note that an additional 260 objects in the 3FHL catalog but without a redshift measurement fall within the SGSO sky coverage. These can be additional promising candidates for long-term study in the VHE band.

A fair number of the objects drawn from the 3FHL have already been studied at TeV energies using IACTs over the past years. A tabulation of IACT observations of 3FHL sources detected above 50~GeV, as listed in TeVCat, is shown in Table~\ref{tab:tevcats}. We note some consistency with the extrapolation from the 3FHL: for example, PKS 2155-304 is the brightest detection, with an observed flux not much different from that estimated in Figure~\ref{fig:3fhl-xtrpol}. Aside from this powerful blazar, and the extreme VHE-detected PKS~0625-35, all sources show fluxes at the level of a few percent of the Crab Nebula and have been detected above energy thresholds around several hundreds of GeV. These sources are thus perfectly accessible for SGSO, which, as indicated in Figure~\ref{fig:3fhl-xtrpol}, will have a sensitivity to fluxes below 1\% Crab at these energies. We note that this study is based on extrapolation of fluxes that are averaged over timescales of several years, thus being relatively insensitive to flaring episodes. It can therefore be considered a rather conservative estimate for the SGSO detection potential: with its large FoV, SGSO will be able to catch flares from previously unknown VHE AGNs across the full southern sky and trigger deep multi-wavelengths (MWL) follow-up observations.

\begin{table}[!t]
\caption{TeV observations of 3FHL Southern AGNs. The bottom four are particular nearby objects of interest.}\label{tab:tevcats}
  \scriptsize
  \centering
\begin{tabular}{llcc}
\hline\hline
3FHL source & Counterpart & Redshift & TeVCat flux and threshold \\
\hline
3FHL J0627.13528 & PKS~0625--35 & 0.055 & 0.04 Crab @ 580 GeV \\
3FHL J0303.4--2407 & PKS~0301--243 & 0.266 & 0.014 Crab @ 200 GeV \\
3FHL J2009.4--4849 & PKS~2005--489 & 0.071 & 0.03 Crab @ 400 GeV \\
3FHL J0238.4--3117 & 1RXS J023832.6--311658 & 0.030 & Not specified \\
3FHL J0449.4--4350 & PKS 0447--439 & 0.233 & 0.03 Crab @ 250 GeV \\
3FHL J0648.7+1517 & RX J0648.7+1516 & 0.179 & 0.033 Crab @ 200 GeV \\
3FHL J1010.2--3119 & 1RXS J101015.9--311909 & 0.143 & 0.008 Crab @ 200 GeV \\
3FHL J1443.9--3908 & PKS 1440--389 & 0.065 & 0.03 Crab @ 220 GeV \\
3FHL J1548.7--2250 & PMN J 1548--2251 & 0.192 & TeV candidate \\
3FHL J2158.8--3013 & PKS 2155--304 & 0.116 & 0.15 Crab @ 300 GeV \\
\hline
3FHL J1325.5--4300 & Cen A & 3.8 Mpc & 0.08 Crab @ 250 GeV \\
3FHL J1230.8+1223  & M87  & 16 Mpc & 0.033 Crab @ 730 GeV \\
3FHL J0047.6--2517 & NGC 253 & 3.5 Mpc & 0.002 Crab @ 220 GeV  \\
3FHL J1517.6--2422 & Ap Librae & 0.049 & 0.02 Crab @ 300 GeV  \\
\hline\hline
\end{tabular}
\end{table}

\subsection{Extreme high-frequency peaked BL Lacs}

\begin{figure}[!t]
  \begin{center}
    \includegraphics[width=0.495\linewidth]{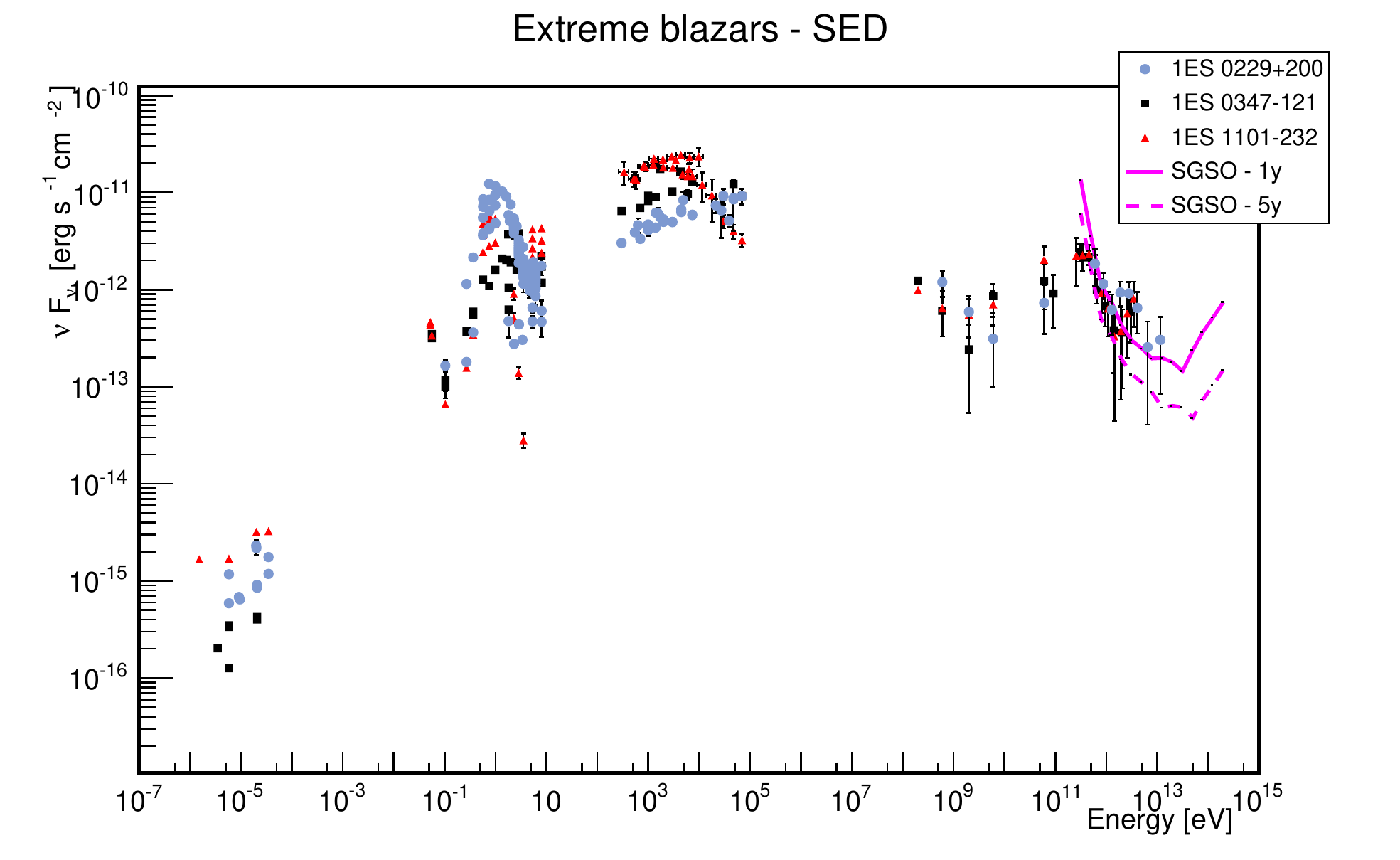}
    \includegraphics[width=0.495\linewidth]{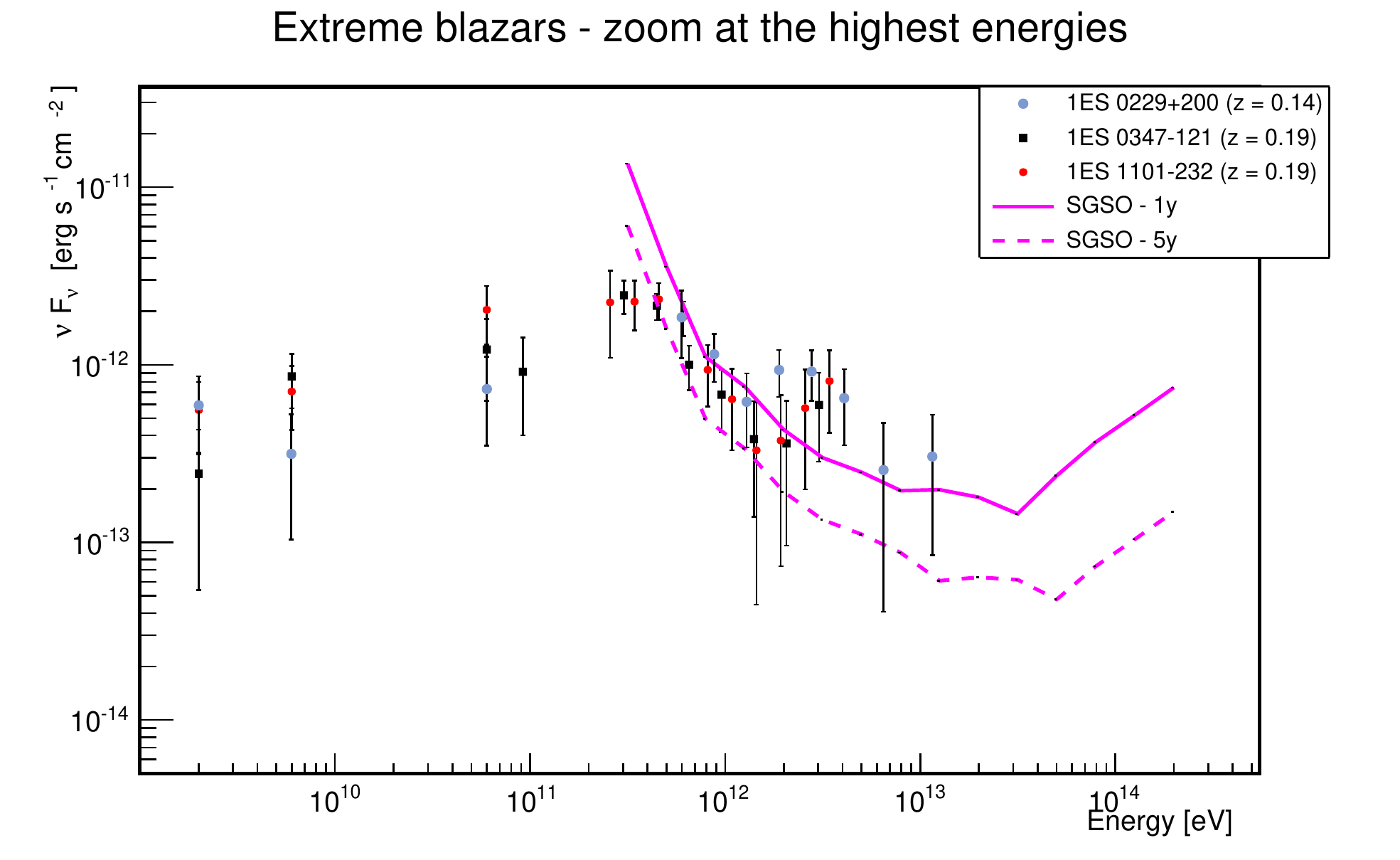}
   \end{center}
  \caption{Broadband SEDs, from radio to TeV energies, for three VHE-detected EHBLs (left) and a closeup at high energies (right). The data are extracted from the SSDC database. The 1-year and 5-year SGSO sensitivity curves appear in both panels.}\label{Fig:EHBL}
\end{figure}

In recent years, increasing attention has been put on a class of blazars known as extreme high-frequency peaked BL Lacs (EHBLs), for which the synchrotron power output peaks at 1 keV or higher~\cite{2015MNRAS.451..611B}.
Figure~\ref{Fig:EHBL} shows the broadband SEDs of three VHE-detected EHBLs from radio to TeV energies.
Among these, 1ES~0229+200 (at a redshift z = 0.140) is considered the prototypical extreme blazar. However, its sky position is not ideal for SGSO, making the target difficult to observe. 1ES~0347-121 and 1ES~1101-232, two EHBLs located at redshift of $\sim$0.19, are instead ideal targets for SGSO. The peak at optical energies represents emission from the host galaxy, which in the EHBL case becomes distinguishable from the jet-emitted continuum emission thanks to its shift in energy. The sharp cutoff at TeV energies, better visible in the right-hand panel of Figure~\ref{Fig:EHBL} where a zoom of the highest energies is presented, is due to absorption on the EBL.

A direct comparison of the three SEDs with the SGSO expected sensitivity shows that these sources are very promising targets. Going beyond these known sources, its large FoV may allow SGSO to detect tens of new TeV-emitting EHBLs at relatively small distances, and for the first time provide an unbiased sample of the closest EHBLs. At large distances (z $>$ 0.3), the EBL absorption effect becomes strong. These observations are therefore also particularly interesting for intergalactic magnetic field studies and searches for physics beyond the standard model such as axion-like particles or Lorentz invariance violation.

\subsection{AGN Emission Models}
Emission of AGN covers the full range of the electromagnetic spectrum (radio, IR/optical/UV, X-ray, GeV, and TeV). At present, and as outlined above, relatively simple models continue to dominate the literature, although astrophysical jets are clearly more complex than originally thought (e.g.~\cite{2015arXiv151107515B}). On the other hand, more complex models involve a large number of degrees of freedom and are thus usually able to explain the data well  (c.f.~\cite{2012arXiv1205.0539B}). Further constraints like exclusion of physically invalid parameter spaces can only be brought about by filling observational gaps. In this context VHE observations are of significant importance as they are able to constrain the location of the emission region, the energetics required in different models, the magnetic fields, etc. 

SGSO will be able to provide the data sets necessary for improved model building and explaining the emission from these complicated sources. Critically, these SGSO observations will take place on a long time scale with a periodic cadence that complements seasonal observations from pointed instruments.

\subsection{Blazar Variability}\label{sec:AGNvariability}
Perhaps the most striking feature of AGNs and especially blazars is their extreme variability over a wide range of time scales ranging from minutes (e.g.~\cite{2007ApJ...664L..71A}) to years~\cite{2001ApJ...559..187K,2007JPhCS..60..318T,2014APh....54....1A}.
The high fluxes of these flares facilitate studies of the mechanisms powering the central engine~\cite{2005A&A...433..479K,2016ARA&A..54..725M} and place stringent constraints on the bulk jet Lorentz factors~\cite{2001MNRAS.325.1559S,2001ApJ...559..187K,2008MNRAS.384L..19B}.
The most rapid VHE variability challenges the conventional models, prompting more complicated explanations such as those involving magnetic reconnection (e.g.~\cite{2016MNRAS.462.3325P}) or jets within a jet~\cite{2012A&A...548A.123B}.
Variability studies help to explain jet formation, composition, and structure, as well as accretion and AGN feedback~\cite{2012SSRv..173..309B}. The most stringent limits come from MWL observations providing spectral information during and after a flare (e.g.~\cite{2008Natur.452..966M,2009Sci...325..444A,2010Natur.463..919A}).

The sensitivity of SGSO would enable dedicated VHE monitoring for a large number of blazars.
The purpose of this monitoring would be twofold: to collect an unbiased population of VHE flares for further study, and to alert other instruments to follow up on exceptional flares in order to provide the needed multi-wavelength monitoring. In the past, typical follow-up observations of flares have commenced only after the flare had entered a decaying state~\cite{2013ApJ...762...92A} or when the starting point of the VHE flaring activity was ambiguous (e.g.~\cite{2017ApJ...836..205A}). SGSO would be able to resolve the ambiguity concerning the starting point of the flare, providing important data to inform models of the emission mechanisms.

While the flux distribution of blazars follows a normal distributions at GeV energies, the fluxes measured by current IACTs have a clear preference for log-normal distributions, suggesting multiplicative rather than additive processes at their origin~\cite{2009A&A...503..797G, 2018RAA....18..141S, 2010A&A...520A..83H}. However, the observations of most IACTs are biased, as the observations are usually not carried out in a monitoring mode but often based on triggers by other instruments. Furthermore, there are often gaps in the observations since the sources can be observed only half of the year due to observations being possible only at night. Additional gaps are introduced by weather conditions and the full moon. Being pointed instruments, IACTs can thus only observe a limited sample of sources and do not provide continuous light curves. SGSO on the other hand will be able to observe a large fraction of the sky and provide regular, periodic light curves which are ideal for characterizing blazar variability and flux states systematically at TeV energies in an unprecedented way.

The variability of blazars can be characterized using methods such as the power spectral density (PSD), structure functions, and probability distribution functions. On one hand this provides information about the nature of the underlying process, and on the other it enables studies of the flux states of the sources. Comparing the measured distributions to those of time-dependent models, conclusions on the emission processes can be drawn and the model parameters can be constrained. As illustrated in Figure~\ref{Fig:AGN-TeVCat-spectra}, a significant set of PSDs will be provided by SGSO for these studies.

In blazar studies, flaring activities are often compared to the so-called quiescent state of the sources. However, as blazars are highly variable and not continuously observed, it is unclear whether blazars actually are continuously active or if there is a baseline, quiescent state. Again, thanks to its continuous observations, SGSO can identify times of low activity and trigger high-sensitivity follow-ups with e.g. CTA to study this quiescent state, if it exists, in detail.

\subsection{Blazar Periodicity}
The typical radiation observed from blazars appears to be stochastic and unpredictable. However, in the optical band, periodicity has been claimed with a period of 12 years from OJ\,287~\cite{1988ApJ...325..628S}, and with a period of 16 years from Mrk 421~\cite{2017ApJS..232....7F}. At present, due to observational gaps and incomplete data samples, no unambiguous periodicity has been observed from blazars in the gamma-ray bands, although some studies have found hints of periodicity in the VHE emission from Mrk 501~\cite{2001ICRC....7.2630K,2006APh....26..209O} and the GeV emission from PG 1553+113~\cite{2015ApJ...813L..41A}. Periodic emission would be expected, for instance, if the central object powering the AGN were a system of binary black holes~\cite{2003ASPC..299...83R}. Long-term observations are essential for disentangling deterministic and stochastic processes, especially on time scales of months to years. Only an unbiased monitoring instrument such as SGSO can convincingly demonstrate the presence of periodic components on these time scales in the VHE emission from blazars.

\subsection{Measuring the IGMF}
Blazars can also be used to measure or limit the strength of the intergalactic magnetic field (IGMF) presumed to exist in the voids of the large scale structure. Since the IGMF is likely formed from processes occurring during phase transitions in the early universe, its detection and characterization would serve as a probe of the universe prior to the formation of the CMB~\cite{2013A&ARv..21...62D}. Measurements of the IGMF based on VHE gamma-ray observations typically rely on the development of intergalactic cascades initiated by pair production interactions of very high energy gamma rays with the EBL and continuing via the subsequent upscattering of CMB photons into the high energy or very high energy bands~\cite{1994ApJ...423L...5A,1995Natur.374..430P}. Recent efforts (e.g.~\cite{2010A&A...524A..77A,2017ApJ...835..288A,2018arXiv180408035F,2014A&A...562A.145H,2010Sci...328...73N,2011A&A...529A.144T}) based on the non-observation of the cascade emission place a lower limits on the IGMF strength on the order of $10^{-16}$ to $10^{-14}$ Gauss. However, these constraints rely on assuming that the VHE observations, frequently taken over a few to several tens of hours, represent the average flux from the source on the time scale $T$ over which the cascade develops, which can be approximated by~\cite{2009PhRvD..80l3012N}
\begin{equation}\label{eqn:igmf:time}
T\approx(2\textnormal{ years})\left(\frac{E_\gamma}{100\textnormal{ GeV}}\right)^{-5/2}\left(\frac{B_\textnormal{IGMF}}{10^{-17}\textnormal{ Gauss}}\right)^2,
\end{equation}
where $E_\gamma$ is the observed gamma-ray energy and $B_\textnormal{IGMF}$ is the strength of the IGMF.
SGSO will perform unbiased monitoring of sources used for these IGMF studies, checking this assumption on time scales of 5 to 10 years. These observations will be a crucial component in assessing the robustness of IGMF limits placed by previous IACTs, as well as by CTA in the future.

\section{Galactic monitor}
In the Galaxy most, if not all, high energy (X and gamma-ray) transient events are associated with violent phenomena involving a compact object (CO): a white dwarf (WD), a neutron star (NS)/pulsar/magnetar, and/or a black hole (BH). A non-exhaustive list includes supernovae, tidal disruption events around the supermassive BH Sgr A$^\star$, thermonuclear bursts on the surface of WDs and NSs, and all phenomena related to accretion and relativistic ejections in binary systems containing a CO. \\   

Their transient nature renders all sky surveys and monitoring of prime importance to catch any peculiar event such as bursts, flares, outbursts, or any change in their emission. This has, in the past, been nicely illustrated by the large amount of results triggered by the discovery of new sources and/or serendipitous detection of specific events by, e.g., the All-Sky Monitor onboard the RXTE observatory, MAXI onboard the ISS, both below 10 keV, or INTEGRAL/IBIS and Swift/BAT above typically 20 keV. The recent detection of some Galactic transient at energies above 1~MeV with OSSE, COMPTEL, and INTEGRAL \cite{OSSEbinaries,COMPTELCygX1,CygX1pol}, and/or in the GeV domain with \emph{Fermi} and AGILE \cite{fermicygs,V404fermi}, and/or (possibly) in the TeV range with the MAGIC and H.E.S.S. telescopes supports the need of VHE monitoring of our own Galaxy to search/study transients at these energies. 

VHE detections will be able to answer questions regarding the media and physical processes at work. These can be addressed by probing the spectral and temporal correlations (or absence thereof) of VHE radiations with those detected at longer wavelengths, from radio to the X-rays. This should, in any case, permit us to probe particles acceleration in jetted sources, or in shocks, particle-matter and particle-particle interactions, and eventually will help us constrain feedback processes in the interstellar medium.

\subsection{X-ray binaries: monitoring microquasars}
X-ray binaries are systems hosting a CO which is accreting matter from a stellar companion. They can be divided into two types: low-mass systems (LMXBs, $M_{\mathrm{star}}\lesssim 1-2 M_\odot$), and high-mass systems (HMXBs). Broadly, accretion proceeds through an accretion disc in LMXBs and directly from the strong stellar wind of the companion in the HMXBs. However, the situation is not completely clear-cut; for example, matter transfer occurs via an accretion disc in some HMXBs, particularly where the companion is a Be star, as such stars do not have particularly strong stellar winds.

Both types of X-ray binaries are capable of launching jets of material at speeds close to the speed of light; in such a case, the system is known as a microquasar, by analogy with AGNs. Such emission is always variable, and difficult to predict. Radio monitoring of these sources has permitted specific phases (or states) of the accretion properties to be connected to the presence of jets. These can be either continuous outflows (``compact jets") produced during the hard state, when the X-ray spectrum is dominated by inverse Compton radiation peaking around 100 keV, or discrete, sometimes ``superluminal'' ejections at transitions to softer states, when the spectrum becomes dominated by the thermal emission from the accretion disc and peaks at $\sim$1 keV. See for example \cite{Mcclintock06,fender09,belloni11,corbel13}.

The analogy with AGNs naturally leads to the expectation of gamma-ray emission, but the simple, standard picture does not predict HE emission from microquasars. However, many BH systems have been detected at MeV energies with OSSE~\cite{OSSEbinaries}, COMPTEL e.g. \cite{COMPTELCygX1} and INTEGRAL e.g.~\cite{CygX1pol, V404integral, Rodriguez17}. More recently, there have been detections of at least 3 microquasars at GeV energies using \emph{Fermi}-LAT \cite{fermicygs,V404fermi}. These observations have raised new questions regarding the emission processes and origin of these spectral components (see \cite{Rodriguez17} for a brief review). While some studies have favoured models based on hybrid Comptonization and/or magnetic coronal emission \cite{Delsanto13, Romero14} the detection of highly polarized $>$400 keV emission in jet states in Cyg~X-1 \cite{CygX1pol,Rodriguez15_cyg}, the independent detection of a variable 511 keV annihilation line \cite{Siegert16} and GeV flares in V404~Cyg \cite{V404fermi} seem to point to the jet as the origin for these radiations. In this respect, the detection of baryonic material associated with jets in SS433~\cite{Kotani94}, and more recently possibly in 4U~1630$-$47~\cite{Diaztrigo13} implies that these sources should be strong sources of gamma rays. In this context, the questions regarding the emission processes are essentially similar to those for AGN detailed in Section~\ref{subsec:AGN}, i.e. whether we have pure leptonic, lepto-hadronic~\cite{Pepe15} or hadronic processes and, connected with this, the details of the energetics of the jets and their interaction with interstellar medium.

\begin{table}[!tbp]
  \caption{Selected microquasars of interest for SGSO monitoring.  BH=black hole, NS=neutron star, BHC=black hole candidate.}\label{table:binaries}
  \scriptsize
    \begin{tabularx}{\textwidth}{lcccX}
    \hline\hline
    \bf Name & \bf Type & \multicolumn{2}{c}{\bf Coordinates} & \bf Notes \\
         &      & RA (hh mm ss) & Dec (dd mm ss)  & \\
    
    \hline\hline
    GRS 1915+105 & LMXB & 19~15~12 & +10~56~45 & BH; IR synchrotron; quasi-periodic in IR, radio, mm.\\
   MAXI J1820+070 & prob. LMXB &18~20~22 & +07~11~07 & BHC; bright X-ray (few Crab) outburst \\
      HESS J0632+057 & HMXB & 06~33~01 & +05~47~39 & Colliding wind binary; Be companion    \\
   SS 443 & HMXB & 19~11~49 & +04~58~58 & Microquasar; super-Eddington accretion ; baryons in jets\\
   LS 5039 & HMXB & 18~26~20 & -14~50~54 & Microquasar?; BHC\\
   HESS J1832-093 & Unid. & 18~32~50 & -09~22~36 & HMXB favoured by X-ray obs.\\
    GRS 1758$-$258 & LMXB & 18~01~12 & -25~44~36 & BHC; X-ray flares    \\
    KS 1731$-$260 & LMXB & 17~34~13 & -26~05~19 & Microquasar; X-ray flares    \\
    XTE J1748$-$288 & LMXB & 17~48~05 & -28~28~26 & BHC; radio jets    \\
    1E1740.7$-$2942 & LMXB & 17~43~55 & -29~44~43 & BH; X-ray flares    \\
    H~1743$-$322 & LMXB & 17~46~16 & -32~14~01 & BHC; recurrent outbursts    \\
    GX 339$-$4 & LMXB & 17~02~49 & -48~47~23 & BHC; recurrent outbursts    \\
    4U1630$-$47 & LMXB & 16~34~02 & -47~23~35 & BHC; accretion disk corona; baryons in jets (?)    \\
     Sco X-1 & LMXB & 16~19~55 & -15~38~25 & NS; brightest persistent X-ray source\\
   Eta Carinae & HMXB & 10~45~04 & -59~41~04 & Colliding-wind binary    \\
    MAXI J1535$-$571 & prob. LMXB &15~35~20 & $-$57~13~47 & BHC; bright X-ray ($\sim2$ Crab) outburst \\
    HESS J1018-589 A & HMXB & 10~18~58 & -58~56~43 & gamma rays up to 20~TeV    \\
    PSR B1259-63 & LMXB & 13~02~49 & -63~49~53 & Pulsar system\\
   LMC P3 & HMXB & 05~36~00 & -67~35~11 & Extragalactic; most luminous known\\
   
    \hline
    \hline
  \end{tabularx}
\end{table}

Detections of XRBs in the TeV regime are so far confined to a handful of high-mass systems, few of which are known to contain a jet. In most cases, it is thought that the emission results from particle acceleration in the shock created by the interaction between the strong stellar wind of the companion star and the compact object. Until recently, the only object in the TeV binary catalogue thought to be a microquasar was LS~5039~\cite{HESSLS5039}; even so, the emission from this object is likely to be due to pulsar/stellar wind interaction rather than from a jet (see, e.g.\cite{Zabalza13}). However, highly relevant to SGSO, the archetypical microquasar SS433 has now been detected with the HAWC array; this emission is associated with the interaction of the system's jets with the supernova remnant W50~\cite{HAWCSS433}.

It is worth noting here that the pulsar/stellar wind emission detected at TeV energies, while variable, is predictably so, typically reaching a maximum around periastron or an orbital conjunction. This is in contrast to the emission from jets, which is unpredictable and variable on short timescales of days or even hours. For the small FoV IACTs to detect the latter emission requires either an extremely rapid response or a very accurate prediction of exactly when the emission is expected. Unsurprisingly, attempts to predict such emission are likely to be unsuccessful~\cite{HESSMicroULs}. This is the great advantage of SGSO: its ability to monitor the overhead sky at all times greatly increases the chances of detecting transient emission from microquasars. The \emph{Fermi}-LAT detection of V404 Cygni has shown that HE flares can be very short in these sources ($\sim6$hr \cite{V404fermi}), and therefore serendipitous detection is not only likely but probably necessary. Indeed, there are many such events occurring each year --- a study of data from a collection of keV--MeV satellites over 20 years detected 132 transient outbursts from black hole systems alone, suggesting there is a rich panorama of phenomena which can be studied with SGSO~\cite{WATCHDOG}.

Table~\ref{table:binaries} lists known GeV/TeV microquasars/binary sources that are promising sources to be studied with SGSO. These objects are recurrent outbursting sources (GX 339$-$4, H1743$-$322, ...); known ``superluminal'', persistently bright, and very active sources (GRS 1915+105, 1E 1740.7$-$2942, ...); ``baryonic'' ejectors (4U 1630$-$47); or the more recently discovered, X-ray bright ($>$few Crab) outbursting sources (MAXI J1820$+$070; MAXI J1535$-$571).

\section{Gamma-Ray Bursts and Gravitational Waves} 
\label{sec:GRB}
Gamma-ray bursts (GRBs) are short and impulsive events of gamma-ray emission lasting from milliseconds up to hundreds of seconds during their prompt phase, sometimes followed by a long lasting afterglow emission at low energies (i.e. radio, optical and X-rays). In this brief time interval, GRBs release as much as $10^{50}-10^{54}$ ergs of isotropic-equivalent energy mainly in the sub-MeV energy range, becoming the most luminous gamma-ray sources of the sky. Since their discovery in 1969, GRBs have been the target of many observational efforts at all wavelengths. However, the origin of these enigmatic objects is still poorly understood. The so-called long GRBs (with a prompt duration longer than $\sim 2$~s) are likely associated with the violent death of very massive stars. In the case of short GRBs (duration shorter than $\sim 2$~s), time scales and energetics are compatible with a merger of two compact objects (like a neutron star-neutron star or black hole-neutron star merger). This latter scenario has been recently confirmed by the first observation of a short GRB as the electromagnetic counterpart of gravitational waves emitted during the coalescence of two compact objects~\cite{GW170817}. 

Although electromagnetic emission from GRBs is widely observed from gamma/X-ray down to radio wavelengths at different times from the event's onset, the VHE domain above 100~GeV remains largely uncharted. Observations by {\it Fermi}-LAT, including the detection of a $\sim 94$~GeV photon from GRB~130427A~\cite{highenergyphoton}, have proven that emission from GRBs can extend up to the VHE regime, possibly observed only for the brightest events due to the poor effective area of space-based detectors. The MAGIC Collaboration reported a hint of a signal above 400 GeV from a close-by GRB~160821B ~\cite{2017ifs..confE..84P} and has recently detected GRB~190114C above energies of 300 GeV~\cite{MAGIC-GRB190114C-ATEL}. These results are clear illustrations of the feasibility of GRB observations in the VHE domain with ground based detectors and shows that the transient sky at few $\sim$100 GeV still hides many surprises.
\begin{figure}[t!]
  \centerline{
  \includegraphics[width=0.95\linewidth]{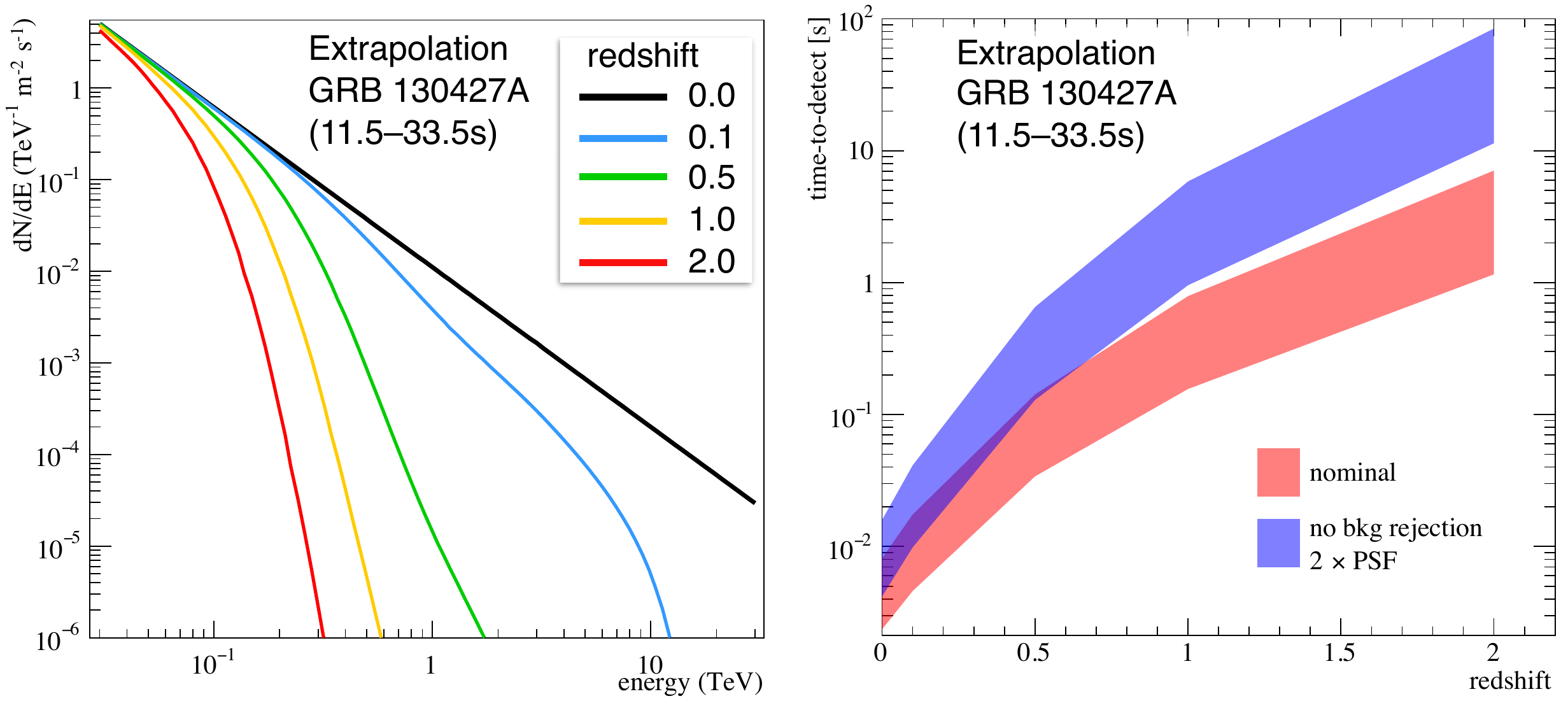}
}
  \caption{{\em Left:} Flux extrapolation of the energy spectrum of GRB~130427A by \emph{Fermi}-LAT~\cite{highenergyphoton} above $>100$~GeV. The spectrum has been attenuated due to interaction of gamma rays with the extragalactic background light for different redshift assumptions according to the model in \cite{2011MNRAS.410.2556D}. {\em Right:} Time needed to obtain a $5\sigma$ detection for the extrapolated energy spectra from the right panel as a function of redshift including uncertainty on the flux normalization (not shown in the left panel). Two scenarios are tested, one with the nominal performance of the straw man design (red) as described in  section~\ref{sec:straw_perf}, and one pessimistic scenario (blue) with no background rejection and point-spread-function that is twice as large as the nominal one.}
  \label{fig:GRBExample}
\end{figure}
In order to assess the feasibility of GRB detection with SGSO, a case study was performed using the observation of GRB~130427A by \emph{Fermi}-LAT~\cite{highenergyphoton}. The energy spectrum of the observation in the period 11.5--33.0 s after the start of the burst follows a power-law with spectral index of -1.66$\pm$0.13 above a few GeV. Figure~\ref{fig:GRBExample}~({\em left}) shows extrapolations of this spectrum to higher energies. In order to probe a wide range of spectral assumptions, it has been attenuated with a model for interaction of gamma rays with the EBL~\cite{2011MNRAS.410.2556D} for different redshifts (note, no additional $1/d^2$ flux-reduction was taken into account). The right panel of Figure~\ref{fig:GRBExample} shows the time needed for a $5\sigma$ detection according to the straw man design if the burst is observed at a zenith angle of 20$^{\circ}$. Most of the signal will come from low-energy events just above the detection threshold of the observatory. In this regime, the performance might be reduced compared to expectations because of the influence of noise contributions. Therefore two scenarios are considered, one nominal scenario (see section~\ref{sec:straw_perf}) and one pessimistic scenario where no background rejection of hadronic induced air showers was applied and the point-spread-function for gamma-ray events is a factor of two worse than nominal. Even in this pessimistic scenario, the detection of a GRB with a flux normalization like GRB~130427A is expected within seconds for redshifts $z<1$.  

Increasing statistics of GRB observations in the VHE range are a vital step for GRB understanding, potentially allowing for a discrimination between different proposed emission scenarios. Indeed, many questions about the physical properties of GRBs still remain unanswered, such as the nature of the central engine and the mechanisms of particle acceleration and radiation, e.g. the magnetic field~\cite{2017ApJ...848...15F}, the bulk Lorentz factor of the jet~\cite{bulklorentz, 2016ApJ...818..190F} and polarization \cite{2017ApJ...848...94F}. GRBs can also be used as a probe of their environment~\cite{2015ApJ...804..105F}. High-redshift GRBs could be a cosmological tool to test Lorentz-invariance violation~\cite{LIV} and may also provide important information on the intergalactic magnetic field~\cite{magnetic}.

\begin{figure}[t!]
  \centerline{
  \includegraphics[width=0.95\linewidth]{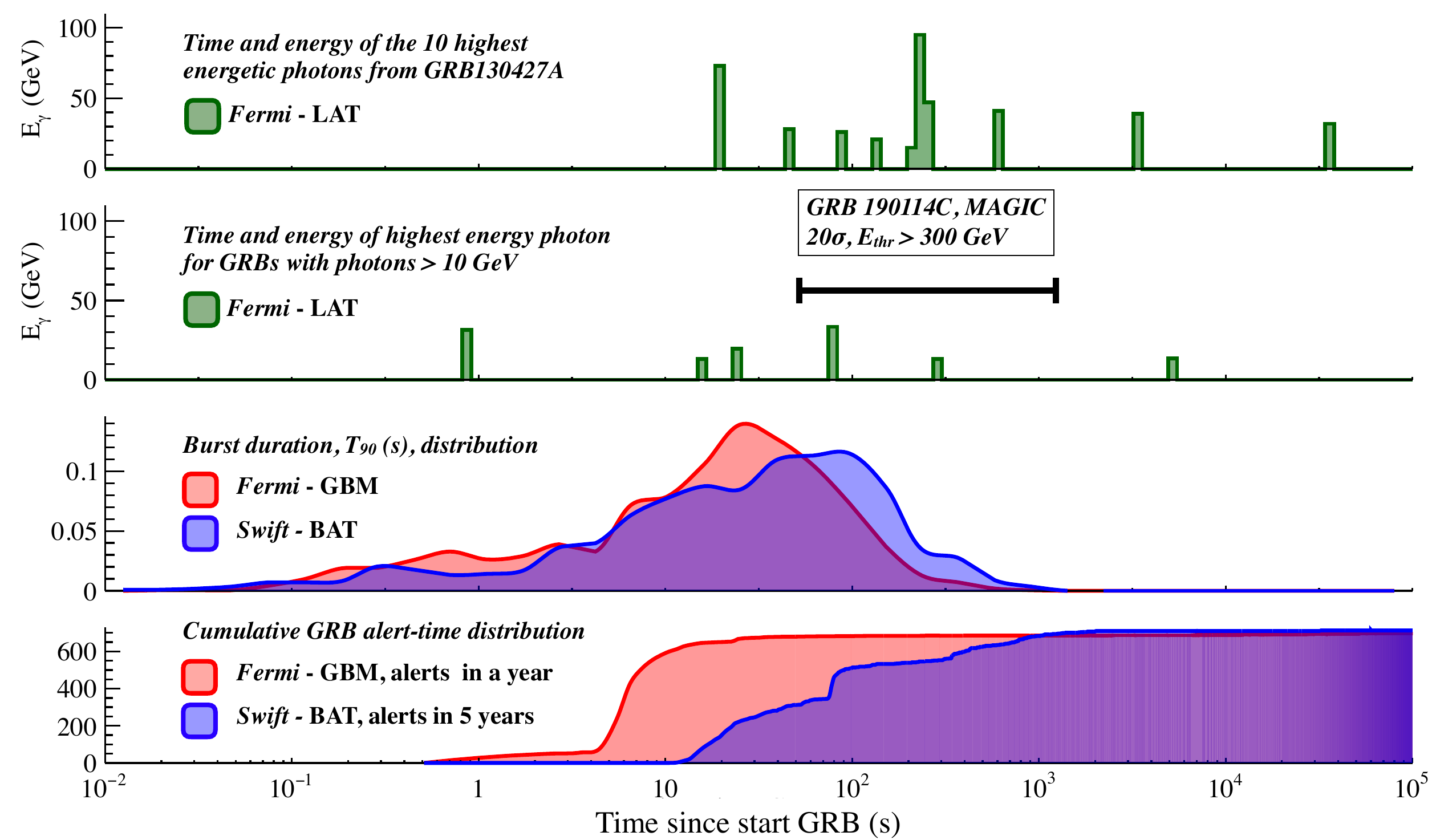}
}
  \caption{Timescales involved in GRB follow-up observations. From top to bottom: arrival time of high energy photons detected by \emph{Fermi}-LAT from GRB130427A, arrival time of photons with E$>$10GeV from \emph{Fermi}-LAT detected GRBs, burst durations in X-rays and delays induced due to the finite alert emission times.}
  \label{fig:GRBs}
\end{figure}

The detection of VHE emission from GRB~190114C~\cite{MAGIC-GRB190114C-ATEL} proves that, under favourable conditions of low redshift and rapid follow-up a firm detection is within the sensitivity range achieved by the current IACT generation. Unfortunately, one of the most limiting factors in GRB observation with ground-based, narrow-field instruments is their low duty cycle. Taking into account the observational constraints, the typical duty cycle for an IACT is $\sim 10\%-15 \%$. It is important to remark that, although achieving better sensitivity and lower threshold than current generation IACTs, the forthcoming Cherenkov Telescope Array (CTA) will be characterized by a fairly similar duty cycle. In addition, the transient and unpredictable nature of GRBs combined with intrinsic alert emission delays by the X-ray satellites makes it difficult for IACTs to observe them rapidly enough to catch the prompt-to-early afterglow phase where certain scenarios favour the emission of VHE radiation. These time ranges are illustrated in Figure~\ref{fig:GRBs}.

For SGSO, a duty cycle of $\sim 100 \%$ is achievable. It would allow the monitoring of all GRBs inside its large FoV and increase the number of observations by a factor of $\sim 10$ ($\gtrsim 10$ GRBs/month) with respect to IACTs. In addition, IACTs have to rely on external triggers provided currently by space-borne X-ray detectors like the instruments onboard the Swift satellite and {\it Fermi}-GBM. SGSO on the other hand will be able to self-trigger on GRBs (and other VHE transients), gathering crucial information that be relayed to the community and especially the southern part of the CTA observatory in real-time. As SGSO will be monitoring the same sky, CTA can fully profit from these alerts.

Furthermore, the possibility to perform an archival data analysis will allow to bypass the delays until the alerts announcing GRB detections are distributed (cf. Figure~\ref{fig:GRBs}). This guarantees the coverage of the important time window from the pre-burst phase up to the very early afterglow phase where IACT observation are strongly affected by delays in the alert chain. This is particularly important for short GRBs, for which the needed reaction time is usually much shorter than the alert emission plus response capability of any pointing (narrow-field) instrument. 

The recently-established link between short GRBs and mergers of binary neutron star systems promises that the number and variety of GRB detections will improve significantly following the enhancements of the current and future gravitational-wave (GW) interferometers. SGSO can play a key role in the Southern hemisphere by observing GW/GRB locations over extended periods of time not interrupted by daylight or moonlight (see below for details). Probing GeV-TeV emission with sufficiently-large photon statistics during this otherwise inaccessible time interval would provide important information on the physics at work in GRBs. In particular, SGSO follow-up would be useful to discriminate between the different proposed radiation mechanisms such as inverse Compton and/or scenarios involving hadronic particles.

\subsection{Gravitational waves} 
Gravitational wave astronomy was launched with the start of operation of the LIGO interferometers in their advanced configuration at the end of 2015, and was further boosted when the European interferometer Virgo joined the global GW network on August 2017. Very rapidly, the detection of GW signals from the mergers of binary black holes became an established window to the nearby universe. Another crucial breakthrough was the first real joint GW multi-messenger observation: the first detection of GWs from the merger of a binary neutron star system (GW170817~\cite{GW170817}). It was also the most extensive astronomical observational campaign up to date~\cite{GW170817_MMA}. The observations of the associated GRB~170817A and the subsequent kilonova emission across the electromagnetic spectrum clearly established GW multi-messenger astronomy as a new and very promising field of astrophysics. 

Observations of GW170817 with H.E.S.S. illustrated the performance and capabilities of IACTs in this domain~\cite{GW170817_HESS}. While H.E.S.S. was able to obtain the first ground-based observations of the region of the event, problems in the LIGO data analysis delayed the publication of the GW localization maps by 5.3 hours. Although the H.E.S.S. observations started only 5 minutes after the publication of the GW uncertainty region by Virgo/LIGO, it becomes clear that only a monitoring instrument like SGSO is able to provide on-source data right at the time of the event regardless of any possible alert emission delays. Since GW data analysis is inherently complex, the time needed to analyze the data recorded by the interferometers to produce first sky localizations of GW events will exceed the minute scale throughout the foreseeable future. Even if the event is received promptly by current or future IACTs, their narrow field of view compared to the GW localization region (ranging from tens to hundreds of square degrees, cf.~\cite{2018LRR....21....3A}) would also introduce a significant delay in covering a sizable fraction of the GW event, or could make the selection of promising targets for follow-up too model-dependent and therefore biased. 
Given that the timescale of the prompt VHE emission of GRBs is expected to be of the same order of magnitude, the capability of SGSO to provide unbiased archival data from this phase is of utmost importance.

Although the continued upgrades of the GW observatories will lead to extending their horizons over the next years, the most promising GW events expected to have an EM counterpart (NS-NS and NS-BH mergers) will still only be detected in the nearby universe (redshift $z<0.1$ for NS-NS). Therefore EBL absorption and thus the higher-energy threshold of SGSO compared to IACTs does not play a dominant role.

During the operation of SGSO, Advanced Virgo and Advanced LIGO will have reached their design sensitivity and new interferometers (e.g. KAGRA and LIGO-India)  may have commenced their first data taking operations. The individual duty cycle of GW detectors will probably remain around $80~\%$ and one can thus expect a significant number of detected events: 4-80 BNS/yr assuming design sensitivity of Advanced LIGO/Virgo (years 2020+) and even 11-180 BNS/yr for the Advanced+ configuration of LIGO/Virgo (years 2024+)~\cite{2018LRR....21....3A}. Even though follow-up observations of GWs are currently considered high-priority ToOs for IACT-based observatories like CTA, the limited amount of available observation time (e.g. 5 h/yr/site for CTA~\cite{CTA_ScienceTDR}) will force them to impose strict selection criteria for the most promising candidates. SGSO would be able to record real-time high-energy gamma-ray data for all GW events falling into its FoV without having to select particular events to observe. Combined with the roughly ten-fold duty cycle of SGSO compared to IACTs, this translates directly into an enormous range of discovery opportunities including new classes of events, like the large domain of currently un-modeled burst-like GW signals. At the same time, the potential discovery of high-energy gamma-ray emission associated with GWs by SGSO also creates increased opportunities for CTA. Being located in close proximity, real-time notifications of SGSO detections will allow CTA to react and perform detailed follow-up observations covering a large energy range with high sensitivity.

While at the time of operation of SGSO most of the GW events will be recorded by at least three interferometers, their median sky localization will still be between $110$ to $180~\mathrm{deg}^2$ (9 to $12~\mathrm{deg}^2$) for observations in the years 2020+ and 2024+ respectively~\cite{2018LRR....21....3A}. SGSO with its large FoV will be one of the very few instruments able to cover these sizable regions instantaneously.

\section{High-energy neutrinos}
\label{sec:neutrinos}
High-energy neutrinos provide a crucial piece of information in the quest for the sources of high-energy cosmic ray accelerators as they are evidence for interactions of high-energy hadronic particles. Significant experimental efforts over the last decades have been crowned by the detection of an astrophysical flux of high-energy neutrinos by the IceCube collaboration~\cite{IceCube_HESE}. On the other hand, no individual neutrino source could be identified so far~\cite{IceCube_SourceSearches}. 

A very promising way to achieve the goal of localising high-energy neutrino and thus cosmic rays sources is the multi-messenger approach: combining the advantages of high-energy gamma-ray observations (i.e. precise localisation of the emission region allowing for the identification of the astrophysical source, study of the overall source energetics and the energy and time dependent power output, etc.) with the link to hadronic origin of the radiation brought by high-energy neutrino observations. 

First attempts were made with the Neutrino Target of Opportunity and Gamma-ray Follow-Up programs, involving IceCube, MAGIC and VERITAS \cite{NToO:2007, 2016JInst..1111009I}. Also searches for persistent gamma-ray emitters in the error box of individual high-energy neutrinos~\cite{HESSICRC15, VERITAS_MS, MAGIC_KS} were performed. All major high-energy neutrino telescopes have now implemented and commissioned automatic data analyses and alert systems able to inform the broader community within seconds about the detection of a promising neutrino candidate~\cite{IceCube_AlertSystem, ANTARES_TAToO}. Follow-up programs of these alerts have been installed with all current IACT collaborations~\cite{IC-IACTs_ICRC2017} and similar plans are outlined for CTA~\cite{CTA_ScienceTDR}. 

\begin{figure}[t!]
  \centerline{
  \includegraphics[width=0.75\linewidth]{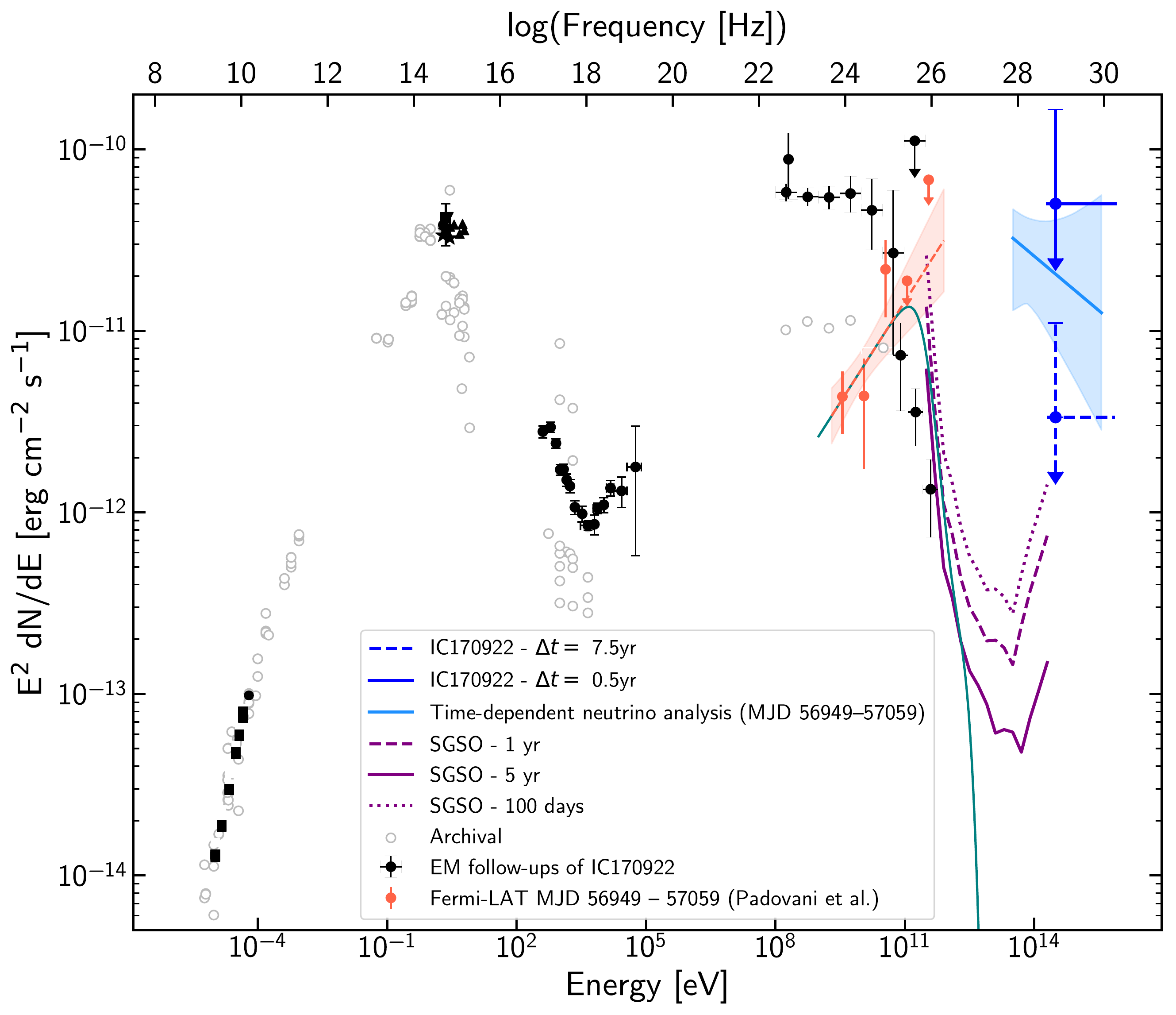}
}
  \caption{Multiwavelength and multi-messenger observations of the blazar TXS 0506+056 illustrating the crucial energy and sensitivity range SGSO will cover. Modified from~\cite{TXS_Science_MM, TXS_Science_Flare}.}
  \label{fig:TXS}
\end{figure}

The searches for transient high-energy gamma-ray emission correlated with high-energy neutrinos already revealed a first promising result: the detection of the flaring blazar TXS 0506+056 in coincidence with the high-energy neutrino IceCube-170922A~\cite{TXS_Science_MM}. The significance of the temporal and spatial correlation between the neutrino event and the blazar, which at the time was flaring in high-energy gamma rays, is estimated to be at the $3\sigma$ level. The analysis of archival data from IceCube revealed an increase in the rate of neutrino events from the direction of TXS 0506+056 over the course of period of about 100 days in 2014-2015~\cite{TXS_Science_Flare}, providing further evidence that TXS 0506+056 is a potential neutrino source. An overview over the MWL observations of TXS 0506+056 and their relation to the neutrino data is given in Figure~\ref{fig:TXS}. This first evidence of a multi-messenger signal involving high-energy neutrinos opened a new window to the violent universe and illustrates one of the likely paths to resolving the century old quest for the sources of cosmic rays. 

On the other hand, this first event also raised many new questions and the need for follow-up studies and confirmations that can only be brought about by SGSO. A crucial one is the need of long-term light-curves of blazars at the highest energies. This is illustrated further by the lack of sensivity VHE gamma-ray observations of the TXS 0506+056 during the 2014-2015 that could help constrain the high-energy emission of the source. As discussed in Sec.~\ref{sec:AGNvariability}, SGSO will be the only instrument to provide these unbiased, quasi-continuous light curves for blazars in the Southern sky. Statistical analysis of light-curves will allow to derive the frequency of occurrence of blazar flares in the TeV energy range, a crucial input to evaluate the significance of neutrino-blazar flare correlations. This is currently not available in the VHE regime. In addition, only SGSO will monitor sources that are spatially consistent with the neutrino direction over a wide range of time scales. 

The MeV-GeV flare of TXS 0506+056 detected by \emph{Fermi}-LAT lasted for several months, but the source was detected above a 100 GeV only 11 days after the neutrino~\cite{MAGIC_TXS}. While H.E.S.S. observations taken 4h after the neutrino detection did not reveal significant gamma-ray emission~\cite{ATEL:HESS170922}, only an extensive campaign over two weeks following the event allowed the detection of TXS 0506+056 by MAGIC~\cite{MAGIC_TXS}, and over a few months for VERITAS~\cite{VERITAS_TXS}. While these extensive campaigns allowed to detect variability at the timescale of days~\cite{MAGIC_TXS}, the necessary observation/monitoring time will be difficult to obtain for a large number of follow-up observations. Although with a higher energy threshold compared to IACTs, SGSO will be the prime instrument to search for high-energy gamma-ray associated to any neutrino alert falling into its field of view. SGSO will be the only instrument able to detect gamma-ray emission at any timescale surrounding the arrival time of neutrino events and will thus be able to provide an unbiased assessment of the correlation between the two messengers. 

The capability of SGSO to provide gamma-ray observations for an unlimited number of neutrino candidates will also be of crucial importance in order to deal with the expected increasing number of neutrino alerts. Existing instruments like IceCube are continuously refining their data filtering strategies with the aim to increase the rate of detections/alerts while keeping the purity at a relatively high level (around $50\,\%$ as for the current alerts\footnote{\url{https://gcn.gsfc.nasa.gov/amon.html}}). More importantly, the future extension of IceCube (IceCube-Gen2~\cite{IceCubeGen2}) and the KM3NeT neutrino telescope~\cite{KM3NeT} currently under construction in the Mediterranean Sea will provide a significant increase in the number of potentially interesting events. While pointing instruments are limited by both their duty cycle and the limited amount of observation time allocated for each science topic (e.g. 5h/year/site are currently allocated for follow-up observations of high-energy neutrinos with CTA~\cite{CTA_ScienceTDR} once the full CTA array is operational), only SGSO will be able to observe all of them for a large amount of time. 

Its location in the Southern hemisphere also allows SGSO to benefit from the large and clean samples of the most interesting very-high-energy neutrinos that will be detected as downgoing or Earth skimming events by IceCube and especially IceCube-Gen2. Combined with lower energy events provided by KM3NeT using neutrinos that cross the Earth, SGSO will be able to provide associated gamma-ray observations that are largely complementary to northern observatories like HAWC and LHAASO.

\section{Multi-messenger and multi-wavelength observations} 
As it is the case for any detector, there will be signal (i.e. gamma-ray) events in SGSO that are indistinguishable from cosmic-ray background events, so called \textit{subthreshold} events. Although, such events do not reach the detection threshold, they can be statistically recovered. They are likely to be spotted through the search and correlation with one of their possible multi-messenger counterparts, which could either lead to individual, combined detections or at least provide a statistical measure of correlated emissions.

The Astrophysical Multimessenger Observatory Network (AMON) is designed specifically to look for this kind of events~\cite{amon}. AMON serves as a cyber-infrastructure that connects several observatories in order to combine their observations, then launches a coincidence analysis and reports significant joint events. Those will be distributed to AMON partners, or publicly in the case of the corresponding data being public, through the Gamma-Ray Coordinates Network (GCN) \cite{gcn}. As of yet, AMON has successfully performed archival coincidence studies of IceCube and \emph{Fermi}-LAT data \cite{icFermi}. Other pipelines are currently in development for both archival and real-time analyses. The ultimate goal is to include all astrophysical messengers: cosmic rays, neutrinos, gamma rays, and the recently detected gravitational waves. Moreover, AMON also serves as a pass-through for the distribution of individual, high-energy IceCube events to the community. 

By the time SGSO will be operative, AMON will reach a mature performance level. The large FoV and expected performance of SGSO will provide extremely valuable input for AMON coincidence analyses searching for transients with expected VHE gamma-ray emission. The main opportunity would be certainly the search for correlated emission of gamma rays and high-energy neutrinos detected by IceCube-Gen2 and KM3Net. Expected in most models of hadronic acceleration, the discovery of joint gamma-ray and neutrino emission would be used to trigger MWL follow-up observations across the electromagnetic spectrum and thus allow detailed studies of the underlying mechanisms.

\begin{figure}
\includegraphics[width=0.52\hsize]{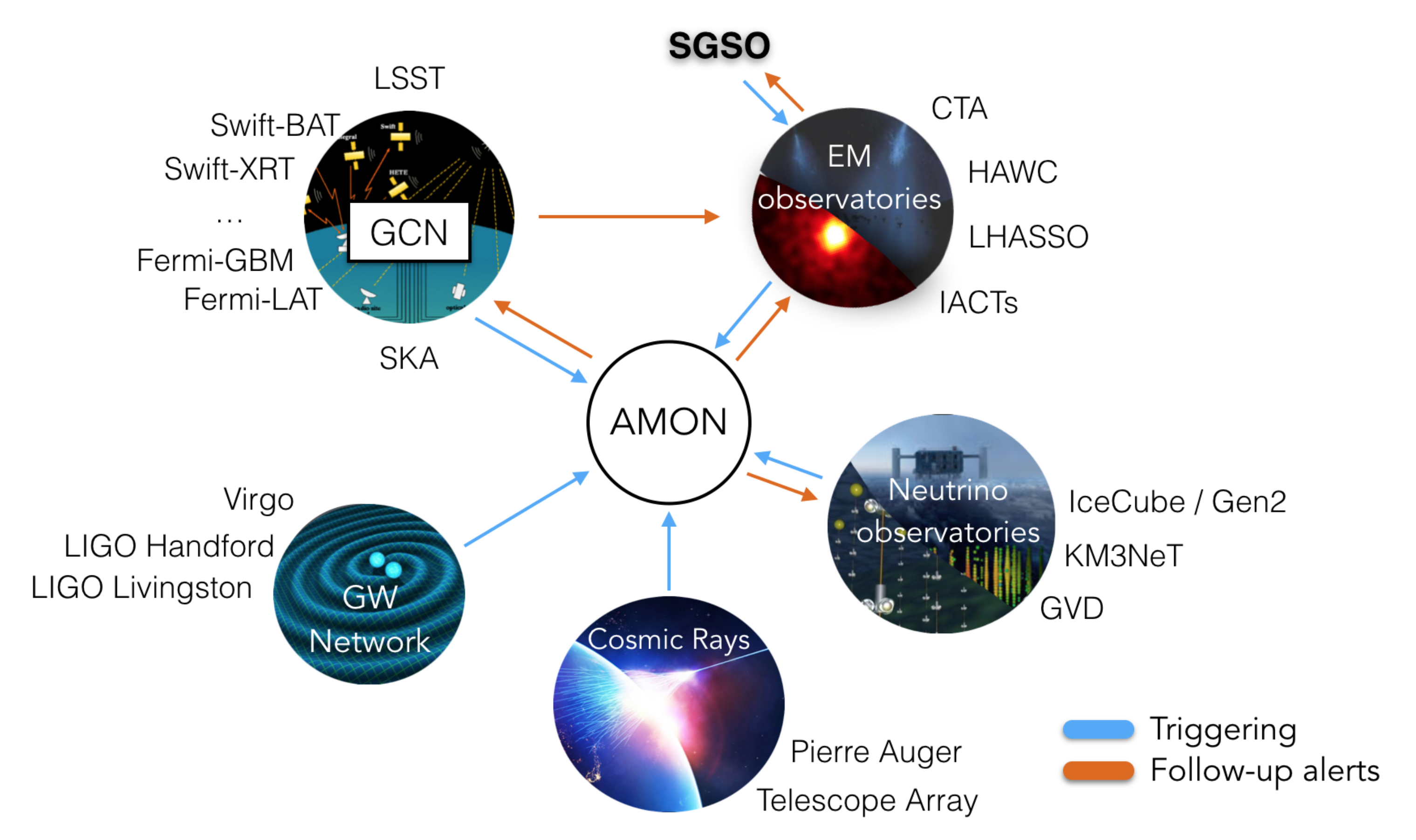}
\includegraphics[width=0.45\hsize]{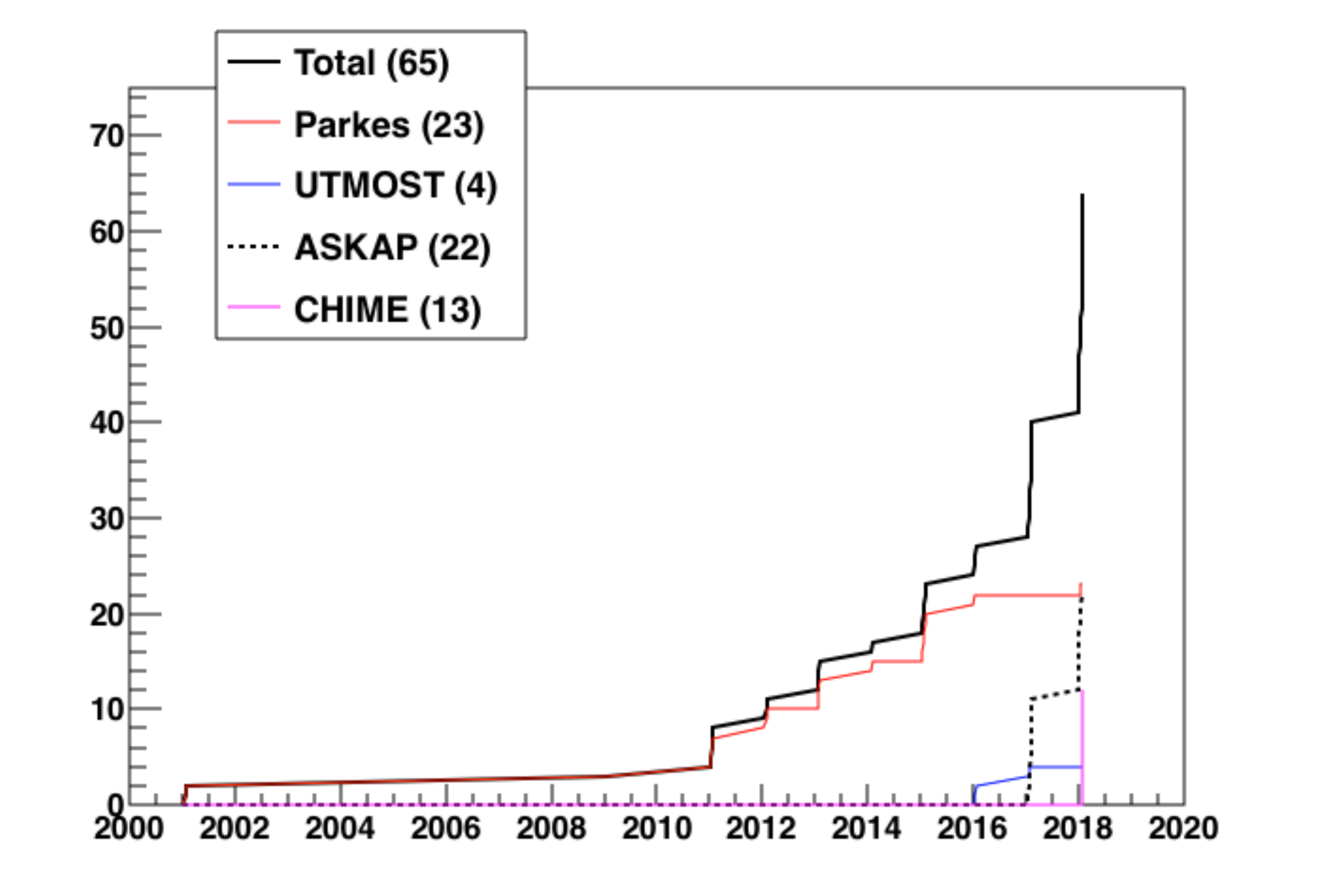}
\caption{{\em Left:} SGSO will be embedded in the Astrophysical Multi-messenger Observatory Network (AMON) and the MWL community enabling real-time searches for multi-messenger transients and follow-up observations. {\em Right:} Over the last years the number of detected Fast Radio Bursts has increased significantly. Data from FRBcat~\cite{Petroff:2016}.}\label{fig:AMON+FRB}
\end{figure}

Due to its design, SGSO is also a perfect observatory to participate both in archival analyses and in real time searches for transients at different timescales. It can thus react to AMON triggers and provide further detailed information or constraints about the potential transient sources detected in AMON analyses. The place of SGSO within the AMON network and in general the worldwide multi-messenger and multi-wavelength community is illustrated in Figure~\ref{fig:AMON+FRB}.

\section{Exploratory searches for new transient phenomena} 
The transient sky keeps providing new surprises. One of the most prominent examples of a major astronomical mystery that has emerged in the last decade and now form a new class of transient objects are Fast Radio Bursts (FRBs). These millisecond-duration bursts were first noticed in 2007 in archival data taken with the Parkes radio telescope~\cite{LorimerFRB}, and over 60 of them have been detected so far. (cf. the online catalog FRBCAT\footnote{\url{http://frbcat.org/}} ~\cite{Petroff:2016}). The rapid evolution of the field over the last years is illustrated in Figure~\ref{fig:AMON+FRB}. Due to their frequency-dependent dispersion properties, FRBs are considered to be of extragalactic origin with their distance being estimated to $z\sim0.1-1$~\cite{Petroff:2016}, which is supported by the localization of the only repeating source of FRBs within a small galaxy at $z=0.19$ \cite{Tendulkar:2017}. Several scenarios link the typical radio energy output of a few $10^{39}D^2_{\rm 1\,Gpc}$\,erg, assuming isotropic emission at distance $D_{\rm 1\,Gpc}$\;=\;$D/ 1\mathrm{Gpc}$, and the millisecond duration of FRBs to cataclysmic events involving compact objects (white dwarfs, neutron stars and/or black holes). A review of potential sources is available for example in~\cite{Kulkarni2014}. Many of these scenarios show similarities with other transients seen also in the X-ray and multi-GeV gamma-ray bands such as short and long GRBs~\cite{Zhang:2014}. Several models have also specifically suggested the existence of flares in the TeV band (e.g., \cite{Lyubarsky2014,Murase:2016}) and proposed follow-ups of FRBs at very high energies.

A first follow-up campaign of FRBs in VHE gamma rays has been conducted by the H.E.S.S. observatory after the detection of FRB150418~\cite{2017A&A...597A.115H}. Limited by the alert emission delay by the SUPERB project on the Parkes radio telescope and the strong constraints on the observation conditions of IACTs, data taking started only within 14.5 hours of the radio burst. No gamma-ray emission was detected and integral upper limits (at $99~\%$ C.L.) at $\Phi_\gamma(E > 350\,\mathrm{GeV}) < 1.33\times 10^{-8}\,\mathrm{m}^{-2} \mathrm{s}^{-1}$ were obtained. 

In the years 2016 and 2017 MAGIC~\cite{2018arXiv180900663M} and VERITAS~\cite{VERITAS_FRB} conducted observations of FRB121102 in simultaneity with the Arecibo radio telescope. FRB121102 is associated with a host galaxy at redshift z = 0.193 and the first known repeating FRB. Several bursts were detected by Arecibo during these observations.  Both MAGIC and VERITAS searched for millisecond-timescale burst emission in VHE gamma rays as well as the optical band. No persistent nor burst-like VHE or optical emission was detected by the IACTs. The average integral flux upper limits set by these observations above 100-200 GeV at 95\%  confidence level is at the level of 5-7 $\times$ 10$^{-12}$ [photons cm$^{-2}$ s$^{-1}$] over the entire observation period, and for the MAGIC coincident observations of 1.2 $\times$ 10$^{-7}$ [photons cm$^{-2}$ s$^{-1}$] for the five detected bursts~\cite{2018arXiv180900663M}. 

The online data analysis of radio telescopes is making significant progress in various projects and will be a major advantage of the upcoming Square Kilometer Array (SKA) and its predecessors. Nevertheless, these analyses and the subsequent alert creation, emission, reception, and reaction of pointing instruments will not be able to reach the sub-second time scales involved in FRBs.  A monitoring approach by IACTs would also have its drawbacks, as the duty cycle of the bursters is poorly constrained and therefore could lead to long coordinated observations during which no FRBs are detected in radio. Therefore, only large FoV, monitoring instruments like SGSO will be able to provide constraints on the gamma-ray emission in coincidence with the bursts detected in the radio domain.

In general, searches for novel transient phenomena are progressing at all wavelengths and for all astrophysical messengers. Similar to the unexpected discovery of FRBs, we may witness other surprises, the establishment of new source classes, and the detection of new phenomena. Very likely, some of these discoveries may happen only after a detailed, off-line analysis of the acquired data (cf. the discovery of FRBs). Only the continuous observations of SGSO will allow to search for strictly contemporaneous VHE gamma-ray counterparts to these novel phenomena. Thanks to the large data archive being built by SGSO, these searches can be performed at any time scale ranging from seconds or minutes necessary to exchange alerts to years after the events have been recorded. 

Another exciting possibility is a blind search for new transient TeV phenomena like low luminosity GRBs~\cite{2018arXiv180807481B}. This kind of search has not been performed by any gamma-ray instrument so far. It can be performed both offline, allowing for a better sensitivity, as well as online, with a goal of alerting the multi-messenger community in real time. Such a dedicated real time analysis (RTA) can also help in pinpointing the location of GW or neutrino emitters which will originally have large localization uncertainties.

\cleardoublepage
\chapter{Probing Physics Beyond the Standard Model}

\section{Dark Matter} 

\subsection{Introduction} 
Although the evidence for astrophysical dark matter (DM) is plentiful, from galactic rotation curves, galaxy cluster dynamics, the cosmic microwave background, and others, the nature of the DM is undetermined and a top science goal for SGSO. One primary direction for DM studies is the indirect search for Weakly Interacting Massive Particles (WIMPs). These are particles with masses in the GeV-TeV range and weak-scale interaction strength, although related models have expanded the mass range to include PeV masses and stronger interactions or even decaying DM (e.g. dark glueballs~\cite{Acharya:2017szw,Boddy:2014yra,Cohen:2016uyg,Faraggi:2000pv,Forestell:2016qhc,Halverson:2016nfq,Soni:2017nlm} and hidden sector DM~\cite{Berlin:2016vnh,Berlin:2016gtr}).

If the DM particle has a mass well above the TeV scale, the only discovery space in the near future may be astrophysical, as these particles would be well above the mass range accessible via production in terrestrial particle colliders and would have number densities too low for direct-detection searches. However, the high-dark-matter-density regions observed astrophysically and the high-energy reach of astrophysical experiments like SGSO make it possible to detect gamma-ray signals from dark matter, even for DM masses much greater than 1 TeV can be identified. In particular, the large field-of-view of SGSO will enable robust searches for consistent dark matter signals from multiple source classes across the sky, including extended emissions such as DM particles decaying in galactic halos, as well as rigorous quantification of signal backgrounds. At the highest energies, where the sensitivity of SGSO peaks, the astrophysical backgrounds are expected to be small and allow for detection of very faint dark matter signals.

\subsection{Synergies} 
Importantly, by combining deep observations of the Galactic Center region with \emph{Fermi}-LAT, CTA and SGSO, a thermal relic cross-section could be probed for DM masses $\lesssim 80$ TeV (see Fig~\ref{fig:DM_sens} and ref.~\cite{Acharya:2017ttl,Fermi-LAT:2016uux}). For masses close to the CTA/SGSO overlap region, greatly increased confidence in a detection could be achieved by measurements in both detectors. In the mass range 10 -- 80 TeV, SGSO would provide the WIMP mass measurement by probing the spectral cut-off, with CTA helping to constrain the morphology (see Fig~\ref{fig:DM_sens}, right panel). We note that this mass range has considerable advantages to the GeV range in terms of astrophysical foreground, with a much shorter list of objects capable of accelerating particles to these energies and in particular avoiding the magnetospheric emission of pulsars whose spectra can mimic an annihilation spectrum in the GeV~\cite{Daylan:2014rsa,Hooper:2010mq}.

\subsection{Sensitivity to dark matter annihilation and decay} 
The gamma-ray flux from the annihilations (${\rm d}\Phi_{Ann}/{\rm d} E_{\gamma}$) and decays (${\rm d}\Phi_{Dec}/{\rm d} E_{\gamma}$) of dark matter particles of mass $M_{\rm DM}$ in a DM halo can be expressed as the product of a particle physics term (left parenthesis) and an astrophysical term (right parenthesis):\\

\begin{equation}
\label{eq:dm_flux_ann}
\frac{{\rm d}\Phi_{Ann}(\Delta\Omega,E_{\gamma})}{{\rm d} E_{\gamma}}\,= \left(\frac12 \frac{1}{4\pi}\, \frac{\langle \sigma v \rangle}{M_{\rm DM}^2}
	\frac{{\rm d} N}{{\rm d}E_\gamma} \right) \,\times\, \left(J(\Delta\Omega)\right) \, ,
\end{equation}

and 

\begin{equation}
\label{eq:dm_flux_dec}
\frac{{\rm d}\Phi_{Dec}(\Delta\Omega,E_{\gamma})}{{\rm d} E_{\gamma}}\,= \left(\frac{1}{4\pi}\, \frac{1}{\tau_{\rm DM} M_{\rm DM}^2}
	\frac{{\rm d} N}{{\rm d}E_\gamma} \right) \,\times\, \left(D(\Delta\Omega)\right) \, .
\end{equation}

The astrophysical factors, also called J-factor for annihilations and D-factor for decays, are defined as\\

\begin{equation}
J(\Delta\Omega) = \int_{\Delta \Omega}  \int_{\rm l.o.s.} {\rm d}\Omega \, {\rm d} s \ \rho_{\rm DM}^2[r(s,\Omega)] \, ,
\label{eq:Jfactors}
\end{equation}

and

\begin{equation}
D(\Delta\Omega) = \int_{\Delta \Omega}  \int_{\rm l.o.s.} {\rm d}\Omega \, {\rm d} s \ \rho_{\rm DM}[r(s,\Omega)] \, ,
\label{eq:Dfactors}
\end{equation}

where $\rho_{\rm DM}$ is the DM density distribution. Both astrophysical factors are given by an integral along the line of sight (l.o.s.) and over the solid angle $\Delta\Omega$, integrating over $\rho_{\rm DM}^2$ for annihilation, and $\rho_{\rm DM}$ for decay. The particle physics terms contain the DM particle mass, $M_{\rm DM}$, the velocity-weighted annihilation cross section, $\langle \sigma v\rangle$, DM lifetime, $\tau_{\rm DM}$, and the differential spectrum of gamma rays in a specific annihilation or decay channel, ${\rm d} N/{\rm d} E_{\gamma}$.

The sensitivity of SGSO to DM annihilations (decays) can be found by comparing the number of observable gamma rays with the expected background (see Section~\ref{sec:straw_perf}). The statistical tool used to derive limits is a 2D (energy and space) joint-likelihood method, where the comparisons between DM and background fluxes are performed in different energy and spatial intervals (or bins)~\cite{Abdallah:2016ygi}. Assuming that the number of detected events follows a Poisson distribution, the likelihood functions are calculated in each individual bin and combined into a joint-likelihood function from which we derive the limits at different confidence levels. This method takes full advantage of differences on the energy and spatial distribution between the expected DM signal and the background. For instance, the former is supposed to follow the J-factor (D-factor), whereas the latter is isotropic in the sky. Hereafter we divide the energy range between 100 GeV and 100 TeV into 40 logarithmically-spaced bins. The division into spatial bins will depend on the DM halo density distribution of the specific target we are discussing.   

\subsection{Dark matter annihilation searches towards the Galactic halo}
One of the advantages of the construction of a Southern wide field-of-view gamma-ray observatory is the possibility of observing the inner region of the Galaxy. The Galactic Center (GC) region is the most interesting location to look for a DM annihilation signal. This area is expected to be the brightest source of DM annihilations in the gamma-ray sky by several orders of magnitude. Even considering the contamination from standard astrophysical sources, it is one of the most promising targets to detect the presence of new massive particles. However, the search for a DM-induced signal toward the GC region has been inconclusive so far, partly complicated by the detection of multiple gamma-ray sources connected with astrophysical particle accelerators, which has resulted in several major discoveries~\cite{HESSGC,Fermi-LAT:2014sfa,Aharonian:2006au,Hooper:2010mq}. As already mentioned before, these astrophysical gamma-ray sources would have a much smaller influence at the SGSO energy range.  

The DM density distribution of the Galactic halo is poorly constrained, so in order to estimate the predicted DM flux, it is important to consider different classes of halos. Here two models are assumed: a peaked Einasto profile~\cite{Pieri:2009je} and a cored Burkert profile~\cite{Burkert:1995yz}. We focus our searches for DM signals to the inner 10$^{\circ}$ of the Galaxy. The spatial regions of interest are defined as circular concentric regions of 0.2$^{\circ}$ width each, excluding a $\pm$0.3$^{\circ}$ band in Galactic latitude to avoid the above-mentioned standard astrophysical background. 

\begin{figure}[!t]
	\includegraphics[width=0.485\linewidth]{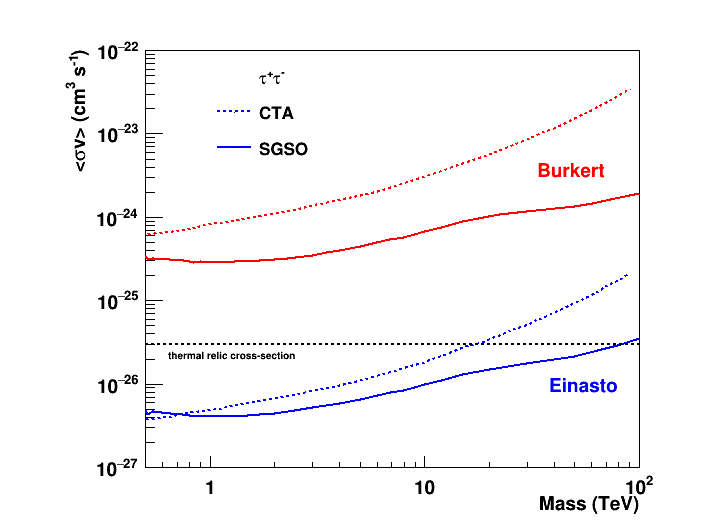}
	\includegraphics[width=0.485\linewidth]{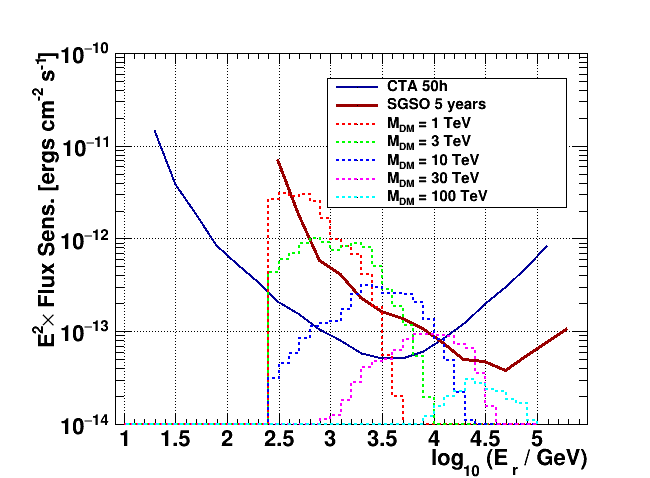}
	\caption{{\em Left:} 95\% C.L. sensitivity upper-limit on the velocity weighted cross section for DM self-annihilation into $\tau^+ \tau^-$ as a function of $M_{\rm DM}$ for Einasto and Burkert profiles of the Galactic halo. SGSO sensitivity is calculated in the inner 10$^{\circ}$, and CTA in the inner 1$^{\circ}$ of the Galaxy, excluding a $\pm$0.3$^{\circ}$ band in Galactic latitude. {\em Right:} SGSO and CTA flux sensitivity curves as function of reconstructed gamma-ray energy, for 5 years and 50 hours of observation of the Galactic halo, respectively. Also plotted are the dark matter annihilation rate into $\tau^+ \tau^-$ per reconstructed energy bin for different DM particle masses in arbitrary units, but keeping $\langle \sigma v\rangle$ and the J-factor the same for all masses.}
	\label{fig:DM_sens}
\end{figure}

Figure~\ref{fig:DM_sens}, left panel, shows the 95\% C.L. sensitivity upper-limits on $\langle \sigma v \rangle$ versus $M_{\rm DM}$ for 10 years of observation of SGSO in the case of a DM particle annihilating into $\tau^+\tau^-$ for Einasto and Burkert profiles of the Galactic halo. The sensitivity of CTA is also plotted for comparison, and it was calculated for 500 hours of observation~\cite{Acharya:2017ttl}. Due to the smaller field-of-view of CTA, the signal extraction region was limited to the inner 1$^{\circ}$ of the Galaxy. A sensitivity smaller than the thermal relic cross-section $\langle \sigma v \rangle \lesssim 3 \times 10^{-26}$ cm$^3$ s$^{-1}$ is reachable for SGSO in the mass range of $\sim$500 GeV to $\sim$80 TeV. It would be more sensitive than CTA for all  masses $\gtrsim$700 GeV. Also, due to the larger signal extraction region of SGSO, the sensitivity difference between the Einasto and Burkert profiles are less pronounced than for CTA (a factor of ~70 for SGSO and ~160 for CTA).      

\subsection{Dark matter searches towards satellite galaxies} 
Dwarf spheroidal galaxies (dSphs) are some of the most dark matter dominated objects known. Dozens are known to exist nearby in the Milky Way dark matter halo. Given their proximity and low astrophysical backgrounds, Milky Way dSphs are excellent targets for searching for gamma-ray emission from dark matter annihilation or decay. For example, HAWC has derived competitive limits on dark matter annihilation and decay using 14 dSphs with known dark matter content~\cite{dSphHAWC}. 

Recent deep observations with wide-field optical imaging surveys have discovered 33 new ultra-faint Milky Way satellites~\cite{DESy1,DESy2,VirIdis,PicIIdis,PegIIIdis,HydIIdis,TriIIdis,DraIIdis,Cardis,HydrusI,CetIIIdis,AquII,CraIIdis,DESJ225dis,PisIIdis} mostly in the Southern Hemisphere. These objects are potentially dark-matter dominated dSphs, but this needs to be confirmed with spectroscopic follow-up observations. 15 of the new satellites have already been spectroscopically confirmed as dSphs~\cite{HydrusI,AquII,RetIIJ,EriIIJ,TucIIJ,HorIJ,TucIIIJ,CarJ,PegIIIJ,HydIIJ,TriIIJ,DraIIJ,CraIIJ} 
In particular, Reticulum II is quite nearby and has a large J-factor ($6.3\times10^{18}$ GeV$^2$ cm$^{-5}$)\cite{RetIIJ}. 
These add to the 18 well characterized dSphs known before these surveys~\cite{LATdSph6yr}, 5 of which are located in the SGSO field of view. Note that many of the previously known dSphs were discovered in the Sloan Digital Sky Survey, which mostly covered the Northern Hemisphere. The potential improvement for SGSO is possibly higher, once surveys of the Southern Hemisphere sky will be available. In addition to newly discovered dSphs, SGSO can search for dark matter signals from the Large and Small Magellanic Clouds.

\begin{figure}[!t]
\includegraphics[width=0.485\linewidth]{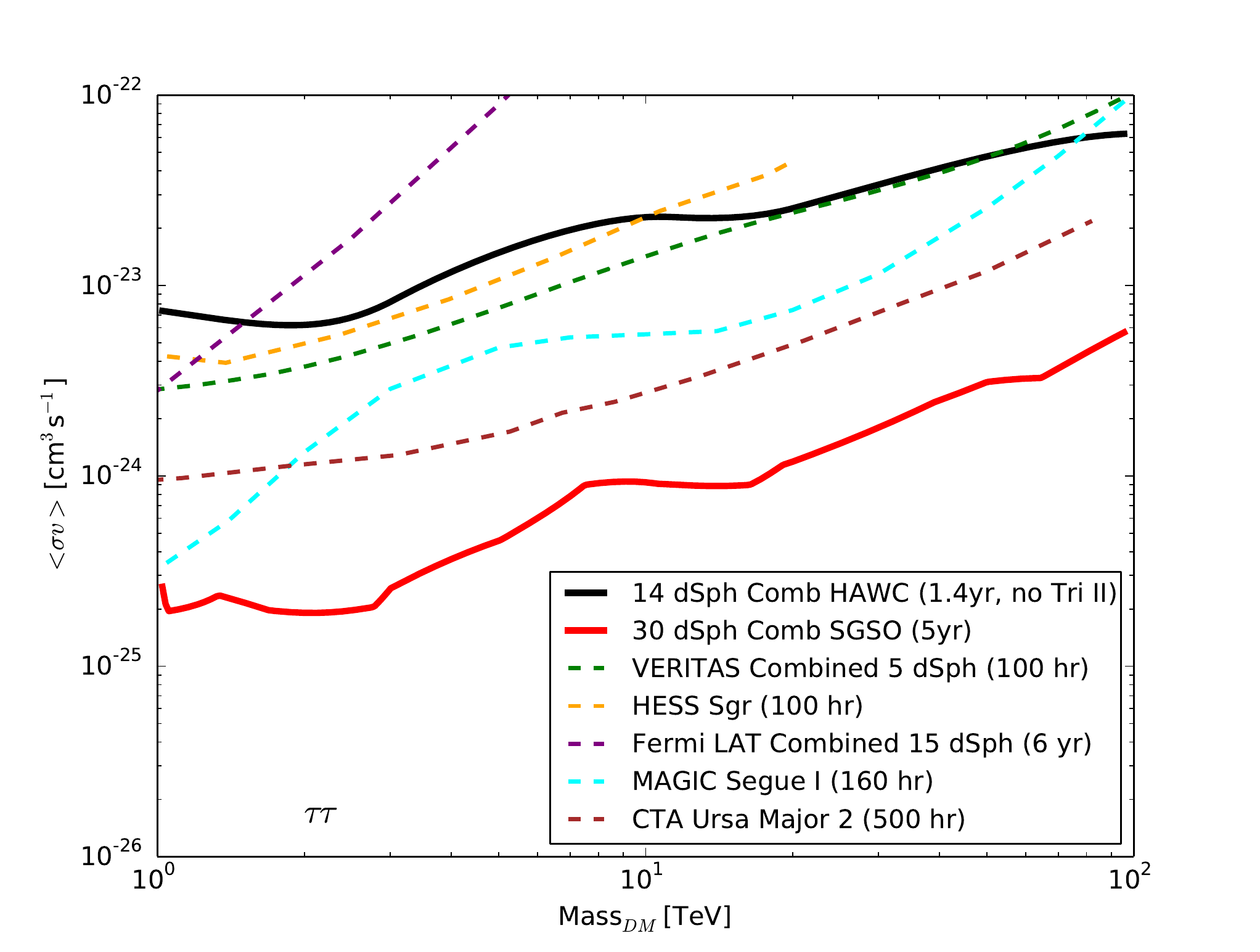}
\includegraphics[width=0.485\linewidth]{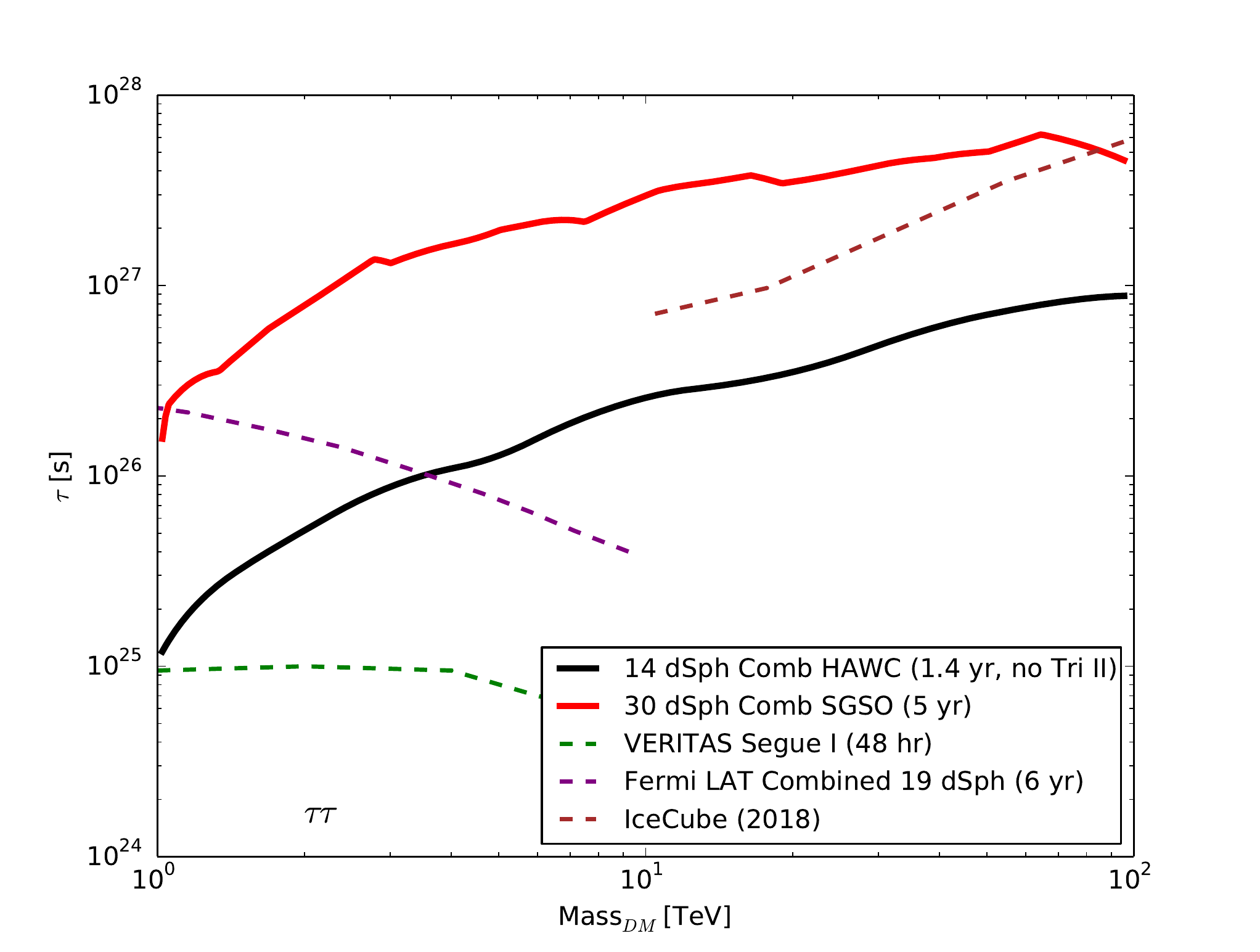}
  \caption{{\em Left:} Expected 95\% confidence level dark matter annihilation cross section upper limits in dSphs. Also shown are the observed dSph limits from VERITAS~\cite{dSphVeritas}, \HESS~\cite{dSphHESS}, \emph{Fermi} LAT~\cite{LATdSph6yr}, and MAGIC~\cite{dSphMagic} and the expected limits from CTA~\cite{dSphCTA}. {\em Right:} Expected 95\% confidence level dark matter decay lifetime lower limits for dark matter decay in dSphs. Also shown are the observed dSph limits from VERITAS~\cite{dSphVERITASSegI}, \emph{Fermi} LAT~\cite{dSphFermiDec}, and IceCube~\cite{dmIceCube}. }
  \label{fig:dSph_comp}
\end{figure}

Figure~\ref{fig:dSph_comp} shows the expected improvement in the dark matter annihilation and decay limits relative to the current HAWC limits given the increased sensitivity of SGSO and an increased number of dSphs. We assume the J-factor and D-factor distributions of the new dSphs matches that of the previously known dSphs. 
With these assumptions, we expect to improve the current limits by nearly an order of magnitude. 

SGSO will be a survey instrument, surveying the southern gamma-ray sky every day. Therefore, any additional southern dSphs that may be discovered in the future will have already be observed by SGSO, unlike the case with IACTs. The Large Synoptic Survey Telescope (LSST) will survey the Southern Hemisphere sky with unprecedented sensitivity and it is expected to find hundreds of new dSphs~\cite{Hargis:2014kaa}. Legacy SGSO data could easily be analysed when new dSphs are found.

\section{Primordial Black Holes} 
Primordial black holes (PBH) are hypothetical black holes which may have been formed in the early stages of the Universe~\cite{2010PhRvD..81j4019C}. Their mass spectrum depends on the formation mechanism and spans a large mass range. They are expected to evaporate by emitting a thermal spectrum at Hawking temperature \cite{1974Natur.248...30H}. The lifetime of a PBH is $\tau\approx4.55\times10^{-28}(M_{BH}/1g)^3$ s \cite{MacGibbon:1991tj}, i.e. a PBH with a mass of $\sim5\times10^{14}$ g would have an evaporation time of about the current age of the universe. In the latest stage of evaporation, the PBH emits high energy particles which are detectable by atmospheric or water Cherenkov detectors \cite{2016APh....80...90U}.

SGSO will search for the bursts of high-energy gamma-rays created by the evaporation of PBHs with an initial mass close to $\sim5\times10^{14}$ g. Bounds on the local density of low mass PBHs  have been reported by MILAGRO \cite{2015APh....64....4A} and are expected to be lowered by an order of magnitude by HAWC. The idea of the analysis is to search for localized and short-term bursts of gamma-ray events unrelated to an astrophysical source. The choice of the time window is a compromise between signal and background, with typical time windows being 1--10 s. 

The sensitivity to PBH bursts is quantified by the  detectable volume, which is the spatial volume over which a PBH burst would give a signal significantly larger than the background more than 50\% of the time. Because of this, the sensitivity to PBHs scales as the 3/2 power of the gamma-ray sensitivity. The sensitivity of SGSO is expected to be roughly ten times better than HAWC. Assuming no PBH burst detection, the upper limits on the local rate of PBH explosion is thus expected to improve by more than a factor of 30 compared to the expected HAWC limits (a factor of 300 better than the MILAGRO limits), reaching the level of $\sim 130 \mathrm{pc}^{-3} \mathrm{yr}^{-1}.$

\section{Axion-like Particles}
Axion-like Particles (ALPs) are hypothesized in many theories beyond the Standard Model~\cite{alp_theory}. They are similar to axions~\cite{axions}, except their mass and coupling strength to photons are independent. ALPs would be produced non-thermally in the early Universe and could account for some or all of the dark matter that exists today~\cite{alp_dm}.

SGSO will have competitive sensitivity at very high energies ($>10$ TeV), making it well-equipped to search for ALP gamma-ray signatures. A potential ALP signature would be the detection of greater than $\sim 30$ TeV photons from a distant, hard-spectrum extragalactic source. The very high energy photons produced at the source are typically not observed at Earth since they have been attentuated by the Extragalactic Background Light (EBL). However, gamma rays produced at the source could convert into ALPs in the intergalatic magnetic field. These ALPs would then travel to us without EBL attenuation. They would convert back into gamma rays in the magnetic field of the Milky Way and then be detected at Earth~\cite{Horns:2012kw,alp_intgal}. In this way, very high energy photons exceeding the expectations including EBL absorption would be detected. Therefore, the detection on greater than $\sim~30$ TeV photons from a high redshift, hard gamma-ray source would be strong evidence for the existence of ALPs. Several such sources are known and will be observed by SGSO, for example PKS 0447-439 (z = 0.343)~\cite{pks0447,pks0447_2} and 1RXS J023832.6-311658 (z = 0.232)~\cite{RXSJ023832}. 

It should be noted that ALP conversions could also produce sharp spectral features in the observed spectrum of extragalactic sources~\cite{alpIrr_fermi,alpIrr_hess}. To resolve these features, a combination of a good energy resolution (cf. Figure~\ref{fig:SGSOperformance}) as well as sizable event statistics are crucial ingredients.

\section{Testing Lorentz invariance with SGSO} 
Very high energy photons can be used as a test of fundamental physics, such as the Lorentz symmetry. As for any other fundamental principle, exploring its limits of validity  has been an important motivation for theoretical and experimental research. Moreover, some Lorentz invariance violation (LIV) can be motivated as a possible consequence of theories beyond the Standard Model (SM), such as quantum gravity or string theory (see for instance Refs.~\cite{Alfaro:2004aa,AmelinoCamelia:2001dy,Ellis:1999uh} and references therein). 

It has been shown that LIV can induce effects that cause photons of sufficient energy to become unstable and decay very fast~\cite{Martinez-Huerta:2016azo}. The latter strongly restricts the propagation of photons to very short distances from the source. Such phenomena can be induced by introducing an isotropic correction to the photon dispersion relation: 
\begin{equation}\label{LIV:eq1}
E_{\gamma}^2 - p_{\gamma}^2 = \pm \frac{E_{\gamma}^{n+2}}{\left(E_{LIV}^{(n)}\right)^n},
\end{equation}
where $E_{LIV}^{(n)}$ is the Lorentz invariance violation energy scale at leading order $n$. Considering only superluminal effects in Eq. (\ref{LIV:eq1}), photon decay, $\gamma \rightarrow e^-e^+$, would be allowed above a certain energy threshold $E_{th}$, so that no photons above that threshold should reach the Earth from astrophysical distances. Hence, observations of astrophysical VHE photons of energy $E_{obs}$ place constraints on the threshold energy ($E_{th} > E_{obs}$) and on the LIV energy scale:

\begin{equation}
E_{LIV}^{(n)} > E_{obs} \left[ \frac{E_{obs}^2 - 4m_{e^-}^2}{4m_{e^-}^2} \right]^{1/n}.
\end{equation}
Constraints to the LIV energy scale have been established by looking at the highest-energy photons from the Crab Nebula and SNR RX J1713.7-3946~\cite{Ahnen:2017wec,Martinez-Huerta:2016azo}. However, stronger limits than these are expected from the Crab source and the absence of photon decay by HAWC in the Northern Hemisphere \cite{Martinez-Huerta:2017gna}. Moreover, new data and limits to $E_{LIV}^{(n)}$ can be obtained from RX J1713.7-3946 measured by SGSO due to its location, energy range, resolution and sensitivity at the highest energies, which will be much better than the limits obtainable with HAWC and improve further with time. 

\begin{figure}[!t]
  \begin{center}
    \includegraphics[width=0.48\linewidth]{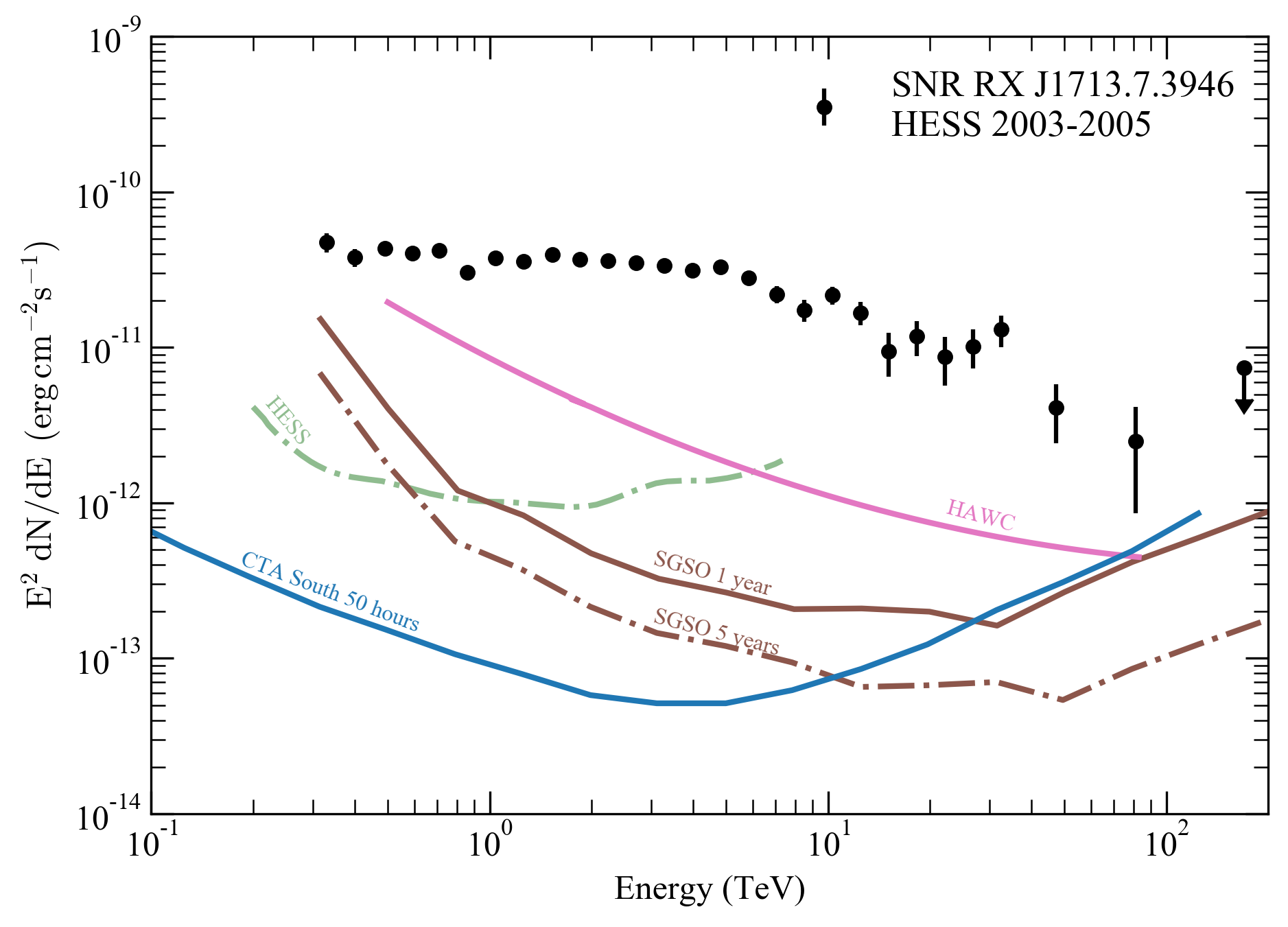}
    \includegraphics[width=0.43\linewidth]{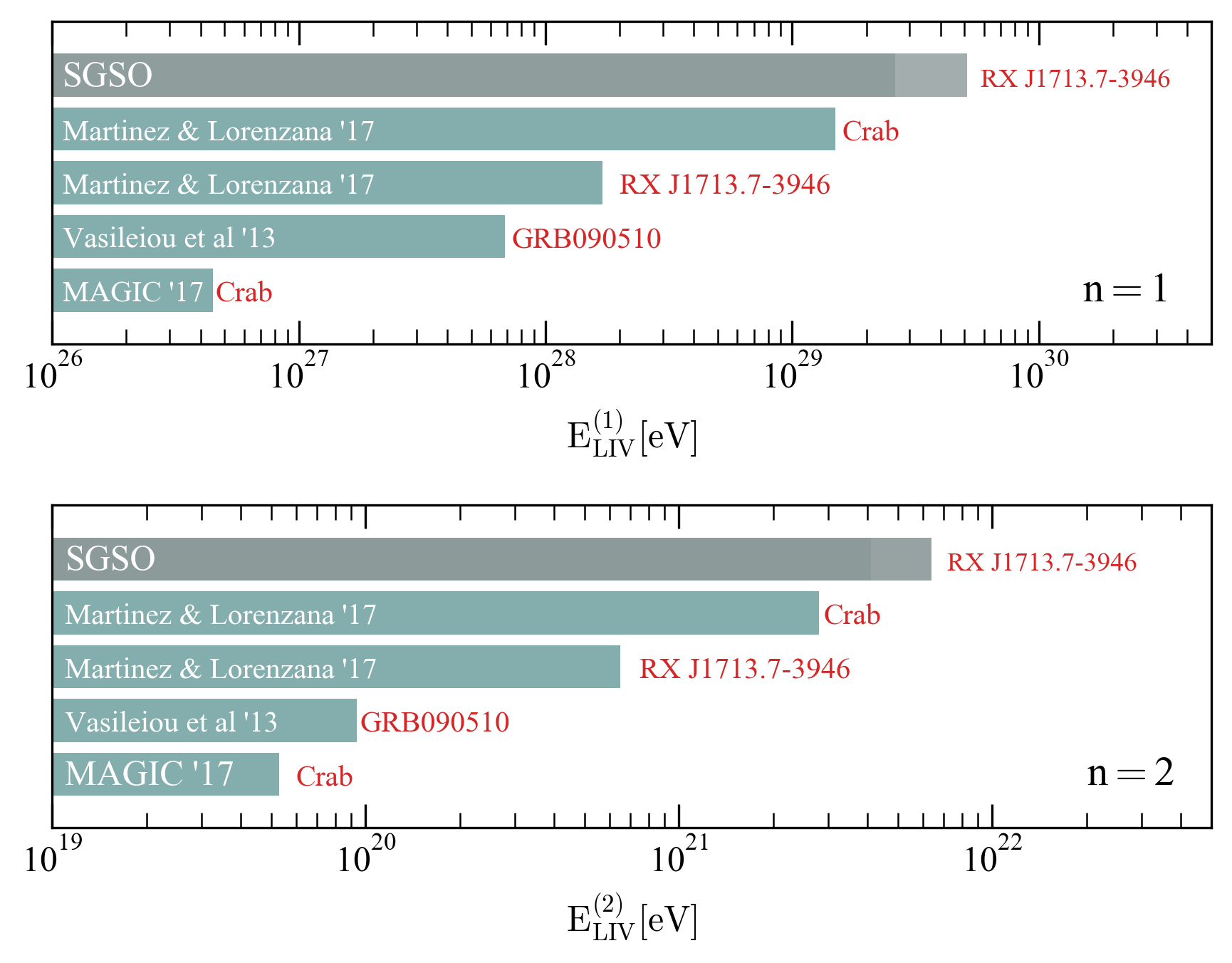}	
   \end{center}
  \caption{{\em Left:} RX J1713.7-3946 source reported by H.E.S.S. \cite{HESS_RXJ1713}, the differential sensitivity threshold of H.E.S.S. and HAWC experiments and the estimated thresholds  of CTA and SGSO at 1 and 5 years. {\em Right:} From bottom to top, $E_{LIV}$ limits ($n$=1 and $n$=2) from superluminal LIV searches for energy dependent time delays, photon decay into electron-positron pairs and the potential reference of SGSO by measuring RXJ1713.7-3946 photons at 80 TeV and 100 TeV. }\label{LIV:fig1}
\end{figure}

The left panel of Figure \ref{LIV:fig1} shows the H.E.S.S. observations for RX J1713.7-3946 from 2003 to 2005~\cite{HESS_RXJ1713}, where the last two energy bins at  47.19 TeV and 81.26 TeV  have been reported with $\sigma=$ 2.5 and 1.5 respectively. There is significant potential for improvements by SGSO. As a reference, potential SGSO limits to LIV energy scale at leading order n=1 and 2 are presented in the right panel of Figure (\ref{LIV:fig1}), by assuming a $\geq5\sigma$ detection at 80 and 100 TeV with 20$\%$ of energy uncertainties. The higher and better the detected gamma-ray energy is, the more stringent the limits on $E_{LIV}^{(n)}$ would be. Thus, instruments optimized at the highest energies, such as SGSO, would be suitable instruments to search for LIV signatures like photon decay. It is worth to notice that already after the first year of observations, SGSO would reach deeper differential sensitivity than current instruments, thus SGSO would be providing more stringent limits on LIV.   

\cleardoublepage
\chapter{Cosmic-ray observations}

\section{Spectrum and composition} 
Understanding the CR origin and propagation at any energy is made difficult in part due to the poor knowledge of the elemental composition of the radiation as a function of the energy. Last generation experiments, measuring with high resolution different EAS components (mainly the number of electrons, N$_e$, and the number of muons, N$_\mu$, at ground level), have reached the sensibility to separate two mass groups (light and heavy) with an analysis technique not critically based on EAS simulations \cite{apel2014} or five mass groups (H, He, CNO, MgSi, Fe) with an unfolding technique that is heavily based on simulations \cite{kascade1, kascade2}.

The main feature in the energy spectrum of Galactic cosmic rays is the so-called  \emph{``knee"}, which is characterized by a steepening of the spectral index from $\sim$-2.7 to $\sim$-3.1 at about 3 PeV (=3$\times$10$^{15}$ eV).
Understanding the origin of the \emph{``knee"} is the key for a comprehensive theory of the origin of cosmic rays up to the highest observed energies. In fact, the knee is clearly connected with the issue of the end of the Galactic CR spectrum and the transition from Galactic to extragalactic CRs. Determining elemental composition in the knee energy region is crucial to understand where the Galactic CR spectrum ends. If the mass of the knee is light, according to the standard model, the Galactic CR spectrum is expected to end around 10$^{17}$ eV. On the contrary, if the composition at the knee energies is heavier and dominated by CNO / MgSi, the interpretation of the CR energy spectrum may not be straightforward.

In the standard picture, mainly based on the results of the KASCADE experiment (located at an altitude of 110 m a.s.l.), the knee is attributed to the steepening of the p and He spectra \cite{kascade}. According to a rigidity-dependent structure (Peter's cycle), the sum of the fluxes of all elements, with their individual knees at energies $E_Z$ proportional to the nuclear charge ($E_Z$ = Z$\times$3 PeV), makes up the CR all-particle spectrum \cite{peters}. With increasing energies, not only does the spectrum become steeper due to these cutoffs, but also heavier.

However, a number of experiments (in particular those obtained by experiments located at high altitudes, like BASJE-MAS and Tibet AS$\gamma$) seem to indicate that the bending of the light component (p+He) occurs well below PeV energies and the knee of the all-particle spectrum is due to heavier nuclei \cite{tibet,casamia,basje-mas}. 

\begin{figure*}[!t]
\centering
\includegraphics[width=0.85\linewidth]{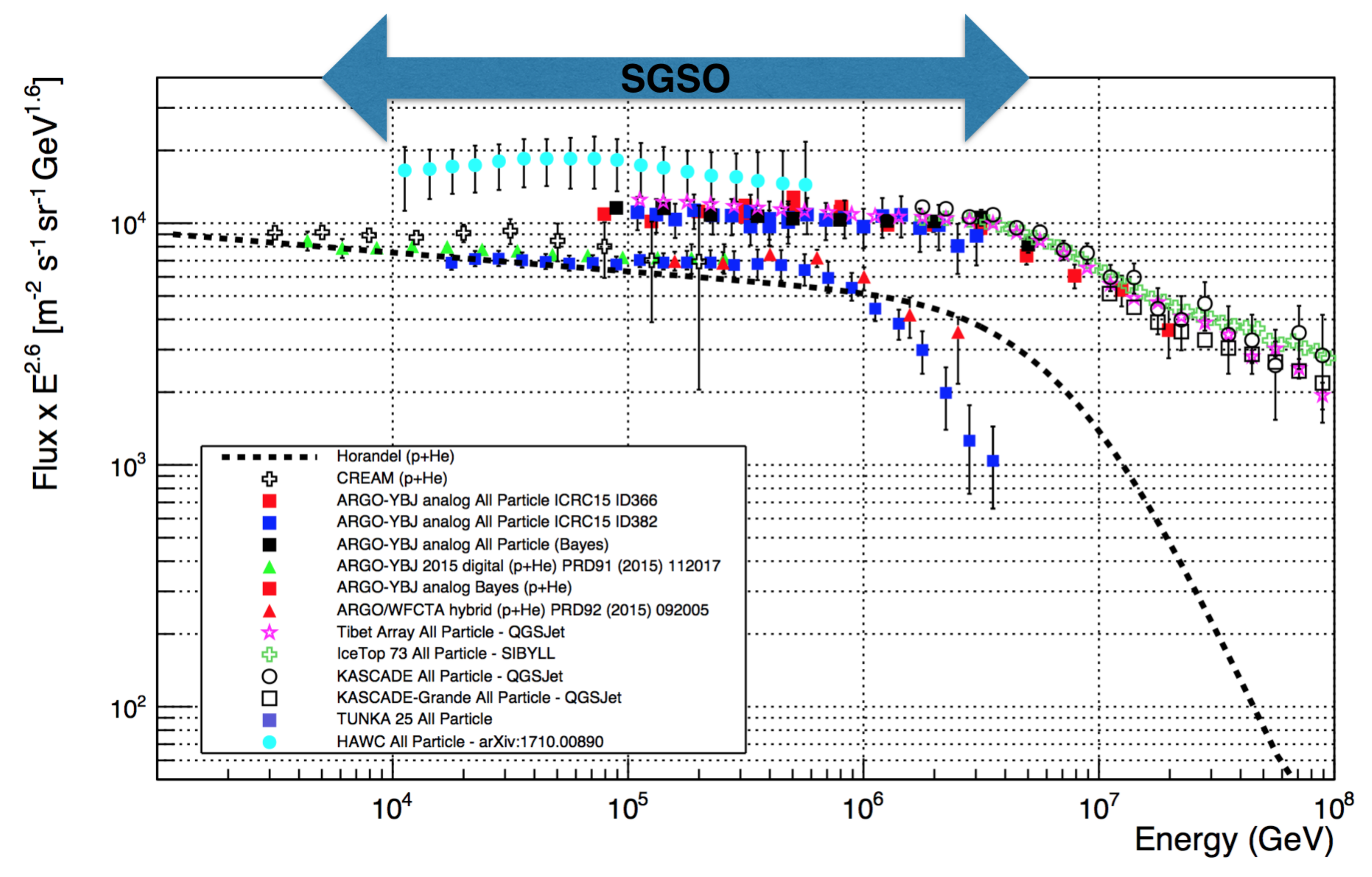}
\caption{All-particle and light (p+He) component energy spectra of primary CR measured by ARGO-YBJ and compared to different experimental results. The parametrization provided by H\"orandel \cite{horandel} for the spectrum of the light CR component is shown for comparison. The systematic uncertainty is shown by the error bars.}\label{fig:argo-phe-knee}
\end{figure*}

Recent results obtained by the ARGO-YBJ experiment (located at 4300 m a.s.l.) clearly show, with different analyses, that the knee of the light component starts at $\sim$700 TeV, well below the knee of the all-particle spectrum that is confirmed by ARGO-YBJ at $\sim$4$\times$10$^{15}$ eV \cite{hybrid15} (see Figure \ref{fig:argo-phe-knee}). 
The observation of a He energy spectrum harder than the proton one, reported by PAMELA \cite{pamelahe}, CREAM \cite{creamhe1,creamhe2} and AMS-02 \cite{ams02he3,ams02he2,ams02he1}, has the interesting consequence that the knee region could be dominated by He and CNO masses with a proton knee below one PeV, as suggested by the ARGO-YBJ results. High quality experimental observations in the energy range between TeV and several PeV, with mass sensitivity to resolve primary masses, would contribute to fill the gap between direct and indirect cosmic ray experiments. SGSO is a key instrument in this context.

On the other hand, in gamma-ray astronomy, observations of photon spectra from young supernova remnants show a clear cutoff around 10 TeV. If the emission is hadronic, that implies a maximum energy of $\sim$100 TeV for protons accelerated in SNRs \cite{magic-casA}.

More than half a century after the discovery of the knee, experimental results are still conflicting with uncertainties on its origin. This is not surprising for several reasons. The reconstruction of the CR elemental composition is often carried out by means of complex unfolding techniques based on the measurements of electronic and muonic sizes, procedures that heavily depend on the hadronic interaction models. The muonic size is much smaller than the electronic one with a wider lateral distribution, but the total sensitive area of muon detectors is typically only few hundred square meters and, due to the poor sampling, large instrumental fluctuations can be added to the stochastic ones associated with the shower development. In addition, the \emph{`punch-through effect'} due to high-energy secondary electromagnetic particles could heavily affect the measurements. Finally, some arrays have been operated close to the sea level and not in the shower maximum region where fluctuations are smaller and all nuclei produce the same electromagnetic size, and hence where the trigger efficiency is the same for all primary particles.

At higher energies, the KASCADE-Grande, IceTop and Tunka experiments observed a hardening slightly above 10$^{16}$ eV and a steepening at log10(E/eV) = 16.92$\pm$0.10 in the CR all-particle spectrum.
A steepening at log10(E/eV) = 16.92$\pm$0.04 in the spectrum of the electron poor event sample (heavy primaries) and a hardening at log10(E/eV) = 17.08$\pm$0.08 in the electron-rich (light primaries) one were observed by KASCADE-Grande even with modest statistical significance \cite{kascadeg-chiavassa}. The absolute fluxes of CRs with different masses measured by KASCADE-Grande are however strongly dependent on the adopted hadronic interaction models \cite{kascadeg-hadron}, thus requiring new high resolution data to clarify the observations.

The measurement of spectra of five mass groups (p, He, CNO, MgSi, Fe) up to 10$^{17}$ eV should be a high priority for future wide-field of view instruments in order to:
\begin{itemize}
\item measure the maximum acceleration energy in Galactic sources through the determination of the knee energy in the proton primary spectrum;
\item investigate the origin of the knee tracing the different components up to 10$^{16}$ eV and measuring the CR anisotropy as a function of rigidity;
\item search for CR factories through a combined mapping of the photon and proton spectra in different regions of the Galactic plane.
\end{itemize}

In order to outperform the current state-of-the-art and clarify the tension between different measurements in the knee energy region, a future experiment should:
\begin{itemize}
\item be located well above 4000 m a.s.l., to approach the atmospheric depth of maximum development of showers in the PeV energy region. In this way fluctuations are smaller and all nuclei produce the same electromagnetic size. As a consequence, the shower size-energy conversion is mass-independent and the trigger efficiency same for all primary particles.
\item be instrumented with detectors able to measure different shower mass-sensitive observables, as the muon component and the lateral distribution in the shower core region. A detector able to study the core region is important also to investigate the hadronic interaction models in the forward region: the pseudo-rapidity corresponding to a region within about 10 m from the shower core is greater than 8;
\item have an energy threshold for CRs of order of TeV for absolute energy calibration purposes exploiting the \emph{`Moon shadow'} technique and to superimpose the measured spectra with direct observations over a wide energy range;
\item have a sufficiently large instrumented area to enable sensitive measurements up to 10$^{16}$ eV. 
\end{itemize}

While conceived primarily as a high-energy gamma-ray observatory, SGSO will fulfill most of these requirements. SGSO will thus be able to collect high-quality cosmic ray data and contribute significantly to the elucidation of the outlined open questions in CR physics.

\section{Anisotropy} 
\label{subsec:cranis}
During the last several decades, a number of observations have provided long-term, statistically significant evidence of a faint anisotropy in the arrival direction distribution of cosmic rays from tens of GeV to over PeV in particle energy (see, e.g.,~\cite{disciascio}). A strong anisotropy was also observed at energy in excess of 10$^{19}$ eV~\cite{auger, ta}, thought to be linked to extragalactic cosmic rays. Large ground-based experiments have contributed to the extension of these observations up to the knee region in the cosmic ray spectrum~\cite{disciascio} and to the indirect probe of possible nearby sources and of propagation properties~\cite{ahlers}. The energy dependency and complex angular structure of the cosmic ray anisotropy hint at complex overlapping processes shaping the particle distribution during their journey to Earth. Disentangling and understanding these processes may prove useful in probing the properties of our local environment and, therefore, understanding diffusion across the interstellar medium.

Cosmic ray anisotropy observations by individual experiments, however, are restricted by limited sky coverage, leading to undesirable biases in the determination of their arrival direction distribution on the sky, especially at angular scales larger than the field of view of the instrument. As a consequence, the angular power spectrum of the anisotropy obtained from any one measurement displays a systematic correlation between different spherical harmonic multipole modes $C_\ell$. Such correlations are eliminated if full sky observations are carried out. In this case, it is possible to better constrain the angular structures with $\ell \leq 3$  of the spherical harmonic expansion, by reducing correlations from degeneracy between $a_{\ell m}$ coefficients.

\begin{figure*}[t!]
\label{fig:hawc-ic}
\centering
\includegraphics[scale=0.7]{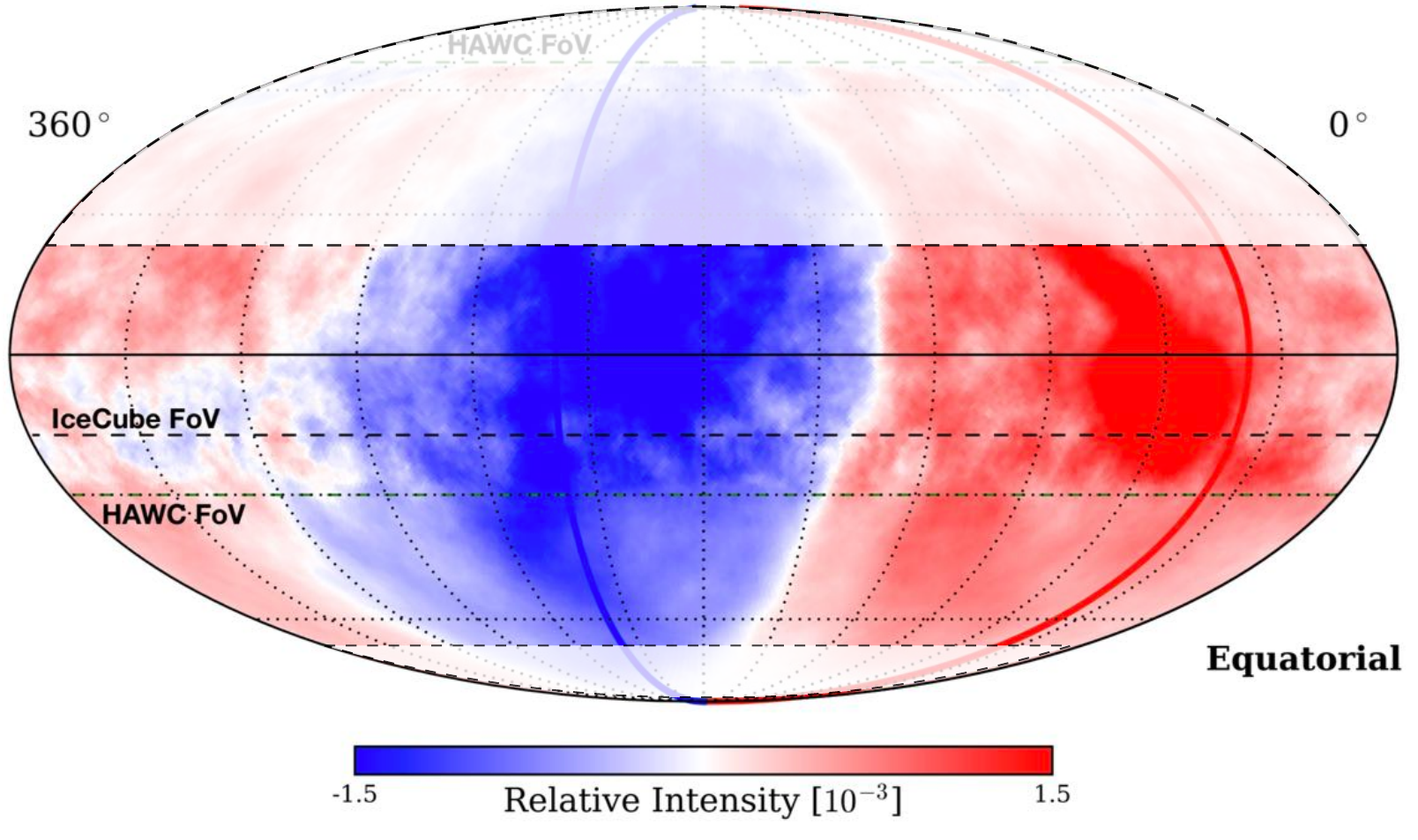}
\caption{Relative intensity of cosmic-ray arrival directions for the combined IC86 and HAWC-300 dataset~\cite{hawc-icecube-cra}, the unmasked region indicates the declination range of SGSO and illustrates the overlap with both IceCube in the south and HAWC in the north.}
\end{figure*}
The first full sky map of 10 TeV cosmic ray anisotropy was obtained by combining HAWC and IceCube data~\cite{hawc-icecube-cra}, even though the individual experiments have different sensitivities in energy and composition (see figure~\ref{fig:hawc-ic}). While IceCube is only sensitive to deep penetrating muons from cosmic-ray showers, HAWC is more sensitive to the electromagnetic component of the shower. A study with similar detection methods in both the North and South would eliminate most of the systematic uncertainties related to these different techniques. The discrepancy in energy response also restricts the study to a single energy range that overlaps between the two observatories. Such combined analysis provides an experimental means to eliminate biases affecting the individual observations, and a new probe into the physical processes that shape the arrival direction distributions of cosmic rays on Earth. With a large Southern ground-based observatory similar to HAWC and with wide cosmic ray energy sensitivity, it would be possible to provide full sky maps of combined data at different energies. A significant boost in science reach is expected to derive from observations of anisotropy as a function of cosmic ray particle rigidity, from 1 TV to several hundreds of TV. 

An important test is to perform a comparative measurement in the overlapping field of view region for the two experiments. However in the HAWC-IceCube study, this region is very small ($\sim 2^\circ$) as the quality of the reconstruction quickly decreases towards the edge of the exposure region. A larger overlapping FoV would be advantageous with a Southern observatory located closer to the equator.

Recent observations from IceCube~\cite{icecube} and Tibet array~\cite{tibet2} shows that the observed anisotropy changes topology in the range of 100-300 TeV particle energy. A Southern hemisphere based experiment would complement the results currently available by IceCube with coverage across the equator. This would provide a clear and complete view of the dip feature observed by IceCube above hundreds of TeV at a right ascension of 2-6 hrs near the celestial equator. The understanding of the high-energy anisotropy pattern will provide a solid basis for probing into its origin, such as the propagation properties in the turbulent interstellar magnetic field.

On the one hand, such a change must be partially associated to the weakening of heliospheric effects above a rigidity of 10 TV (where the particle gyro-radius is about the size of the heliosphere). On the other hand, a modification in the density gradient and pitch angle distributions of cosmic rays in the interstellar medium must occur as well to explain the observations. Improving the quality of cosmic ray observations in energy, composition, and direction will provide a powerful probe into the understanding of the heliospheric and interstellar medium influence~\cite{helio,ming}, and, ultimately, into the understanding of diffusion processes in magnetic turbulence and the effects of large scale coherent magnetic structures affecting the local cosmic ray observations~\cite{giacinti}. Correlation between anisotropy and spectral anomalies of cosmic rays is a key aspect of such a program.

\section{Electron spectrum and anisotropy}
Unlike hadronic cosmic rays, cosmic ray electrons suffer from strong energy losses as they travel through the interstellar medium, severely limiting their range to just a few hundred parsecs above 1~TeV. As the highest-energy electrons can only travel short distances, this affords us a unique opportunity to study the particle acceleration originating only from the most local cosmic ray accelerators. Current measurements show a steep E$^{-3}$ spectrum, with a steepening to around E$^{-3.8}$ at 1~TeV reaching to around 20~TeV \cite{2017Natur.552...63D, PhysRevLett.120.261102, HESS_electrons}.
As electron induced air showers behave almost identically to those induced by gamma rays, VHE gamma-ray observatories provide an excellent resource for their study, as proven by the measurements of the current generation of IACTs in the 500~GeV - 20~TeV energy range. With its wide aperture and its almost 100\% duty cycle, SGSO will be able to gather significantly larger electron event statistics than Cherenkov telescopes, which may help in extending the electron spectrum further than the 20~TeV already measured, potentially revealing more details of the local source distribution. Additionally, the large sky coverage of SGSO would enable the search for anisotropy in the electron arrival direction at different angular scales, which would provide additional very relevant information for the understanding of Galactic particle acceleration and cosmic ray transport. So far, anisotropy has only been constrained at sub-TeV electron energy by \emph{Fermi}-LAT \cite{2017PhRvL.118i1103A} where most likely multiple sources contribute to the total flux. SGSO will probe the multi-TeV region, where the flux might be dominated by a single local source.  
The major difficulty in the detection of cosmic ray electrons originates from their flux, which is about a factor of 100 lower than the hadronic cosmic ray flux, as well as their diffuse nature. Through a variety of different ways like increasing the array size in comparison to current arrays, increasing the array altitude or the introducing of dedicated muon detection instrumentation (cf. Sec.~\ref{sec:design}), SGSO will provide optimized separation power and thus be able to provide the necessary performance for these measurements. 

\section{Space weather and heliospheric physics}
Space weather involves several physical domains, with special interest on fluxes of energetic particles in the terrestrial space environment. The global structure of the heliosphere and the typical turbulent diffusion conditions, both determining propagation of galactic particles in the solar system, significantly change along the solar cycle on time scales of years to decades.
These changes produce the well-known anti-correlation between the solar activity and galactic cosmic ray flux observed at ground level.

Transient solar magnetic activity can produce strong perturbations to the solar wind conditions. One specific type of solar transient, the so-called Coronal Mass Ejection (CME), is linked to large amounts of magnetized plasma ejected from the solar corona and usually accompanies a solar flare. The interplanetary manifestation of a CME is called Interplanetary CME (ICME). When a CME is launched in the direction of the Earth, its ICME counterpart interacts with the geomagnetic field and can cause perturbations in the geo-space, with possible consequences on technologies, such as interfering with the correct interpretation of signals from the Global Navigation Satellite System (GNSS), satellite operation, and electric power grids on the ground. The propagation and evolution of a CMEs/ICMEs have been predicted using simulations based on coronal magnetic field models and/or Magneto-Hydro-Dynamics (MHD) \cite{Shiota10}.  This kind of numerical simulation plays a very important role for short and long-term space weather forecasts.

These perturbations can also significantly affect the transport of low-energy Galactic cosmic rays in the interplanetary medium in shorter time scales (hours-days). These interplanetary transients of solar origin can change the magnetic topology-connectivity and turbulent properties of the solar wind at meso-scales, and as consequence of these changes a significant decrease of flux of cosmic rays can be produced. When these decreases are observed at ground level, they are called Forbush decreases \cite{Forbush46}. Ground level neutron monitors have been systematically observing variations of galactic cosmic rays as consequence of these interplanetary transients (at both, large and small scales) since the 1960s, counting secondary neutrons produced in the atmospheric shower.

In 2011, it was shown that short term interplanetary modulation can be also observed from low energy modes of particle detectors like water Cherenkov detectors (WCDs). It was found that the time profiles of ground level fluxes correlate well with neutron monitor observations \cite{Auger_scalers_11}. One major advantage of WCDs with respect to neutron monitors is that WCDs can observe fluxes of different secondary particles. A comparison between neutron monitors and two particle types (electromagnetic and muons) has been done during a Forbush decrease~\cite{Asorey_icrc_lago_15}.

On another hand, the Sun blocks a part of isotropic galactic cosmic rays and casts a shadow in the cosmic-ray intensity on the Earth, which is called the ``Sun shadow" ~\cite{Clark57}.  The Tibet air shower array observed a clear solar-cycle variation of the Sun shadow at 10~TeV \cite{Amenomori13}.  The intensity deficit of the Sun shadow is sensitive to the coronal magnetic field structure, which varies on an 11-year cycle. Therefore, the measurement of the variation of the intensity deficit provides a powerful tool for evaluating the solar coronal magnetic field quantitatively. On the other hand, the center of the Sun shadow is displaced from the optical center of the Sun, since the interplanetary magnetic field (IMF) between the Sun and the Earth deflects TeV cosmic rays following the simple Lorentz force law.  This displacement of the Sun shadow can be used to measure the strength/direction of the IMF \cite{Aielli11,Amenomori18a}. Recently, the experimental evidence of Earth-directed CMEs affecting the Sun shadow was found in a few TeV energy region \cite{Amenomori18b}.
 
Illustrating the techniques, HAWC was able to derive the energy dependence and solar cycle variation of the Sun shadow in the 2-50~TeV energy region~\cite{Enriquez15a}. Since lower-energy ($<$TeV) cosmic rays are more sensitive to the IMF, SGSO will be able to significantly expand these observations and be thus able to provide very useful input to space weather sciences. While the Sun-Earth transit time of a CME is 2-5 days after the solar eruption, cosmic rays take only 8 minutes to reach the Earth. Therefore, if we can resolve individual CMEs on a daily or hourly basis, the Sun shadow will be applicable to space weather forecasts.  In addition, HAWC has already shown that transient variations of cosmic-ray intensities in the heliosphere can be detected, such as Ground Level Enhancements (which corresponds to increase of the flux of cosmic rays of solar origin that can be observed at ground level) and Forbush decreases, by using a dedicated data taking mode (e.g. single-counting scaler mode)~\cite{Enriquez15b}. These phenomena are also thought to be reliable indicators of a geomagnetic storm. The employed analysis techniques will be available within SGSO, which will thus be able to provide crucial information on these interesting and potential harmful events.

In order to monitor such transient phenomena, a continuous observation is crucial. Having detectors in both hemispheres enables us to observe the Sun at a high elevation throughout the year. In the southern hemisphere, IceCube at the South Pole has successfully observed the temporal variation of the Sun shadow between 2011 and 2015 at a mean energy of $\sim$40~TeV, which is much higher energy as compared with the Tibet air shower array and HAWC \cite{Aartsen18}. Low-energy-threshold cosmic-ray detectors in the southern hemisphere and especially SGSO will be able to contribute to space weather sciences together with northern ground arrays in operation (e.g. HAWC and Tibet) or under construction (e.g. LHAASO).

Several open questions with respect to the specific physical mechanisms affecting the transport and modulation of galactic cosmic rays in the heliosphere have to be answered in the next years. These mechanisms strongly depend on the energy. The lower energy threshold of SGSO will therefore allow for significant progress on these topics.

\cleardoublepage
\chapter{Design Considerations for a Southern Hemisphere VHE gamma-ray Observatory}
\label{sec:design}

\section{Detector Unit Design} 
Given a high-altitude site, the effective energy threshold for gamma-ray sources will likely be dominated by the detector's sensitivity to low-energy shower particles and the hadron rejection efficiency. To achieve precision angular resolution, the cell size and array fill factor need to be optimised and nanosecond-level inter-cell timing needs to be established.

Viable detector technologies include particle counting (scintillators, resistive plate chambers) and calorimetry (WCDs---water Cherenkov detectors instrumented with large-area photo-multiplier tubes). Thin particle counters would likely need to be augmented by a conversion layer for the abundant photons present in the gamma-ray shower front to decrease the overall energy threshold. Further stacking may aid their gamma-hadron separation power. For WCDs, muon tagging and thus gamma-hadron separation may be improved at moderate additional cost by vertical segmentation~\cite{Letessier-Selvon:2014sga} or using a hybrid detector design (see Refs.~\cite{Assis:2016gjo,Takita:2017pem,Thoudam:2017hgj} for current design efforts). Another option may be to replace the outer structure of the WCDs by installing segmented, light-tight enclosures in a natural environment like a high-altitude lake. 

A thorough analysis of the total construction and operation costs will be needed to converge upon a cost-effective detector design reaching the envisaged science goals.

\section{Site Considerations: Latitude and Altitude} 
\label{subsec:site}

Complementing the current and future ground-based gamma-ray observatories using particle detectors, HAWC and LHAASO, both located in the Northern hemisphere, SGSO will be located in the Southern hemisphere to allow for an all-sky coverage. A location in the south also allows to fully exploit the synergies with the southern site of CTA and provides access to the central region of our Galaxy with its wealth of known high-energy gamma-ray sources. Maximizing the exposure to Galactic sources (e.g. having the Galactic Center not far from zenith) and considering a field of view of $\sim 45^{\circ}$ from zenith, latitudes in the range $20^{\circ}-30^{\circ}$ South are desirable for the SGSO site. 

\begin{figure}[!t]
  \begin{center}	
    \includegraphics[width=0.95\linewidth]{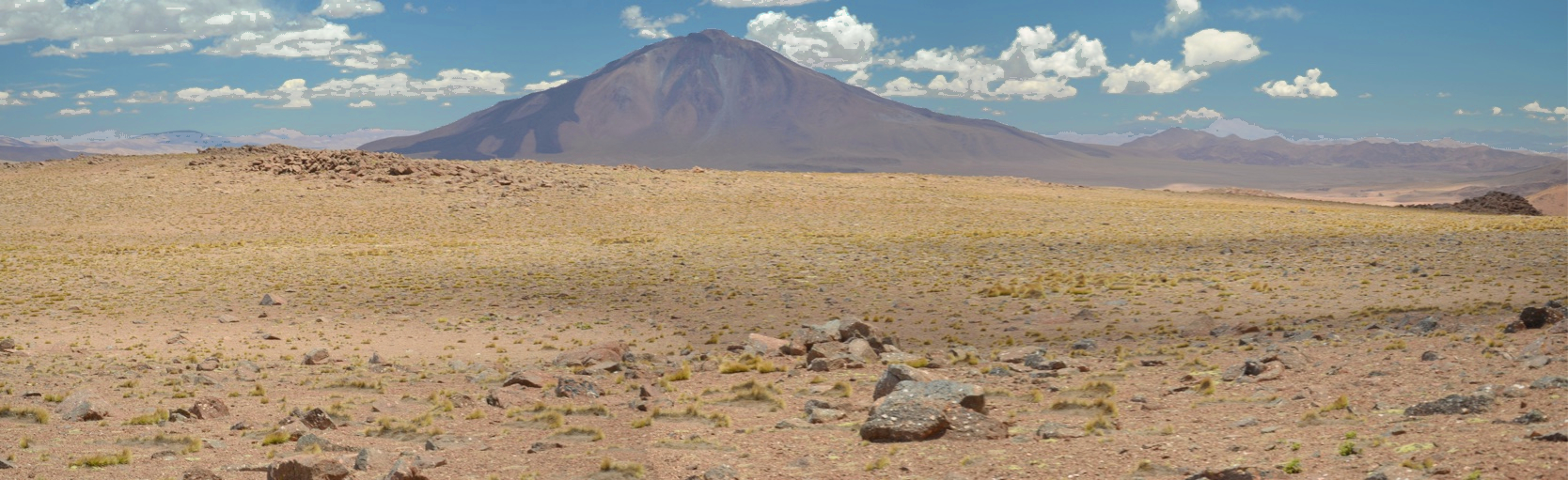}	
   \end{center}
  \caption{Panoramic view of the site in Cerro Vecar, province of Salta, Argentina. The Tuzgle volcano is seen on the background.}
  \label{site:arg}
\end{figure}

\begin{figure}[!t]
  \begin{center}	
   \includegraphics[width=0.47\linewidth]{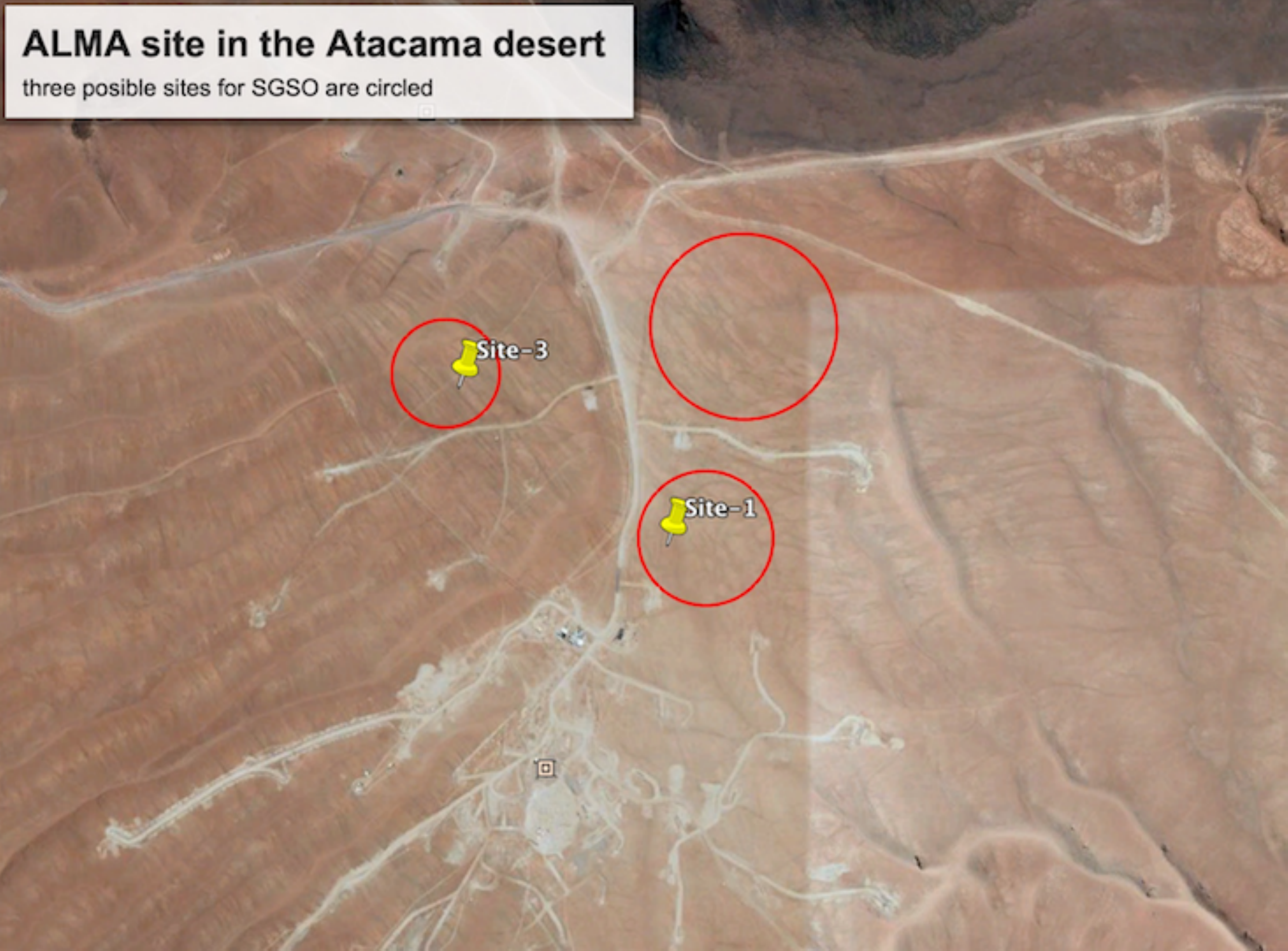}
   \includegraphics[width=0.52\linewidth]{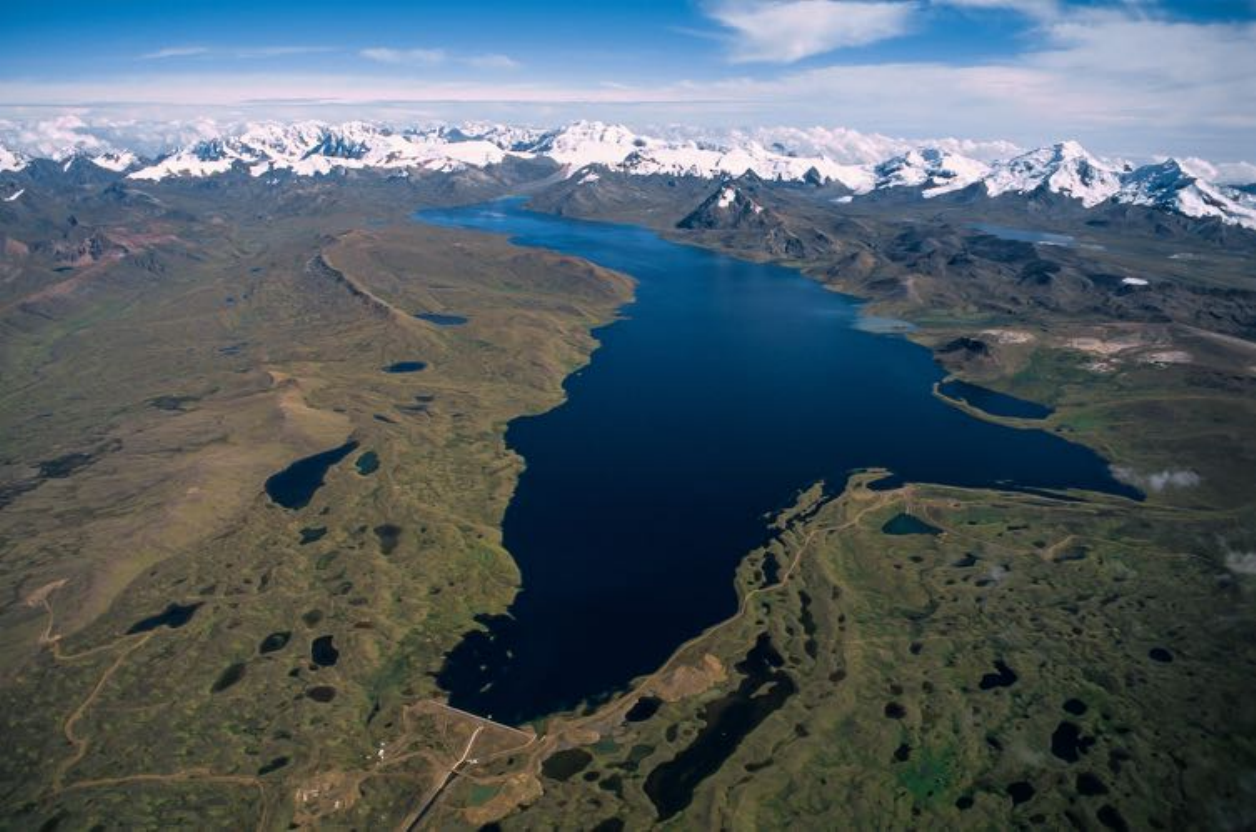}
   \end{center}
  \caption{{\em Left:} Google Earth image of the ALMA site indicating three possible sites for SGSO. {\em Right:} Aerial view of lake Sibinacocha (Peru), one of several high-altitude lakes being considered for SGSO (from \url{www.sibinacocha.org}).}
  \label{site:alma}
\end{figure}

The altitude of the observatory is one of the main variables to improve the detector sensitivity, effective area and to lower the energy threshold. The main effect is the increase in the flux of shower particles at ground level. The benefits of a higher detector site have to be balanced against the increasing difficulties to install and operate an observatory at very high altitudes. For example, low oxygen levels prohibit long-term manual labour. Depending on the availability of the necessary infrastructure and easy site access, one might prepare the detector units at low altitudes and bring them up to the final site largely complete. If relying on the water Cherenkov technique, water availability nearby is desirable as this could be one of the most budget consuming aspects of the observatory construction and freezing of the water tanks may have to be dealt with. The expected high rates at high altitudes have to be taken into account in the design of the readout and trigger electronics and/or the design of the tanks and the array (e.g. a larger number of smaller tanks with local coincidence readouts). These considerations will be discussed in detail during the development and definition of the final technical design of SGSO. Currently sites at altitudes around 5000\,m.a.s.l. are considered. Due to the absence of appropriate mountain ranges with these altitudes in Africa or Australia, several plausible sites in South America are under investigation:

\begin{description}
\item[Sites in Argentina] Several suitable high-altitude locations are available in the NW of the country. One of them is in Cerro Vecar (24:11:04 S, 66:28:32 W), at the site of the radio telescope LLAMA\footnote{\url{https://www.llamaobservatory.org}} and the CMB telescope QUBIC\footnote{\url{http://qubic.in2p3.fr}}, accessible by a wide road. Figure~\ref{site:arg} shows a panoramic view of the site located at 4800 m.a.s.l.. Temperature measurements from a weather station installed on site in the period 2012-2013 show a mean daily temperature distribution centered at $0^\circ$C.
The average during 3 months in Summer is $3^\circ \pm 3^\circ$C, and $-4^\circ \pm 3^\circ$C in Winter. The site provides good access conditions, several nearby water sources and nearby available infrastructure at lower altitudes (San Antonio de los Cobres, 3775\,m.a.s.l.).

\item[Sites in Chile] A suitable site is the plateau of the ALMA radio telescope array (23:01:29 S and 67:45:24 W) at 5000\,m.a.s.l. near San Pedro de Atacama. With a complete infrastructure of wide roads as well as power and optical fiber networks, the extensive plateau provides several potential sites for the installation of SGSO (cf. Figure~\ref{site:alma}). The ALMA Operations and Support Facility (OSF) is located nearby at 3000\,m.a.s.l. The ALMA collaboration is open to the installation of other non-ALMA projects at the site. Large quantities of high-quality water are trucked each day to the OSF and a similar operation can be contemplated to fill the water Cherenkov detectors in case this solution is chosen. Temperature measurements from a weather station installed on site since 2004 show a mean daily temperature distribution centered at $-3^\circ$C. The average during 3 months in Summer is $0^\circ \pm 3^\circ$C and $-7^\circ \pm 3^\circ$C in Winter\footnote{\url{https://almascience.eso.org/about-alma/weather}}.

\item[High altitude lakes] Several suitable lakes at high altitudes have been identified in South America and especially in Peru. Further detailed investigations of them are ongoing. An example is Lake Sibinacocha (cf. Figure~\ref{site:alma}), located in the Cusco Region (Canchis Province) at an altitude of 4.870m and with a size of $2.86 \times 15.2\;\mathrm{km}^2$, largely sufficient for the requirement of SGSO. 
\end{description}

\chapter{Summary} 
The Southern Gamma-ray Survey Observatory (SGSO), a next-generation high-energy gamma-ray observatory in the Southern hemisphere, aims to provide unprecedented observations of high-energy phenomena in the universe. 

SGSO will allow studying particle acceleration in the most violent sources of our Galaxy. It will provide an unbiased and deep survey of the Milky Way at multi-TeV energies, localize sources able to accelerate particles at least up to the knee of the cosmic rays spectrum and study large scale and extended emission throughout our cosmic neighbourhood.

Thanks to its large field-of-view and duty-cycle, the SGSO continuous monitoring capability of the Southern sky will be unrivaled. It will collect long-term measurements of variable high-energy emitters ranging from Galactic binaries to active galactic nuclei and will thus provide crucial input to the understanding the 
underlying phenomena that give rise to gamma-ray emission from these objects. As a truly multi-messenger observatory, SGSO will contribute to our understanding of compact objects via searches for VHE counterparts to gravitational waves and the study of gamma-ray bursts at VHE energies. SGSO will elucidate links between astrophysical sources of high-energy radiation and high-energy neutrinos and play a leading role in searches for VHE emission associated to novel transient phenomena like fast radio bursts.

SGSO will provide new observational impulses for searches of phenomena beyond the Standard Model of particle physics. It will enable detailed searches for dark matter in a large variety of regions like the center of our Galaxy or newly discovered satellite galaxies. SGSO will also enable searches of Lorentz-invariance violating effects, the evaporation of primordial black holes and axion-like particles. 

Complementing the searches for the sources of high-energy cosmic rays using gamma rays as tracers, SGSO will record an enormous amount of high-precision data on hadronic cosmic rays. It will therefore be able to study the cosmic ray energy spectrum in the crucial energy range around the knee and provide novel insights into cosmic ray anisotropy at various angular scales and across a large range of energies. SGSO will help study the Sun and fluxes of energetic particles in the heliosphere. 

SGSO will be a key player in the multi-wavelength and especially the multi-messenger community. Its unique monitoring capabilities will allow to alert observers around the world, across the full electromagnetic spectrum and all known messengers of new detections and phenomena. These alerts will be provided in real-time and will thus allow triggering detailed follow-up observations. While broad-band MWL information will be an important ingredient to SGSO science, the expected performance is to a significant extent complementary to and beneficial for the science of the upcoming Cherenkov Telecope Array (CTA). SGSO will for example act as a high-duty cycle, large FoV finder scope for deep and high-resolution CTA observations for many sources, including transient phenomena in various multi-wavelength and multi-messenger contexts. In addition to providing real-time ``triggers", SGSO may also provide important input for detailed CTA analyses on extended sources like PWNe, their TeV halos, the \emph{Fermi} Bubbles or the diffuse Galactic emission.

The ambitious goals of SGSO will be made possible by important developments and design studies. Various detector and array designs are currently studied with simulations and validated with prototypes. In parallel, several candidates for an optimal site for the future observatory have been identified and are being assessed.

Sensitive to astrophysical gamma rays and cosmic rays in the VHE energy range and combining a large field-of-view with a high duty-cycle, the Southern Gamma-ray Survey Observatory will become a major player within the next-generation observatories for the high-energy astrophysics and astroparticle communities.

\vspace*{10cm}
\section*{Acknowledgements}
\label{sec:Acknowledgements}
\addcontentsline{toc}{chapter}{Acknowledgements}

\input{acknowledgements.tex}


\clearpage

\bibliographystyle{etal}
\addcontentsline{toc}{chapter}{References}
\bibliography{references}

\end{document}

%% file: authors.tex

\begin{center}
A.~Albert$^{1}$,
R.~Alfaro$^{2}$
H.~Ashkar$^{3}$,
C.~Alvarez$^{4}$,{4}
J.\'{A}lvarez$^{5}$,
J.C.~Arteaga-Vel\'{a}zquez$^{5}$,
H.~A.~Ayala~Solares$^{6}$,
R.~Arceo$^{4}$,
J.A.~Bellido$^{7}$,
S.~BenZvi$^{8}$
% ...

\end{center}


\begin{center}
\small
$^{1}$ Physics Division, Los Alamos National Laboratory, Los Alamos, NM, USA \\

$^{2}$ Instituto de F\' isica, Universidad Nacional Aut\'onoma de M\'exico, Circuito de la Investigaci\'on Cient\'ifica, C.U., A. Postal 70-364, 04510 Cd. de M\'exico,  M\'exico\\

$^{3}$ IRFU, CEA, Universit\'e Paris-Saclay, F-91191 Gif-sur-Yvette, France\\

$^{4}$ Facultad de Ciencias en F\'isica y Matem\'aticas, Universidad Aut\'onoma de Chiapas, C. P. 29050, Tuxtla Guti\'errez,  Chiapas, M\'exico\\

$^{5}$ Universidad Michoacana de San Nicol\'{a}s de Hidalgo, Morelia, Michoac\'{a}n, M\'{e}xico\\

$^{6}$ Department of Physics, Pennsylvania State University, University Park, PA, USA\\

$^{7}$ School of Physical Sciences, University of Adelaide, Adelaide, SA 5005, Australia\\

$^{8}$ Department of Physics and Astronomy, University of Rochester, 500 Wilson Boulevard, Rochester NY 14627, USA\\

% ...
\end{center}

%% file: acknowledgements.tex

This work was supported by the Programme National des Hautes Energies of CNRS/INSU with INP and IN2P3, co-funded by CEA and CNES.%
This work was supported by the U.S. Department of Energy Office of High Energy Physics under grant number DE-SC0008475.%
This work was supported by the National Science Foundation, University of Wisconsin institutional funds, and the Wisconsin IceCube Particle Astrophysics Center.%
HMH and VdS acknowledge FAPESP support numbers 2015/15897-1 and 2017/03680-3 and the National Laboratory for Scientific Computing (LNCC/MCTI, Brazil) for providing HPC resources of the SDumont supercomputer (http://sdumont.lncc.br).%
AR acknowledges support from CONICYT (Chile) project Basal AFB-170002, as well as FONDECYT (Chile) projects 1150411 and 1171421.%
DD acknowledges the support from BMBF (Germany) within the project 05A17WW1. %
LN acknowledges support by UNAM PAPIIT IN112218.%
MMG acknowledges support by UNAM PAPIIT IG100317.%
NF acknowledges support by UNAM PAPIIT IA102917 and IA102019.%
ACR acknowledges support from CONICET (Argentina) by a dedicated grant for SGSO 
(RD968; 10-05-2017) and partial support from ANPCyT (Argentina), PICT 2016-3180.